
\input graphicx
\graphicspath{{PDF_files/}}

\input color
\makeatletter

\headline{}



%

%
 
\ifx\tenpoint\undefined\let\loadedfrommacro=Y
\ifx\loadedfrommacro Y\else
         \message{10point.TeX must be loaded from a macro package.}
         \message{Input terminated.}
          \fi
 
\font\tencsc=cmcsc10
 
\newfam\scfam
 
\def\tenpoint{\def\rm{\fam0\tenrm}
    \textfont0=\tenrm  \scriptfont0=\sevenrm  \scriptscriptfont0=\fiverm
    \textfont1=\teni   \scriptfont1=\seveni   \scriptscriptfont1=\fivei
    \textfont2=\tensy  \scriptfont2=\sevensy  \scriptscriptfont2=\fivesy
    \textfont3=\tenex  \scriptfont3=\tenex    \scriptscriptfont3=\tenex
    \textfont\itfam=\tenit   \def\it{\fam\itfam\tenit}%
    \textfont\slfam=\tensl   \def\sl{\fam\slfam\tensl}%
    \textfont\ttfam=\tentt   \def\tt{\fam\ttfam\tentt}%
    \textfont\bffam=\tenbf   \scriptfont\bffam=\sevenbf
    \scriptscriptfont\bffam=\fivebf  \def\bf{\fam\bffam\tenbf}%
    \textfont\scfam=\tencsc  \def\sc{\fam\scfam\tencsc}%
    \normalbaselineskip=12pt
    \setbox\strutbox=\hbox{\vrule height8.5pt depth 3.5pt width0pt}%
    \normalbaselines\rm}

         \let\loadedfrommacro=N\fi
 
\font\ninerm=cmr9            \font\sixrm=cmr6
\font\ninei=cmmi9            \font\sixi=cmmi6
\font\ninesy=cmsy9           \font\sixsy=cmsy6
\font\ninebf=cmbx9           \font\sixbf=cmbx6
\font\ninesl=cmsl9           \font\ninett=cmtt9      \font\nineit=cmti9
\font\ninecsc=cmcsc10
 
\ifx\ninepoint\undefined
   \def\ninepoint{\def\rm{\fam0\ninerm} 
       \textfont0=\ninerm  \scriptfont0=\sixrm  \scriptscriptfont0=\fiverm
       \textfont1=\ninei   \scriptfont1=\sixi   \scriptscriptfont1=\fivei
       \textfont2=\ninesy  \scriptfont2=\sixsy  \scriptscriptfont2=\fivesy
       \textfont3=\tenex   \scriptfont3=\tenex  \scriptscriptfont3=\tenex
       \textfont\itfam=\nineit   \def\it{\fam\itfam\nineit}%
       \textfont\slfam=\ninesl   \def\sl{\fam\slfam\ninesl}%
       \textfont\ttfam=\ninett   \def\tt{\fam\ttfam\ninett}%
       \textfont\bffam=\ninebf   \scriptfont\bffam=\sixbf
        \scriptscriptfont\bffam=\fivebf   \def\bf{\fam\bffam\ninebf}%
       \textfont\scfam=\ninecsc  \def\sc{\fam\scfam\ninecsc}%
       \normalbaselineskip=11pt
       \setbox\strutbox=\hbox{\vrule height8pt depth3pt width0pt}%
       \normalbaselines\rm}
   \fi

 
\ifx\tenpoint\undefined\let\loadedfrommacro=Y
\ifx\loadedfrommacro Y\else
         \message{10point.TeX must be loaded from a macro package.}
         \message{Input terminated.}
          \fi
 
\font\tencsc=cmcsc10
 
\newfam\scfam
 
\def\tenpoint{\def\rm{\fam0\tenrm}
    \textfont0=\tenrm  \scriptfont0=\sevenrm  \scriptscriptfont0=\fiverm
    \textfont1=\teni   \scriptfont1=\seveni   \scriptscriptfont1=\fivei
    \textfont2=\tensy  \scriptfont2=\sevensy  \scriptscriptfont2=\fivesy
    \textfont3=\tenex  \scriptfont3=\tenex    \scriptscriptfont3=\tenex
    \textfont\itfam=\tenit   \def\it{\fam\itfam\tenit}%
    \textfont\slfam=\tensl   \def\sl{\fam\slfam\tensl}%
    \textfont\ttfam=\tentt   \def\tt{\fam\ttfam\tentt}%
    \textfont\bffam=\tenbf   \scriptfont\bffam=\sevenbf
    \scriptscriptfont\bffam=\fivebf  \def\bf{\fam\bffam\tenbf}%
    \textfont\scfam=\tencsc  \def\sc{\fam\scfam\tencsc}%
    \normalbaselineskip=12pt
    \setbox\strutbox=\hbox{\vrule height8.5pt depth 3.5pt width0pt}%
    \normalbaselines\rm}

         \let\loadedfrommacro=N\fi
 
\font\eightrm=cmr8           \font\sixrm=cmr6
\font\eighti=cmmi8           \font\sixi=cmmi6
\font\eightsy=cmsy8          \font\sixsy=cmsy6
\font\eightbf=cmbx8          \font\sixbf=cmbx6
\font\eightsl=cmsl8          \font\eighttt=cmtt8    \font\eightit=cmti8
\ifx\eightpoint\undefined
   \def\eightpoint{\def\rm{\fam0\eightrm} 
       \textfont0=\eightrm  \scriptfont0=\sixrm  \scriptscriptfont0=\fiverm
       \textfont1=\eighti   \scriptfont1=\sixi   \scriptscriptfont1=\fivei
       \textfont2=\eightsy  \scriptfont2=\sixsy  \scriptscriptfont2=\fivesy
       \textfont3=\tenex    \scriptfont3=\tenex  \scriptscriptfont3=\tenex
       \textfont\itfam=\eightit   \def\it{\fam\itfam\eightit}%
       \textfont\slfam=\eightsl   \def\sl{\fam\slfam\eightsl}%
       \textfont\ttfam=\eighttt   \def\tt{\fam\ttfam\eighttt}%
       \textfont\bffam=\eightbf   \scriptfont\bffam=\sixbf
        \scriptscriptfont\bffam=\fivebf   \def\bf{\fam\bffam\eightbf}%
       \normalbaselineskip=10pt
       \setbox\strutbox=\hbox{\vrule height7.5pt depth2.5pt width0pt}%
       \normalbaselines\rm}
   \fi

\definecolor {red}{rgb}{0.5,0,0.0}
\definecolor {red}{rgb}{0,0,0.0}
\definecolor {blue}{rgb}{0,0,0.0}
\def \eitwo         {1}
\def \eiitwo        {2}
\def \eiifour       {3}
\def \eiisix        {4}
\def \eiieight      {5}
\def \eiiten        {6}
\def \eiitwelve     {7}
\def \eiifourteen   {8}
\def \eiififteen    {9}
\def \eiisixteen    {10}
\def \eiitwentyeight{11}
\def \eiiifour      {12}
\def \eiiisix       {13}
\def \eiiiten       {14}
\def \eiiitwelve    {15} 
\def \eiiifourteen  {16}
\def \eiiisixteen	{17}
\def \eivtwo       	{18}
\def \eivthree     	{19}
\def \eivthreeA		{20}
\def \eivthreeB		{21}
\def \eivthreeBB	{22}
\def \eivthreeC     {23}
\def \eivthreeD     {24}
\def \eivthreeE     {25}
\def \eivthreeF     {26}
\def \eivthreeG     {27}
\def \eivthreeH     {28}
\def \eivthreeI     {29}
\def \eivfour       {30}
\def \eivsix        {31}
\def \evsix         {32} 
\def \evten         {33}
\def \evtwelve      {34}
\def \evfourteen    {35}
\def \evtwenty      {36}
\def \evtwentytwo   {37}
\def \evtwentyfour  {38}
\def \evtwentysix   {39}
\def \evtwentysixa  {40}
\def \evtwentysixb  {41}
\def \evtwentysixc  {42}
\def \evtwentysixd  {43}
\def \evthirty      {44}
\def \evthirtytwo   {45}
\def \evthirtyfour  {46}
\def \evthirtyeight {47}
\def \evfourtytwo   {48}
\def \evfourtyfour  {49}
\def \evfourtysix   {50}
\def \evfifty       {51}
\def \evfiftyA       {52}
\def \evfiftyone    {53}
\def \evfourtyeight {54}
\def \evfiftytwo    {55}
\def \evfiftyfour    {56}
\def \evfiftysix    {57}
\def \eviten        {58}
\def \eDtwo			{59}
\def \eDthree		{60}
\def \eDfour		{61}
\def \eDfourA		{62}
\def \eDfive		{63}
\def \eDfiveA		{64}
\def \eDfiveB		{65}
\def \eDfiveD		{66}
\def \eDfiveE		{67}
\def \eDfiveF		{68}
\def \eDsix			{69}
\def \eDseven		{70}
\def \eDeight		{71}
\def \eFastB		{72} 
\def \eDnine		{73}
\def \eDnineA		{74} 
\def \eDten			{75}
\def \eDtenA		{76}
\def \eDsixteen		{77}
\def \eDeighteen	{78}
\def \eDtwenty		{79}
\def \eDtwentytwo	{80}
\def \eDtwentyfour	{81}
\def \eDeleven		{82}
\def \eDelevenA		{83}
\def \eDtwelve		{83}
\def \eDfourteen	{85}

\def \eVIIIii		{86}
\def \eExactPu		{87} 
\def \eAm			{88} 
\def \eBm			{89} 
\def \eEstimatePu	{90} 

\def \evithirtyone  {91}
\def \evithirtytwo  {92}
\def \evithirtysix 	{93}
\def \evithirtyeight{94}
\def \evifourty		{95}

\def \fiitwo    {~1}
\def \fiithree  {~2}
\def \fiifour   {~3}
\def \fiifive   {~4}
\def \fiiitwo   {~5}
\def \fivone    {~6} 
\def \fAtwo   	{~7}
\def \fAfour  	{~8}
\def \fivtwo    {~9}
\def \fvtwo     {~10}
\def \fAone   	{~11}
\def \fAthree 	{~12}
\def \fvfour    {~13}
\def \fvfive    {~14}
\def \fAtwop   	{~15}
\def \fAfourp  	{~16}
\def \fAfive  	{~17}
\def \fvsix     {~18}
\def \fAonni   {~19}
\def \fAtwze    {~20}
\def \fvtwelve  {~21}
\def \fvfourteen{~22}
\def \fAsix   	{~23}
\def \fSawadaTable {~24} 		
\def \fvitwo    {~25}
\def \fvifour   {~26}
\def \fDfour    {~27}			
\def \fvisix    {~28}			

\def	\fweightEqualSix	{~29}	
\def	\fCCodeCRCxvi	{~30}	
\def	\fCCodeCRCThirtytwo	{~31}	

\def	\fCCodeOptimalThirtytwoBitwise	{~32}	

\def	\fCCodeOptimalThirtytwoTable	{~33}	

\def	\fweightGreaterSix	{~34}	
\def	\fweightEqualFive	{~35}	
\def	\fweightGreaterFive	{~36}	
\def	\fweightEqualFivePrimitive	{~37}	
\def	\fweightEqualEight	{~38}	
\def	\fweightGreaterEight	{~39}	

\def	\fweightDistCRCeight		{~40} 
\def	\fweightDistCRCsixteen		{~41} 
\def	\fweightDistCRCthirtytwo	{~42} 
\def	\fweightDistCRCsixtyfour	{~43} 

\def \rtwo      	{~1}
\def \rfour     	{~2}
\def \rsix     		{~3}
\def \rseven     	{~4}
\def \reight		{~5} 
\def \rnine     	{~6}
\def \rten      	{~7}
\def \releven   	{~8}
\def \rtwelve   	{~9}
\def \rfourteen 	{~10}
\def \rSawadaTable 	{~11}
\def \rfourteenA 	{~12}
\def \rfourteenB 	{~13}
\def \rSawadaISO 	{~14}
\def \rfifteen  	{~15} 
\def \rIIii  		{~16}
\def \rIIiv  		{~17} 
\def \rIIvi  		{~18}
\def \rIIviii  		{~19}
\def \rsevenA     	{~20}

\def \Ber{1}	
\def \BiR{2}
\def \Cha{3}
\def \Far{4}
\def \Fel{5}
\def \Fle{6}
\def \JiK{7}
\def \KlK{8}
\def \Knu{9}
\def \Koo{10}
\def \KoB{11}
\def \LiC{12}
\def \MaS{13}
\def \McA{14}
\def \Ngu{15}
\def \NguA{16}
\def \Per{17}
\def \RaG{18}
\def \RaK{19}
\def \Sar{20}	
\def \SGP{21}
\def \Zie{22}

\def\monthname#1{\ifcase#1 \errmessage{0 is not a month}
    \or January\or February\or March\or April\or May\or June\or 
    July\or August\or September\or October\or November\or
    December\else \errmessage{#1 is not a month}\fi}
    

\def \MOD #1#2{{\rm R}_{#2}\left[#1\right]} 

\def \frac#1#2{{#1\over#2}} 

\def \QED{\nobreak \hfill{$\sqcup \kern -0.66em  \sqcap$}\medskip}


\def \section #1#2 
{\goodbreak
\bigskip
\par 
\parindent = 0ex 
{%
\sectionfont
\setbox0=\hbox{#1} \dimen0=\wd0 
\setbox1=\hbox{\ \ \ } \dimen1=\wd1 
\advance \dimen0 by \dimen1 
\hangindent=\dimen0 
\rightskip=0pt plus 1fill
\leftskip=0pt 
{\hyphenpenalty=99999 \uppercase{#1}\ \ \ {\ignorespaces#2}} 
\medskip \nobreak \ignorespaces 
} %
\parindent = 0.2in 
\noindent
}
\def \subsection #1#2 
{\goodbreak
\bigskip
\par 
\parindent = 0ex 
{%
\sectionfont
\setbox0=\hbox{#1} \dimen0=\wd0 
\setbox1=\hbox{\ \ \ } \dimen1=\wd1 
\advance \dimen0 by \dimen1 
\hangindent=\dimen0 
\rightskip=0pt plus 1fill
\leftskip=0pt 
{\hyphenpenalty=99999 {#1}\ \ \ {\ignorespaces#2}} 
\parindent = 0.2in
\par
\rightskip=0pt 
\medskip \nobreak \ignorespaces \noindent 
}%
\parindent = 0.2in
\noindent
}

\def \subsubsection #1#2
{\goodbreak
\bigskip
\noindent
{{\subsectionfont {#1\ \ \ }{\ignorespaces#2}}}
\smallskip \nobreak \noindent
}
 
\def \remark #1
{
\smallskip
\noindent
{\bf #1}
\nobreak
}

\def \figure #1#2#3
{
\bigskip
\setbox0=\vbox
{%
\hsize=\figureboxsize
\centerline{#1}
\smallskip 
\noindent
\rightskip=0pt plus 1fil
\leftskip=\rightskip
{\FigureCaptionFont {Fig.#2\ \ \ }{\ignorespaces#3}}
\parfillskip=0pt plus 0fil
}%
\box0
\par
\rightskip=0in
\leftskip=0in
\parfillskip=0pt plus 1fil
\smallskip
}

\tolerance = 10000
\def \figureFont{\ninepoint}

\font \titlefont=cmssbx10 at 18pt 
\font \sectionfont=cmssbx10 at 10pt
\font \subsectionfont=cmssi10 at 10pt
\font \FigureCaptionFont=cmss10 at 10pt
\font \FigureCaptionFont=cmss10 at 9pt
\newdimen \figureboxsize
\figureboxsize=\hsize

\newcount\secI
\newcount\secII
\newcount\secIII
\newcount\secIV
\newcount\secV

\newcount\secA
\newcount\secB
\newcount\secC
\newcount\secD
\newcount\secE
\newcount\ref

\baselineskip=15 true pt
\baselineskip=11 true pt


\centerline {\titlefont  Fast CRCs } \medskip
\centerline {\sectionfont (Extended Version) }
\bigskip
\centerline{Gam D. Nguyen}
\medskip
\centerline{Naval Research Laboratory}
\centerline{Washington DC 20375}
\bigskip

\remark{Abstract.} CRCs have desirable properties for effective error detection. But their software implementation, which relies on  many steps of the polynomial division, is typically slower than other codes such as weaker checksums. A relevant question is whether there are some particular CRCs that have fast implementation. In this paper, we introduce such fast CRCs as well as an effective technique to implement them. For these fast CRCs, even without using table lookup, it is possible either to eliminate or to greatly reduce many steps of the polynomial division during their computation. 

\remark{Index Terms.} Fast CRC, low-complexity CRC, checksum, error-detection code, Hamming code, period of polynomial, fast software implementation.

\section {1}{INTRODUCTION} \secI = \pageno
This paper considers cyclical redundancy checks (CRCs), which are effective for detecting errors in communication and computer systems. An $h$-bit CRC is typically generated by a binary polynomial of the form
$$
M(X) = (X+1)M_1(X) \eqno (\eitwo) 
$$
where $M_1(X)$ is a primitive polynomial of degree $h-1$. Existing CRCs include the CRC-16 generated by $X^{16}+X^{15}+X^2+1 = (X+1)(X^{15}+X+1)$, and the CRC-CCITT generated by $X^{16}+X^{12}+X^5+1 = (X+1)(X^{15}+X^{14}+X^{13}+X^{12}+X^4+X^3+X^2+X+1)$. 

The CRC generated by (\eitwo) has the following desirable properties: (a) its maximum length is $2^{h-1}-1$ bits, (b) its minimum distance is $d=4$, i.e., all single and double errors are detected, (c) its burst-error detecting capability is $b=h$, i.e., all error bursts of length up to $h$ bits are detected, and (d) its codewords have even weights, i.e., all odd numbers of errors are detected. These properties are called the {\sl guaranteed} error-detecting capability. The CRC may detect other errors, but not guaranteed, e.g., it can detect a large percentage of error bursts of length greater than $h$ [\BiR, \LiC, \RaG]. {\color{red}General theory and applications of error-detection codes are presented in [\KlK].}

General-purpose computers and compilers are increasingly faster and more sophisticated. Software algorithms are commonly used in operations, modeling, simulations, and performance analysis of systems and networks. CRC implementation in software is desirable, because many computers do not have hardware circuits dedicated for CRC computation. However, software implementation of typical CRCs is slow, because it relies on  many steps of the polynomial division during CRC computation. It is this speed limitation of CRCs that leads to use of checksums (which are fast and typically do not rely on table lookup) as alternatives to CRCs in many high-speed networking applications, although checksums are weaker than CRCs. For example, the 16-bit one's-complement checksum is used in Internet protocol and the Fletcher checksum is used in ISO [\Fle, \SGP]. There are also other fast error-detection codes [\Far, \Fel, \McA, \Ngu], but they do not have all the desirable properties of CRCs.

A relevant question is whether there is a new family of CRCs that are faster the existing CRCs. In this paper, we introduce such CRCs, as well as a technique for their efficient implementation. For these fast CRCs, it is possible either to eliminate or to greatly reduce many steps of the polynomial division during their computation.  

A common existing technique for reducing the many steps during CRC computation is to use table lookup, which requires extra memory [\KoB, \Per, \RaG, \Sar]. In contrast, even without table lookup, our fast CRCs require only a small number of steps for their computation. Algorithms that do not rely on table lookup have an advantage of being less dependent on issues such as cache architecture and cache miss. In particular, it is possible to use as low as 1.5 operations per input message byte to encode our fast 64-bit CRC (which is implemented in C and requires no table lookup).

This paper, an extension of [\NguA], is organized as follows. In Section~2, we review known facts about CRCs, which serve as the background for our discussions. We present several different algorithms for computing CRCs, some of which are designed especially for our fast CRCs. In Section~3, we identify the form of the generator polynomials for the fast CRCs, and introduce a new technique for their implementation. We then determine their {\sl guaranteed} error-detecting capability: the minimum distance, the burst-error-detecting capability, and the maximum code length. In Section~4, we discuss CRC software complexity and show that our fast CRCs are typically faster than other CRCs. In Section~5, we present summaries and extensions of the paper.

\subsection{1.1}{Notation and Convention}  
In this paper, we consider polynomials that have binary coefficients~0 and~1. Thus, all polynomial operations are performed in the binary field GF(2), i.e., by using polynomial arithmetic modulo 2. Let $A(X)$ and $M(X)$ be 2 polynomials, then $\MOD{A(X)}{M(X)}$ denotes the remainder polynomial that is obtained when $A(X)$ is divided by $M(X)$. We must have ${\rm degree}(\MOD{A(X)}{M(X)}) < {\rm degree}(M(X))$. 

An $s$-tuple denotes a block of $s$ bits $A = (a_{s-1}, a_{s-2}, \dots, a_1,a_0)$, which is also presented by the binary polynomial $a_{s-1}X^{s-1}+a_{s-2}X^{s-2}+\cdots+a_1X+a_0$ of degree less than $s$.  We use the closely related notation $A(X)$ to denote this polynomial, i.e., $A$ is composed of the binary coefficients of $A(X)$. Thus, the tuple~$A$ and the polynomial $A(X)$ are equivalent and can be used interchangeably. Typically, the polynomial notation is used to describe the mathematical properties of codes, whereas the tuple notation is used to describe the algorithmic properties (such as pseudocodes and computer programs) of codes. 
If $Q_1(X)$ and $Q_2(X)$ are $s_1$-tuple and $s_2$-tuple, respectively, then the $(s_1 + s_2)$-tuple $(Q_1(X), Q_2(X)$) denotes the polynomial $Q_1(X)X^{s_2}+Q_2(X)$, which is the concatenation of $Q_2(X)$ to $Q_1(X)$.

In this paper, we are interested in CRCs that have low software complexity. {\it Software} complexity of an algorithm refers to the number of operations (i.e., operation count) used to implement the algorithm (whereas {\it hardware} complexity refers to the number of gates used to implement the algorithm). Suppose that we have 2 CRCs that operate under similar environments and use similar types of operations, but one CRC requires lower operation count (e.g., having a smaller loop) than the other. It is likely that the CRC with lower operation count (i.e., lower software complexity) will result in faster encoding. Thus, complexity correlates with speed.  However, the amount of the correlation also depends on many other complicating factors such as memory speed, cache size, compiler, operating system, pipelining, and CPU architecture. A CRC is called ``fast" if it has low software complexity and low memory requirement (e.g., it requires no lookup table or only a small lookup table).  A CRC is called ``faster" than another if, for a similar level of memory requirement, it has lower software complexity. 

An algorithm (or implementation) is called {\sl bitwise} if it does not use table lookup. Note that a bitwise algorithm does not necessarily involve only bit-by-bit manipulation or computation. Fast checksums are typically bitwise. Bitwise algorithms, which do not rely on table lookup, have an advantage of being less dependent on issues such as cache architecture, cache miss, and software code space. Ideally, fast CRC algorithms should have low complexity and be bitwise. Thus, unless explicitly stated, we focus on bitwise algorithms in this paper. Table-lookup algorithms are presented in Appendix A.

The notation  $(k, l, d)$ denotes a systematic code with $k$ = the total bit length of the code, $l$ = the bit length of the input message, and $d$ = the minimum distance of the code. The burst-error detecting capability of a code is denoted by $b$. To facilitate cross-references, we label some blocks of text as ``Remarks," which are an integral part of the presentation and should not be viewed as isolated observations or comments. 

\section{2}{CRC ALGORITHMS} \secII = \pageno
In this section, we review some known facts about software CRC implementation (e.g., see [\BiR, \Fel, \JiK, \KoB, \McA, \Per, \RaG, \Sar]). To lay a firm foundation for our later discussions, we present these facts in more precise and general forms than those often seen in the literature. Our presentation is a straightforward generalization of the results in [\RaG]. 

\subsection {2.1}{General CRC Theory} 
Suppose that we use an $h$-bit CRC, generated by  a polynomial $M(X)$ of degree $h$, to protect an input message $U(X)$, which has $l$ bits. By definition, the check polynomial $P(X)$ is the remainder that is obtained by dividing $U(X)X^h$ by $M(X)$, i.e., $P(X) = \MOD{U(X)X^h}{M(X)}$. Because computers can process tuples of bits (e.g., bytes or words) at a time, codes having efficient software implementation should be encoded on tuples. Typical modern processors can efficiently handle tuples of 8, 16, 32, and 64 bits. 
 
Let $s>0$ be any positive integer. We can write $l=r+(n-1)s$, for some $n>0$ and $0<r \le s$. We then process the CRC by dividing the input message $U(X)$ into $n$ tuples. The first tuple has $r$ bits, and all the other tuples have $s$ bits.
Because $r \le s$, we can then insert $(s-r)$ zeros to the left of $U(X)$ to increase its length from $l$ to $l^\prime=l+s-r = ns$, without affecting the CRC computation, because $\MOD{(0,0, \dots, 0, U(X))X^h}{M(X)} = \MOD{U(X)X^h}{M(X)} = P(X)$. That is, the first tuple now also has $s$ bits, the $(s-r)$ left-hand bits of which are always zeros.

Because each tuple $i$ has $s$ bits, it can be represented by a polynomial $Q_i(X)$ of degree~$< s$. Thus, the input message is represented by $U(X)=(Q_0(X), Q_1(X), \dots, Q_{n-1}(X)) $. We emphasize that, for given $h$ and $l$, we are free to choose the value of $s$ (commonly chosen values are $s=8$, 16, 32, and 64 bits). As shown later, the choice of $s$ can have significant impact on CRC speed.

Define $U_i(X) = (Q_0(X), Q_1(X), \dots, Q_i(X))$ to be the first $i+1$ input tuples, i.e.,
$$ 
\eqalign{U_0(X) &= Q_0(X)  				\cr
              U_1(X) &= (Q_0(X), Q_1(X))  \cr
             	&\cdots 				\cr
              U_{n-1}(X) &= (Q_0(X), Q_1(X), \dots, Q_{n-1}(X)) \cr
              &= U(X) }
$$
Thus,  for $i = 1, 2, \dots, n-1$, $U_i(X)$ is determined from $U_{i-1}(X)$ and $Q_i(X)$ by   
$$ \eqalignno{U_{i}(X) &= (U_{i-1}(X), Q_i(X)) \cr
&= U_{i-1}(X)X^{s} + Q_i(X) & (\eiitwo) \cr}
$$ 

For $i = 0, 1, \dots, n-1$, let $P_i(X)$ be the CRC check polynomial for the partial input message $U_i(X)$, i.e.,  
$$ P_i(X) = \MOD{U_i(X)X^h}{M(X)}  \eqno(\eiifour)$$ 
In particular, we have $P_0(X)=\MOD{Q_0(X)X^h}{M(X)}$, and
$$ 
\eqalign{P_{n-1}(X) &= \MOD{U_{n-1}(X)X^h}{M(X)} \cr
					&= \MOD{U(X)X^h}{M(X)} \cr
					&= P(X) }
$$ 
which is the CRC check polynomial for the entire input message $U(X)$. 

Substituting (\eiitwo) into (\eiifour), we have
$$ 
\eqalign{P_i(X) &= \MOD{U_i(X)X^h}{M(X)}         \cr 
                        &= \MOD{(U_{i-1}(X)X^{s} + Q_i(X))X^h}{M(X)}  \cr
                        &= \MOD{(U_{i-1}(X)X^h)X^{s}}{M(X)} + \MOD {Q_i(X)X^h}{M(X)}}  
$$ 
Using (\eiifour), we then have
$$
\eqalignno{P_i(X) &= \MOD{P_{i-1}(X)X^{s}}{M(X)} + \MOD{Q_i(X)X^h}{M(X)}  \cr
                &= \MOD{P_{i-1}(X)X^{s} + Q_i(X)X^h}{M(X)} & (\eiisix) \cr}
$$      
for $i=1,2,\dots,n-1$.  
Note that (\eiisix) is a straightforward generalization of a result in [\RaG], which deals with the special cases $h=16$ and $s \in \{8,16\}$. Thus, the check tuple $P_i(X)$ is computed from  $Q_i(X)$ and the previous check tuple $P_{i-1}(X)$.  Recall that $P_0(X)=\MOD{Q_0(X)X^h}{M(X)}$ and $P(X)=P_{n-1}(X)$ is the CRC check tuple for $U(X)$. Using (\eiisix), $P(X)$ is then computed via the following pseudocode:

$$
\vbox
{
\offinterlineskip 
\halign
{
\strut 
\vrule \ \hss #  \vrule & \ # \hss \vrule \cr
\noalign{\hrule}
1&	$P=0$;	\cr
2&	for $(0 \le i < n)$ \cr
3& \kern 4ex $P = \MOD{PX^{s} + Q_iX^h}{M}$;\cr
4&	return $P$;	\cr
\noalign{\hrule}
}
}
$$


\remark{Remark\rtwo.}  We now review the computational complexity of the polynomial division, which is needed in CRC computation. Given 2 polynomials $W(X)$ and $Y(X)$, let $V(X) = \MOD{Y(X)}{W(X)}$ be the remainder polynomial that is obtained when $Y(X)$ is divided by $W(X)$. Let $w$ and $y$ be the degrees of $W(X)$ and $Y(X)$, respectively. If $y<w$ (i.e., $y-w+1 \le 0$), then $V(X) = Y(X)$, i.e., no polynomial division is needed to obtain the remainder $V(X)$. If $y \ge w$, we then need the polynomial division that requires a loop of $y-w+1$ iterations to obtain the remainder $V(X)$ (see [\Knu], p.\ 421). To summarize, the polynomial ``long division'' for  computing $\MOD{Y(X)}{W(X)}$ requires a loop of ${\rm max}(0, y-w+1)$ iterations. \QED

\subsection {2.2}{Two CRC Algorithms} 
From (\eiisix), we have 
$$
P_i(X) = \MOD{(P_{i-1}(X) + Q_i(X)X^{h-s})X^{s}}{M(X)} \eqno(\eiieight) 
$$
if $s < h$, and 

$$
P_i(X) = \MOD{(P_{i-1}(X)X^{s-h} + Q_i(X))X^{h}}{M(X)} \eqno(\eiiten) 
$$ 
if $s \ge h$. 
The CRC algorithms based on (\eiieight) and (\eiiten), called Algorithm 1 and Algorithm 2,  are shown in Figs.\fiitwo \ and\fiithree, respectively.

\figure{
{\ninepoint
\vbox
{
\offinterlineskip 
\halign
{
\strut 
\vrule \ \hss #  \vrule & \ # \hss \vrule \cr
\noalign{\hrule}
1&	$B=0$;	\cr
2&	for $(0 \le i < n)$ \cr
3& \kern 4ex $\{$		\cr
4& \kern 4ex $A = B + Q_iX^{h-s}$;	\cr
5& \kern 4ex $B = \MOD{AX^{s}}{M}$;\cr
6& \kern 4ex $\}$				\cr
7&	$P = B$;	\cr
8&	return $P$;	\cr
\noalign{\hrule}
}
}
}}{\fiitwo}{CRC Algorithm 1 for computing the check $h$-tuple $P$ from the input $s$-tuples $Q_0,\dots,Q_{n-1}$ ($s<h$).}

\figure{
{\ninepoint
\vbox
{
\offinterlineskip 
\halign
{
\strut 
\vrule \ \hss #  \vrule & \ # \hss \vrule \cr
\noalign{\hrule}
1&	$B=0$;	\cr
2&	for $(0 \le i < n)$ \cr
3& \kern 4ex $\{$		\cr
4& \kern 4ex $A = BX^{s-h} + Q_i$;	\cr
5& \kern 4ex $B = \MOD{AX^{h}}{M}$;\cr
6& \kern 4ex $\}$				\cr
7&	$P = B$;	\cr
8&	return $P$;	\cr
\noalign{\hrule}
}
}
}
}{\fiithree}{CRC Algorithm 2 for computing the check $h$-tuple $P$ from the input $s$-tuples $Q_0,\dots,Q_{n-1}$ ($s \ge h$).}

\subsection {2.3}{Two Alternative CRC Algorithms} 
We now present 2 alternative CRC algorithms, which will be applied to our fast CRCs (see Section 3).

Case 1: $s < h$. The CRC check polynomial $P_j(X)$ for the partial input message $U_j(X)$ can be divided into 2 parts as
$$
P_j(X) = (P_{j,1}(X), P_{j,2}(X)) = P_{j,1}(X)X^{h-s} + P_{j,2}(X) \eqno(\eiitwelve) 
$$
where $P_{j,1}(X)$ and $P_{j,2}(X)$ are polynomials with degree$(P_{j,1}(X)) < s$ and degree$(P_{j,2}(X)) < h-s$. That is, $P_{j,1}(X)$ and $P_{j,2}(X)$ are the $s$ left-hand bits and $(h-s)$ right-hand bits of $P_j(X)$, respectively. Substituting (\eiitwelve) into (\eiisix), we have
$$
\eqalign{P_i(X) &= \MOD{(P_{i-1,1}(X)X^{h-s} + P_{i-1,2}(X))X^{s}}{M(X)} + \MOD{Q_i(X)X^h}{M(X)} \cr
                          &=\MOD{(P_{i-1,1}(X) + Q_i(X))X^h}{M(X)} + \MOD{P_{i-1,2}(X)X^{s}}{M(X)}}  
$$
Because ${\rm degree}(P_{i-1,2}(X)X^{s}) < h = {\rm degree}(M(X))$, we have $\MOD{P_{i-1,2}(X)X^{s}}{M(X)} = P_{i-1,2}(X)X^{s}$. Thus,
$$
P_i(X) = \MOD{(P_{i-1,1}(X) + Q_i(X))X^h}{M(X)} + P_{i-1,2}(X)X^{s} \eqno(\eiifourteen) 
$$ 
The CRC algorithm based on (\eiifourteen), called Algorithm~3, is shown in Fig.\fiifour.

Case 2: $s \ge h$. Multiplying both sides of (\eiiten) by $X^{s-h}$, we have
$$
 \eqalignno{P_i(X)X^{s -h} &= \left(\MOD{(P_{i-1}(X)X^{s -h}+Q_i(X))X^h}{M(X)}\right)X^{s -h} \cr
                                        &= \MOD{(P_{i-1}(X)X^{s -h}+Q_i(X))X^h X^{s -h}}{M(X)X^{s -h}} \cr
                                        &= \MOD{(P_{i-1}(X)X^{s -h}+Q_i(X))X^{s}}{M(X)X^{s-h}} & (\eiififteen) \cr }
$$
Define $L_j(X) = P_j(X) X^{s - h}$. From (\eiififteen), we then have
$$
L_i(X) =\MOD{(L_{i-1}(X) + Q_i(X))X^{s}}{N(X)} 
\eqno(\eiisixteen) 
$$
where $N(X)=M(X)X^{s-h}$. Thus, $L_i(X)$ is computed from $L_{i-1}(X)$ and $Q_i(X)$. 

Note that $L_0(X) = P_0(X)X^{s-h}$, where $P_0(X) = \MOD{Q_0(X)X^h}{M(X)}$.
We then have 
$$
\eqalign{L_0(X) &= (\MOD{Q_0(X)X^h}{M(X)})X^{s-h} \cr
				&=\MOD{Q_0(X)X^hX^{s-h}}{M(X)X^{s-h}} \cr
				&= \MOD{Q_0(X)X^s}{N(X)} }
$$ 

Because $L_i(X) = P_i(X) X^{s - h}$, the term $P_i(X)$ is obtained by shifting $L_i(X)$ to the right by $(s -h)$ bits. Note that degree$(L_i(X)) < s$. We will show in Remark\rfour \ that computing $P_i(X)$ via (\eiisixteen) is slightly faster than via~(\eiiten).
The CRC algorithm based on (\eiisixteen), called Algorithm~4, is shown in Fig.\fiifive, where $N(X)=M(X)^{s-h}$. 

\figure{
{\ninepoint
\vbox
{
\offinterlineskip 
\halign
{
\strut 
\vrule \ \hss #  \vrule & \ # \hss \vrule \cr
\noalign{\hrule}
1&	$P=0$;	\cr
2&	for $(0 \le i < n)$ \cr
3& \kern 4ex $\{$		\cr
4& \kern 4ex $P_1 = s$ left-hand bits of $P$;	\cr
5& \kern 4ex $P_2 = (h-s)$ right-hand bits of $P$;	\cr
6& \kern 4ex $A = P_1 + Q_i$;	\cr
7& \kern 4ex $B = \MOD{AX^{h}}{M}$;\cr
8& \kern 4ex $P = B + P_2X^s$;				\cr
9& \kern 4ex $\}$				\cr
10&	return $P$;	\cr
\noalign{\hrule}
}
}
}}{\fiifour}{CRC Algorithm~3 for computing the check $h$-tuple $P$ from the input $s$-tuples $Q_0,\dots,Q_{n-1}$ ($s < h$).}

\figure{
{\ninepoint
\vbox
{
\offinterlineskip 
\halign
{
\strut 
\vrule \ \hss #  \vrule & \ # \hss \vrule \cr
\noalign{\hrule}
1&	$B=0$;	\cr
2&	for $(0 \le i < n)$ \cr
3& \kern 4ex $\{$		\cr
4& \kern 4ex $A = B+ Q_i$;	\cr
5& \kern 4ex $B = \MOD{AX^{s}}{N}$;\cr
6& \kern 4ex $\}$				\cr
7&  $P = h$ left-hand bits of $B$;	\cr
8&	return $P$;	\cr
\noalign{\hrule}
}
}
}

}{\fiifive}{CRC Algorithm 4 for computing the check $h$-tuple $P$ from the input $s$-tuples $Q_0,\dots,Q_{n-1}$ ($s \ge h$).}

\remark{Remark\rfour.} Suppose that $s \ge h$. The check polynomial $P(X) = P_{n-1}(X) = \MOD{U_{n-1}(X)X^h}{M(X)}$ can then be computed by Algorithm 2 (Fig.\fiithree) or by Algorithm 4 (Fig.\fiifive).  We now show that, for bitwise implementation, Algorithm 4 is slightly faster than Algorithm 2.
By comparing these 2 algorithms, we observe the following. First, the computation of $\MOD{A(X)X^h}{M(X)}$ (in Fig.\fiithree) and the computation of $\MOD{A(X)X^s}{N(X)}$ (in Fig.\fiifive) have the same complexity, because each requires $s$ iterations (by Remark\rtwo). Next, the factor $X^{s-h}$ at line 4 of Fig.\fiithree \ disappears from line 4 of Fig.\fiifive. Finally, one extra operation is required at line 7 of Fig.\fiifive \ to extract the $h$ left-hand bits of the final $B(X)$. The above observations imply that Algorithm 4 requires $n-1$ fewer operations than Algorithm 2. Thus, for bitwise implementation, we will use Algorithm 4 when $s \ge h$. \QED

\subsection {2.4}{Basic CRC Algorithms} 
Given an input message $U(X)$ and a generator polynomial $M(X)$ of degree $h$, Algorithms 1-4 produce the same CRC check tuple $P(X)$. That is, they are 4 different ways for accomplishing the same thing. The main difference among these algorithms is how the input message is divided into $s$-tuples $Q_i(X)$. Algorithms~1 and 3 are for $s<h$, whereas Algorithms 2 and 4 are for $s \ge h$. As shown later, CRC speed depends on the choice of $s$. For flexibility, we allow the possibility that the same CRC is used by computers that have different architectures and capabilities. For example, one computer can choose a value of $s$ for encoding a message to transmit to another computer (with different capabilities), which can choose a different value of~$s$ for detecting the errors in the received message.

The above CRC algorithms require the polynomial divisions. In particular, Algorithm 1 requires the polynomial division $\MOD{A(X)X^s}{M(X)}$, Algorithms 2 and 3 require the polynomial division $\MOD{A(X)X^h}{M(X)}$, and Algorithm 4 requires the polynomial division $\MOD{A(X)X^s}{N(X)}$. To simplify the presentation, we will use the single notation $B(X)$ to denote all these polynomial divisions, i.e., we define
$$
B(X) = \cases {\MOD{A(X)X^s}{M(X)} &(Algo.~1) \cr
                       \MOD{A(X)X^h}{M(X)}  &(Algos.~2 and 3) \cr
                       \MOD{A(X)X^s}{N(X)} &(Algo.~4) }   \eqno(\eiitwentyeight)                    
$$
where $N(X) = M(X)X^{s-h}$. Note that degree$(A(X))<h$ in Algorithm 1, and degree$(A(X))<s$ in Algorithms 2-4. As seen in Figs.\fiitwo-4, CRC computation using any of the above 4 algorithms requires the computation of $B(X)$ for $n$ times. 

A known technique for computing $B(X)$ is to use the polynomial long division algorithm mentioned in Remark\rtwo. 
For example, consider Algorithms 2 and 3. We then have $B(X) = \MOD{A(X)X^h}{M(X)}$, where degree$(A(X))<s$. Because degree$(A(X)X^h)  \le s+h-1$ and degree$(M(X))=h$, from Remark\rtwo, $B(X)$ can be computed via the polynomial long division that requires a loop of $s$ iterations. Similarly, it can be shown that computing $B(X)$ in Algorithms~1 and~4 also requires  a loop of $s$ iterations. That is, the computational complexity for computing $B(X)$ is $O(s)$. 

\remark{Definition 1.} The technique for computing the polynomial $B(X)$ as given in (\eiitwentyeight) is called the {\sl basic} technique. Using the polynomial long division, $B(X)$ can be computed in $s$ iterations. An algorithm (or a CRC) is basic if it uses the basic technique for computing $B(X)$.

\section {3}{FAST CRCS} \secIII = \pageno
Recall that we are given an input message $U_{n-1}(X) = (Q_0(X), Q_1(X), \dots, Q_{n-1}(X))$, where $Q_i(X)$ is an $s$-tuple. We  protect this message by an $h$-bit CRC generated by a polynomial $M(X)$ of degree $h$. The check $h$-tuple 
$$
P(X) = P_{n-1}(X) = \MOD{U_{n-1}(X)X^h}{M(X)}
$$
can be computed by Algorithm~1 or~3 (if $s<h$), or by Algorithm~2 or~4 (if $s \ge h$). We emphasize that each of these algorithms requires the calculation of $B(X)$ defined in (\eiitwentyeight), which involves the polynomial division.

\subsection {3.1}{Fast $h$-Bit CRCs} 
Our goal is to find some CRCs that have fast implementation, i.e., to find a new family of generator polynomials $M(X)$ for CRCs that have low complexity. Recall that the CRC algorithms (Figs.\fiitwo-4) depend on the term $B(X)$. Computation of $B(X)$ is also the most expensive step in the algorithms. Thus, finding fast CRCs requires  finding the polynomials $M(X)$ that yield fast computation of $B(X)$.  

The first technique for computing $B(X)$ is the {\sl basic} technique in Definition~1. Using the polynomial division, we can compute $B(X)$ by a loop of $s$ iterations. In the following, we present the second technique, called the {\sl new} technique, for computing $B(X)$. While the new technique is applicable to any generator polynomial $M(X)$, it is more effective for some special CRC generator polynomials, called the {\it fast} polynomials. Recall that the basic CRCs can use Algorithm 1 or 3 (if $s<h$), or by Algorithm 2 or 4 (if $s \ge h$). However, as seen in the following, the fast CRCs use only Algorithms 3 (for $s < h$) and 4 (for $s \ge h$) for their bitwise implementation.   

We now introduce a new family of CRCs, which are generated by the following polynomials
$$
F_h(X) = X^h+X^2+X+1  \eqno(\eiiifour)
$$
for all $h \ge 4$. We ignore the case $h=3$, which yields the trivial repetition code $\{(0000),(1111)\}$. We call $F_h(X)$ the ``fast polynomial," which can be factored into
$$
X^h+X^2+X+1  = (X+1)G_{h-1}(X) 
$$
where
$$
G_m(X) = X^m+X^{m-1}+\cdots+X^3+X^2+1  \eqno(\eiiisix)
$$
i.e., $G_m(X)$ includes all the terms except $X$.  At first, it is not clear why this particular polynomial $F_h(X)$ will speed up the computation of $B(X)$. We now introduce a technique that is applied to $F_h(X)$ to yield fast computation of $B(X)$.

By considering Algorithms 3 and 4, we have from (\eiitwentyeight)
$$
B(X) = \cases {\MOD{A(X)X^h}{M(X)}  &if $s < h$  \cr
                       \MOD{A(X)X^s}{N(X)}   &if $s \ge h$}       \eqno(\eiiiten) 
$$ 
where $N(X) = M(X)X^{s-h}$, and $A(X)$ is a polynomial of degree less than $s$. We now transform $B(X)$ into a new form that will be used by the fast CRCs. First, note that
$$
\eqalign {\MOD{A(X)(X^h + M(X))}{M(X)} &= \MOD{A(X)X^h}{M(X)} + \MOD{A(X)M(X)}{M(X)} \cr
                                                         &= \MOD{A(X)X^h}{M(X)} }
$$                                                         
because  $\MOD{A(X)M(X)}{M(X)}= 0$.  Similarly, we have
$$                                                 
 \MOD{A(X)(X^s + N(X))}{N(X)} = \MOD{A(X)X^s}{N(X)}  
$$
Thus, (\eiiiten) becomes
$$
B(X) = \cases {\MOD{A(X)(X^h + M(X))}{M(X)}  &if $s < h$  \cr
                       \MOD{A(X)(X^s + N(X))}{N(X)}   &if $s \ge h$}       \eqno(\eiiitwelve) 
$$
where $N(X) = M(X)X^{s-h}$. 

\remark{Definition 2.} Using Algorithms 3 and 4, the technique (\eiiitwelve) for computing the polynomial $B(X)$ is called the {\sl new} technique. The CRC that is generated by the fast polynomial $F_h(X)=X^h+X^2+X+1$ and uses the new technique for computing $B(X)$ is called the {\sl fast} $h$-bit CRC.
 
\remark{Theorem 1.} Using Algorithms 3 and 4, the polynomial $B(X)$ for the fast CRC generated by $F_h(X) = X^h+X^2+X+1$ is given by 
$$
B(X) = \cases {\MOD{A(X)(X^2 + X +1)}{F_h(X)}  &if $s < h$  \cr
               \MOD{A(X)X^{s-h}(X^2 + X +1)}{N(X)}   &if $s \ge h$}       \eqno(\eiiifourteen) 
$$ 
where $N(X) = F_h(X)X^{s-h}$, and $A(X)$ is a polynomial of degree less than $s$. Further, using the polynomial division, $B(X)$ can be computed with $\max(0, s-h+2)$ iterations. 
\remark{Proof.}  Relation (\eiiifourteen) follows by using (\eiiitwelve) with $M(X) = F_h(X)$ and $N(X) = F_h(X)X^{s-h}$. First, suppose that $s<h$. Then $B(X) = \MOD{A(X)(X^2 + X +1)}{F_h(X)}$. Because degree$(F_h(X)) = h$ and degree$(A(X)(X^2+X+1)) < s+2$, from Remark\rtwo, $B(X)$ can be computed with $\max(0, s-h+2)$ iterations. Next, suppose that $s \ge h$. Then $B(X) = \MOD{A(X)X^{s-h}(X^2 + X +1)}{N(X)}$. Because degree$(N(X)) = s$ and degree$(A(X)X^{s-h}(X^2+X+1)) < 2s-h+2$, Remark\rtwo \ implies that $B(X)$ can also be computed with $\max(0, s-h+2)$ iterations.  \QED
  
Let us briefly compare the computational complexity of $B(X)$ for (a) the basic $h$-bit CRC generated by $M(X)$ and (b) the fast $h$-bit CRC generated by $F_h(X)$. For the basic CRC, by Definition~1, $B(X)$ is computed via a loop of $s$ iterations, regardless of the form of $M(X)$. However, for the fast CRC, by Theorem~1, $B(X)$ is computed via a loop of only max$(0,s-h+2)$ iterations. Thus, the fast CRC is much faster than the basic CRC if $s$ is chosen such that $s-h+2$ is much small than $s$. Further, if $s+1 < h$, then max$(0,s-h+2) = 0$ and $B(X) = A(X)(X^2+X+1)$,  i.e., the polynomial division is eliminated. Section 4 presents CRC software complexity in more detail.

We emphasize that the fast $h$-bit CRC denotes a CRC that meets the following 2 conditions: (a) it is generated by the {\sl fast} polynomial $F_h(X) = X^h+X^2+X+1$, and (b) the polynomial $B(X)$ is computed via Theorem~1 by applying the {\sl new} technique (\eiiitwelve) to $F_h(X)$. That is, the fast CRC refers to a CRC that is generated by a specific polynomial and is implemented by a specific technique. Note that a CRC that meets only one of the above 2 conditions may not have any speed advantage over a basic CRC. For example, suppose that, instead of the new technique (\eiiitwelve), the basic technique (in Definition 1) is applied to the CRC generated by the fast polynomial $F_h(X)$. This CRC is then not different from a basic CRC in terms of computational complexity.  Application of the new technique to polynomials other than $F_h(X)$ is considered in Appendix C.

To summarize, the fast $h$-bit CRC is generated by $F_h(X) = X^h+X^2+X+1$. Under bitwise implementation, the fast CRC uses Algorithm 3 if $s < h$ and Algorithm 4 if $s \ge h$. The term $B(X)$ in these algorithms is given in Theorem 1.   

\subsection {3.2}{A Fast 16-Bit CRC} 
We now consider the important case $h=16$.  Many CRCs (as well as weaker checksums) used in practice have 16 check bits, e.g., the CRC-16 and CRC-CCITT mentioned in Section~1. With a small amount of overhead, these CRCs can have length up to $2^{15}-1$ bits $\approx$ 4,096 bytes. Our goal here is to present a concrete example of a new 16-bit CRC that is not only much faster than but also as good as existing 16-bit CRCs. 

Our new 16-bit CRC is generated by 
$$
F_{16}(X) = X^{16}+X^2+X+1 \eqno(\eiiisixteen) 
$$
which can be factored into
$$
F_{16}(X) = (X+1)G_{15}(X) 
$$
where 
$$
G_{15}(X) = X^{15}+X^{14}+\cdots+X^3+X^2+1 
$$
It can be shown that $G_{15}(X)$ is a primitive polynomial, i.e., $F_{16}(X)$ is a product of $X+1$ and a primitive polynomial (however, as seen later, this is not true for many values of $h$). Thus, this fast 16-bit CRC also has length up to $2^{15}-1$ bits. Although the polynomial (\eiiisixteen) is different from the generator polynomials for existing 16-bit CRCs, it does generate a CRC that has the same guaranteed error-detecting capability as existing 16-bit CRCs. From Theorem 1, we have
$$
B(X) = \cases {\MOD{A(X)(X^2 + X +1)}{F_{16}(X)}  &if $s < 16$  \cr
               \MOD{A(X)X^{s-16}(X^2 + X +1)}{N(X)}   &if $s \ge 16$}        
$$
where $N(X)=F_{16}(X)X^{s-16}$.

In the following, we consider 2 cases: $s=8$ and $s=16$. First, assume that $s=8$, i.e., the input message is organized in 8-bit bytes. Because  $s<16$, we have $B(X) = \MOD{A(X)(X^2 + X +1)}{F_{16}(X)} $. Because degree$(A(X))<s=8$, we have degree$(A(X)(X^2+X+1)) < 10$, which is smaller than degree$(F_{16}(X)) =16$. From Remark\rtwo, we have
$$
\eqalign{B(X) &= A(X)(X^2+X+1)  \cr
                      &= A(X)X^2+A(X)X + A(X)}
$$                      
i.e., $B(X)$ is simply the sum of $A(X)$ and its translations. Thus, computing $B(X)$ via the new technique requires no polynomial division. In contrast, computing $B(X)$ via the basic technique requires the polynomial division that has a loop of $s=8$ iterations (see Definition 1).

Next, assume that $s=h=16$, i.e., the input message is organized in 16-tuples. Because $s=16$, we have degree$(A(X))<16$ and degree$(A(X)(X^2+X+1))<18$. Thus, by Remark\rtwo, $B(X)$ is computed by the polynomial division that has a loop of 2 iterations. This contrasts with computing $B(X)$ via the basic technique, which requires a loop of $s=16$ iterations (see Definition 1). Thus, the loop iteration count of our new technique is less than that of the basic technique by the factor of $16/2=8$.

To summarize, when the input message is organized in $s$-tuples, it is possible to have a fast 16-bit CRC that requires no polynomial division (when $s=8$), or that requires the polynomial division that has only~2 loop iterations (when $s=16$). Further, this fast 16-bit CRC has the same guaranteed error-detecting capability as existing 16-bit CRCs.

When computing $B(X)$ via the new technique, although the case $s=16$ requires more loop iterations than the case $s=8$, we will see later in Section 4 that the case $s=16$ has lower {\sl overall} computational complexity (i.e., lower overall operation count per input byte). This is because, when $s=16$, there is no need to compute $P_{j,1}(X)$ and $P_{j,2}(X)$ as defined in (\eiitwelve). Further, the overhead processing cost per input byte when $s=16$ is lower than when $s=8$. The C programs for the fast 16-bit CRC are shown in Fig.\fAfour \ and in Fig.\fAthree \ of Appendix A.

\subsection {3.3}{Error-Detection Capability of Fast CRCs} 
{\color{red} Recall that the maximum length of the $h$-bit CRC generated by (\eitwo) is $2^{h-1}-1$ bits, i.e., this CRC has minimum distance $d=4$ if its total bit length $\le 2^{h-1}-1$. In general, we define the {\sl maximum length} of an error-detection code to be the total bit length at or below which its minimum distance is $d \ge 3$, and beyond which its minimum distance will reduce to $d\le2$.  In the following, we determine the maximum lengths of the fast CRCs.} 

By definition, the {\sl period} of a polynomial $G(X)$ is the smallest positive integer $i$ such that \hbox{$\MOD{X^i}{G(X)} = 1$}. {\color{red}In particular, it can be shown that the period of $M(X)$ in (\eitwo), which is the product of $X+1$ and a primitive polynomial of degree $h-1$, is $2^{h-1}-1$}.  Note that some polynomials, such as $X^2$, do not have periods. 

The period of the fast polynomial $F_h(X) = X^h+X^2+X+1$ can be computed directly from the definition (for small $h$) or from the technique in [\Ber, Section 6.2]. The periods of $F_h(X)$, $h \ge 4$, are shown in Fig.\fiiitwo. The following theorems, which are slight variations of well-known results from cyclic codes [\LiC, Chapter~4], show that the maximum length of a CRC equals the period of its generator polynomial.

\remark{Theorem 2.} Let $C$ be a CRC generated by a polynomial $M(X)$ of degree $h \ge 3$.  Assume that $M(X)$ is not a multiple of $X$. Let $n_{\rm b}$ and $d$ be the bit length and minimum distance of $C$, respectively. We then have \nobreak

\noindent 1. $d \ge 3$ \ if \ $n_{\rm b} \le$ period of $M(X)$. \nobreak

\noindent 2. $d = 2$ \ if \ $n_{\rm b} > $ period of $M(X)$. \nobreak

\noindent 3. $C$ detects all error bursts of length up to $h$ bits, i.e., $b=h$.

\remark{Proof.} Let $t$ be the period of $M(X)$. We must have $t \ge h$. By definition, each codeword of $C$ has the form
$$
V(X) = U(X)X^h + P(X)
$$
where $U(X)$ is the polynomial representing the input message, and $P(X)$ is the check polynomial. Because 
$ P(X) = \MOD{U(X)X^h}{M(X)} $, we have 
$$
U(X)X^h =  K(X)M(X) + P(X)
$$
for some polynomial $K(X)$. Thus, we have
$$
V(X) = U(X)X^h + P(X) = K(X)M(X)
$$
i.e., $C$ is a linear code. If $d=1$, then $X^i = K(X)M(X)$, for some $i$. This implies that $M(X) = X^j$ for some~$j$, which contradicts our assumption that $M(X)$ is not a multiple of $X$. Thus, $d \ge 2$. 

\noindent 1. We now prove, by contradiction, the statement $d \ge 3$ \ if \ $n_{\rm b} \le$ period of $M(X)$. Thus, suppose that there is a codeword $V(X)$ with length $n_{\rm b} \le t$ and weight 2. Then $V(X) = X^j + X^i$ for some $i$ and $j$ such that $n_{\rm b}>j>i \ge 0$. Thus, $V(X) = X^i(X^{j-i} + 1)$.

We also have $V(X) = K(X)M(X)$ for some polynomial $K(X)$. Thus, $X^i(X^{j-i} + 1) = K(X)M(X)$. Because $M(X)$ is not a multiple of $X$ by assumption, $M(X)$ must divide $X^{j-i}+1$, i.e., $\MOD{X^{j-i}}{M(X)}=1$. Thus, $j-i \ge t$ = period of $M(X)$. Then $j \ge t \ge n_{\rm b}$, which contradicts the condition $n_{\rm b}>j$. Thus, all the codewords of length $n_{\rm b} \le t$ must have weight $\ge 3$, i.e., $d \ge 3$.

\noindent 2. We construct a codeword with length $>t$ and weight 2 as follows. Let $U(X) = X^{t-h}$. Then $ P(X) = \MOD{U(X)X^h}{M(X)} $ = $\MOD{X^t}{M(X)}$. We have $P(X) = 1$ because $t$ is the period of $M(X)$. Thus, the codeword $V(X) = U(X)X^h+P(X)= X^t+1$ has length $t+1$ and weight 2. That is, $d=2$ if $n_{\rm b}>t$. 

\noindent 3. The fact that $C$ detects all error bursts of length up to $h$ bits (i.e., $b=h$) is well-known [\LiC]. \QED

\remark{Theorem 3.} Let $C$ be the CRC generated by the fast polynomial $F_h(X) = X^h+X^2+X+1$. Let $n_{\rm b}$ and~$d$ be the bit length and minimum distance of $C$, respectively.  We then have

\noindent 1. $d =4$ \  if \ $n_{\rm b} \le$ period of $F_h(X)$.

\noindent 2. $d = 2$ \  if \ $n_{\rm b} > $ period of $F_h(X)$.

\noindent 3. $C$ detects all error bursts of length up to $h$ bits, i.e., $b=h$. 

\remark{\bf Proof.} Let $t$ be the period of $F_h(X)$. From the proof of Theorem 2, every codeword of $C$ has the form $V(X) = K(X)(X^h+X^2+X+1)$ for some polynomial $K(X)$. Thus, the codewords of $C$ have even weight, i.e., $d$ is even.

Suppose now that the input message is $U(X) = 1$. Then $P(X) = \MOD{X^h}{F_h(X)} = X^2+X+1$, which implies $V(X) = U(X)X^h+P(X) = X^h+X^2+X+1$. That is, the codeword $V(X)$ has weight 4. Thus, $d$ is either 2 or 4. From Theorem 2.1, we must have $d=4$ if $n_{\rm b} \le t$. From Theorem 2.2, we must have $d=2$ if $n_{\rm b} > t$. The fact that $C$ detects all error bursts of length up to $h$ bits is well-known [\LiC]. \QED

{\color{red} Parts 1 and 2 of Theorem 2 show that the maximum length of a CRC equals the period of its generator polynomial (which, by assumption, is not a multiple of $X$). Theorem 2 also explains why, as seen above, both the maximum length of the $h$-bit CRC generated by (\eitwo) and the period of its generator polynomial equal  $2^{h-1}-1$. 
Similarly, parts 1 and 2 of Theorem 3 show that the maximum length of the fast $h$-bit CRC equals the period of its generator polynomial $F_h(X) = X^h+X^2+X+1$. } 

Fig.\fiiitwo \ shows that the maximum length of the fast $h$-bit CRC is also $2^{h-1}-1$ in many important cases, namely when $h=8, 16, 24, 48, 64, 128$. In fact, $F_h(X)=X^h+X^2+X+1$ is also the product of $X+1$ and a primitive polynomial at these values of $h$, i.e., the polynomial $G_{h-1}(X)$ in (\eiiisix) is primitive when $h=8, 16, 24, 48, 64, 128$. 
Fig.\fiiitwo \ also shows that the maximum lengths of many fast $h$-bit CRCs are substantially less than the upper bound $2^{h-1}-1$ (e.g., when $h=12$ and $h=32$). However, in Appendix C, we apply our new technique to more general generator polynomials to yield other fast CRCs whose maximum lengths can approach the upper bound.
\figure{$$
\vbox
{\eightpoint
\offinterlineskip 
\halign
{
\strut
\vrule \quad # \hfill \vrule    &\quad # \hfill\vrule &  \quad# &  \vrule  #  \cr
\noalign{\hrule} 
$h$ & 	period	& $2^{h-1}-1\over{\rm period}$ & \cr
\noalign{\hrule} 	
4       &7      &1              &       \cr
5       &14     &1.07143                &       \cr
6       &31     &1              &       \cr
7       &60     &1.05           &       \cr
8       &127    &1              &       \cr
9       &254    &1.00394                &       \cr
10      &465    &1.09892                &       \cr
11      &868    &1.17857                &       \cr
12      &595    &3.44034                &       \cr
13      &4094   &1.00024                &       \cr
14      &8191   &1              &       \cr
15      &3276   &5.00092                &       \cr
16      &32767  &1              &       \cr
17      &9362   &7.00011                &       \cr
18      &38227  &3.42875                &       \cr
19      &229348 &1.14299                &       \cr
20      &516033 &1.016          &       \cr
21      &1048574        &1              &       \cr
22      &126945 &16.5202                &       \cr
23      &803148 &5.22233                &       \cr
24      &8388607        &1              &       \cr
25      &917490 &18.286         &       \cr
26      &584073 &57.449         &       \cr
27      &65011588       &1.03226                &       \cr
28      &87381  &1536.01                &       \cr
29      &268435454      &1              &       \cr
30      &5013351        &107.088                &       \cr
31      &1900428        &565            &       \cr
32      &2097151        &1024           &       \cr
33      &4194302        &1024           &       \cr
34      &408944445      &21.0051                &       \cr
35      &5637144492     &3.04762                &       \cr
36      &270532479      &127.008                &       \cr           
37      &$2.080\times10^{10}$  &3.30323  & \cr
38      &$2^{37}-1$      &1  & \cr                         
39      &4831838172      &56  & \cr 
40      &$3.006 \times 10^{10}$      & 18.2857 & \cr
48      &$2^{47}-1$      &1  & \cr           
56      &$ 3.573\times 10^{16}$      & 1.00837 & \cr
64      &$2^{63}-1$      &1  & \cr   
128      &$2^{127}-1$      &1  & \cr        
\noalign{\hrule} 	
}
}
$$
}{\fiiitwo}{The period of $F_h(X) = X^h+X^2+X+1$ \hbox{[= the maximum length of the fast $h$-bit CRC generated by $F_h(X)$]}.}

\section {4}{CRC SOFTWARE COMPLEXITY} \secIV = \pageno
We now analyze and compare CRC software complexity. {\it Software} complexity of an algorithm refers to the number of operations (i.e., operation count) used to implement the algorithm. Our goal in this paper is to compute the CRC check $h$-tuple $P(X)$ for an input message that consists of $n$ tuples $Q_0(X), Q_1(X), \dots, Q_{n-1}(X)$. Each tuple $Q_i(X)$ has $s$ bits. This CRC can be either a basic CRC generated by a polynomial $M(X)$ of degree $h$, or the fast CRC generated by $F_h(X) = X^h+X^2+X+1$. For bitwise implementation, while Algorithm 1, 2, 3, or 4 can be used for the basic CRC, only Algorithm 3 or 4 are used for the fast CRC. The check tuple $P(X)$ is computed by using a loop that computes $B(X)$ for $n$ times, where $B(X)$ is given in Definition~1 for the basic CRC and in Theorem~1 for the fast CRC. 

In this section, we compute $e_b$ and $e_f$, which denote the software operation counts per input byte required for computing the check tuple $P(X)$ for the basic CRC and the fast CRC, respectively. These operation counts will then be used to compare the complexity among our fast CRCs, the basic CRCs, the other fast CRCs in [\Fel], and the block-parity checksum. An error-detection code is said to be ``faster" than another if, for a similar level of memory requirement, it has lower software complexity.

\subsection {4.1}{General Complexity Analysis}
We now provide the complexity analysis for the important case $s=h$  for the basic CRC and the fast CRC (other cases can be analyzed similarly). Both Algorithms 2 and 4 (shown in Figs.\fiithree \ and\fiifive) then reduce to Fig.\fivone. Here, we have
$$
B(X) = \MOD{A(X)X^s}{M(X)} \eqno(\eivtwo)
$$
for the basic CRC (see Definition 1), and
$$
B(X) = \MOD{A(X)(X^2+X+1)}{F_h(X)} \eqno(\eivthree)
$$
for the fast CRC (see Theorem 1), where $A(X)$ is a polynomial of degree less than $s$. Note that different CRC algorithms refer to different techniques for computing $B(X)$. In particular, a CRC algorithm is called {\sl table lookup} or {\sl bitwise}, depending on whether the term $B(X)$ in the algorithm is computed with or without table lookup. The bitwise technique is presented in this section. The table-lookup technique is presented in Appendix A.

\figure{
{\ninepoint
\vbox
{
\offinterlineskip 
\halign
{
\strut 
\vrule \ \hss #  \vrule & \ # \hss \vrule \cr
\noalign{\hrule}
1&	$B=0$;	\cr
2&	for ($0 \le i < n$)	\cr
3&	\hskip 3ex $\{$	\cr	
4&	\hskip 3ex $A = B + Q_i$;	\cr
5&	\hskip 3ex $B = \cases {\MOD{AX^s}{M};& {\kern -1ex for basic CRC}\cr
            \MOD{A(X^2+X+1)}{F};	& {\kern -1ex for fast CRC} }$ \cr
6&	\hskip 3ex $\}$	\cr
7& $P=B$;	\cr
8&	return $P$;				\cr
\noalign{\hrule}
}
}
}}{\fivone}{CRC algorithm ($s=h$).}
\medskip
\remark{Remark\rsix.} The term $B(X) = \MOD{A(X)X^s}{M(X)}$ in (\eivtwo) can be computed as follows. First, we write $A(X)X^s = (\cdots (A(X)X) \cdots)X $. Thus, $B(X)$ can be computed in $s$ iterations via the following pseudocode: 
\bigskip

{
$$
\vbox
{
\offinterlineskip 
\halign
{
\strut 
\vrule \ \hss #  \vrule & \ # \hss \vrule \cr
\noalign{\hrule}
1&	for $(0 \le j < s)$ \cr
2&  \kern 4ex $A = \MOD{AX}{M}$;\cr
3&	$B = A$;	\cr
\noalign{\hrule}
}
}
$$
}

\noindent where $\MOD{AX}{M}$ is computed by 
$$
\MOD{AX}{M} = \cases{AX + M    &if msb$(A)=1$ \cr
                 AX        &if msb$(A)=0$}
\eqno(\eivthreeA)                 
$$
where msb($A$) denotes the most significant bit of $A$. The term $\MOD{AX}{M}$ in (\eivthreeA) can also be computed by using a table $T[\ ]$ of only 2 entries defined by $T[0]=0$ and $T[1] = M$. We then have
$$
\MOD{AX}{M}=AX + T[{\rm msb}(A)] \eqno(\eivthreeB)
$$ \QED

Let $u$ be the operation count required for computing $\MOD{A(X)X}{M(X)}$. Using Remark\rsix, the operation count required for computing $B(X)$ in (\eivtwo) for the basic CRC is then $s(u+l_s)$, where $l_s$ denotes the operation count for the loop overhead shown at line 1 of the pseudocode in Remark\rsix \ (in particular, $l_s=0$ if loop unrolling is used).

Let us now consider the term $B(X)$ in (\eivthree) for the fast CRC. We have
$$
\eqalignno{B(X) &= \MOD{A(X)X^2}{F_h(X)} + \MOD{A(X)X}{F_h(X)} + A(X) \cr
&= \MOD{B_1(X)X}{F_h(X)} + B_1(X) + A(X) & (\eivthreeBB) \cr }
$$
where $B_1(X) = \MOD{A(X)X}{F_h(X)}$, which has operation count $u$. After $B_1(X)$ is computed, $\MOD{B_1(X)X}{F_h(X)}$ also has operation count $u$. There are also 2 binary additions (i.e., 2 XOR operations) in (\eivthreeBB). Thus, the operation count required for computing $B(X)$ in (\eivthreeBB) for the fast CRC is $2u+2$. 

Let us now determine the total operation counts $t_b$ and $t_f$ for computing the check tuple $P(X)$ for the basic CRC and the fast CRC, respectively. The CRC algorithm for computing $P(X)$, which is shown in Fig.\fivone, has a loop of $n$ iterations. In addition to the operation count for $B(X)$, there is also one addition as indicated in line~4 of Fig.\fivone. Let $l_n$ be the operation count for the loop overhead shown at line~2 of Fig.\fivone. We then have 
$$ 
t_b = n[l_n + 1 + s(u+l_s)]	\eqno (\eivthreeC)
$$
$$
t_f = n(l_n + 3 + 2u) 	\eqno (\eivthreeD)
$$

The basic CRC and the fast CRC require $t_b$ and $t_f$ operations, respectively, to compute the check tuple $P(X)$ for the input message that has $ns$ bits, i.e., $t_b/(ns)$ and $t_f/(ns)$ operations are required per input bit. Recall that $e_b$ and $e_f$ denote the operation counts per input 8-bit byte required for computing the check tuple $P(X)$, for the basic CRC and the fast CRC, respectively. We then have $e_b = 8t_b/(ns)$ and $e_f = 8t_f/(ns)$. Using (\eivthreeC) and (\eivthreeD), we have 
$$
e_b = \frac{8t_b}{ns} = \frac{8[l_n + 1 + s(u+l_s)]}{s}	\eqno (\eivthreeE)
$$
$$
e_f = \frac{8t_f}{ns} = \frac{8(l_n + 3 + 2u)}{s}			\eqno (\eivthreeF)
$$

$$
\frac{e_b}{e_f} = \frac{t_b}{t_f} = \frac{l_n + 1 + s(u+l_s)}{l_n + 3 + 2u} \eqno(\eivthreeG)	
$$

Simple estimates are $t_b \approx nsu$ [by ignoring $l_n+1$ and $l_s$ in (\eivthreeC)] and $t_f \approx n2u$ [by ignoring $l_n+3$ in (\eivthreeD)]. Substituting these into (\eivthreeG), we have
 $$ 
\frac{e_b}{e_f} = \frac{t_b}{t_f} \approx \frac{s}{2} = \frac{h}{2} \eqno (\eivthreeH)
$$
i.e., the fast CRC is approximately $h/2$ times faster than the basic CRC.

\subsection {4.2}{CRC Complexity Under C Implementation}
Figs.\fAtwo \ and\fAfour \ show the C programs for the basic CRC and the fast CRC, respectively, which are based on Fig.\fivone \ $(s=h)$. For illustration, we let $s=16$ in the figures, and $M(X) = X^{16}+X^{15}+X^2+1$ (which generates the CRC-16) in Fig.\fAtwo. However, the following results are also valid for other values of $s$ and other generator polynomials. 

We use the following 2 rules to count the number of software operations Appendix A:
(R1) The {\it operation count} of a program statement is defined as the number of operations, other than the equal sign (=), that appear in that statement. (R2) For an if-statement, we average the operation count of the if-statement and the operation count of its alternative (e.g., an else-statement). 

The non-zero operation count for each C program statement is recorded between the comment quotes \hbox{(/* */)}. The programs show that $l_n=l_s=2$. Using (\eivthreeA) of Remark\rsix, we have $u$ = 3 if msb$(A) = 0$ and $u$ = 4 if msb$(A) = 1$. Using rule (R2), we have $u=3.5$ (which is the average of 3 and 4), as recorded in Figs.\fAtwo \ and\fAfour. Substituting these values of $l_n$, $l_s$, and $u$ into (\eivthreeE) and (\eivthreeF), we obtain
$e_b = 8(3 + 5.5s)/s$ and $e_f = 96/s$. Thus, we have   
$$
{e_b \over e_f} = {8(3+5.5h) \over 96} = 0.25+0.458h \eqno (\eivthreeI)
$$
which is within 10\% of (\eivthreeH).  For example, let $s=h=16$. Then $e_b = 8(3+5.5\times 16)/16 = 45.5$ and $e_f = 96/16=6$. Thus, $e_b/e_f = 45.5/6 = 7.58$, i.e., the fast CRC is 7.58 times ``faster" than the basic CRC. Further, if $s=h=64$, then $e_f=1.50$, $e_b=44.4$, and $e_b/e_f=29.6$, i.e., the fast 64-bit CRC is 29.6 times faster than the basic 64-bit CRC. These results are recorded in Fig.\fivtwo.

We now briefly present the complexity results for  $s, h \in \{8,16,32,64 \}$, but without the restriction $s =h$. From (\evtwentytwo) and (\evtwentysix) of Appendix A, we have
$$
e_b = \cases {8(4+5.5s)/s     & if $s<h$  \cr   
                     8(3+5.5s)/s       & if $s \ge h$}   \eqno(\eivfour)              
$$

$$
e_f = \cases {80/s                & if $s <h-1$  \cr
              100/s               & if $s =h-1$  \cr 
              96/s                & if $s =h$  \cr            
              8[12+5.5(s-h)]/s    & if $s > h$}   \eqno(\eivsix)
$$

As an example, consider a basic 16-bit CRC and the fast 16-bit CRC, which are used to protect an input message consisting of 8-bit bytes, i.e., $h=16$ and $s=8$. From the above formulas, we have $e_b = 8(4+5.5\times 8)/8 = 48$ and $e_f = 80/8 = 10$. That is, the basic CRC and the fast CRC use 48 and 10 operations per input byte, respectively, to compute their check tuples. Thus, we have $e_b/e_f = 48/10 = 4.8$, i.e., the fast CRC is 4.8 times faster than the basic CRC. The values of $e_b$, $e_f$, and $e_b/e_f$ for various $(h,s)$ pairs are recorded in Fig.\fivtwo. The results show that the complexity of the basic CRCs is rather insensitive to the values of $h$ and $s$, namely, $e_b$ varies from 44.4 to 48 (the variation is only 8.1\%). In contrast, the complexity of the fast CRCs is very sensitive to the values of $h$ and $s$, namely, $e_f$ varies from 1.50 up to 40.0.

For a given $h$, recall from Section 2.1 that we are free to choose the value of $s$. The complexity of the basic CRCs is rather insensitive to the choice of $s$. As seen in Fig.\fivtwo, when $h \in \{8,16,32,64 \}$, the complexity of the fast CRCs is fairly low when $s < h$, and is minimized when $s=h$. When $h \notin \{8,16,32,64 \}$, it is shown in Appendix A that the complexity of the fast CRCs is minimized (i.e., $e_f$ is minimized) either at $s=h$ or at $s=h-2$. 

To summarize, we introduce the new family of CRC generator polynomials that have the explicit form $F_h(X)=X^h+X^2+X+1$, for all $h \ge 4$, as well as the new technique (\eiiitwelve) for their implementation. This family includes  $F_8(X)$, which generates the ATM CRC-8. For this particular CRC, by choosing  $s=h=8$, our new technique provides a new bitwise implementation that is 3.92 times faster than the basic bitwise technique (see Fig.\fivtwo).

\figure{\includegraphics[height=7cm]{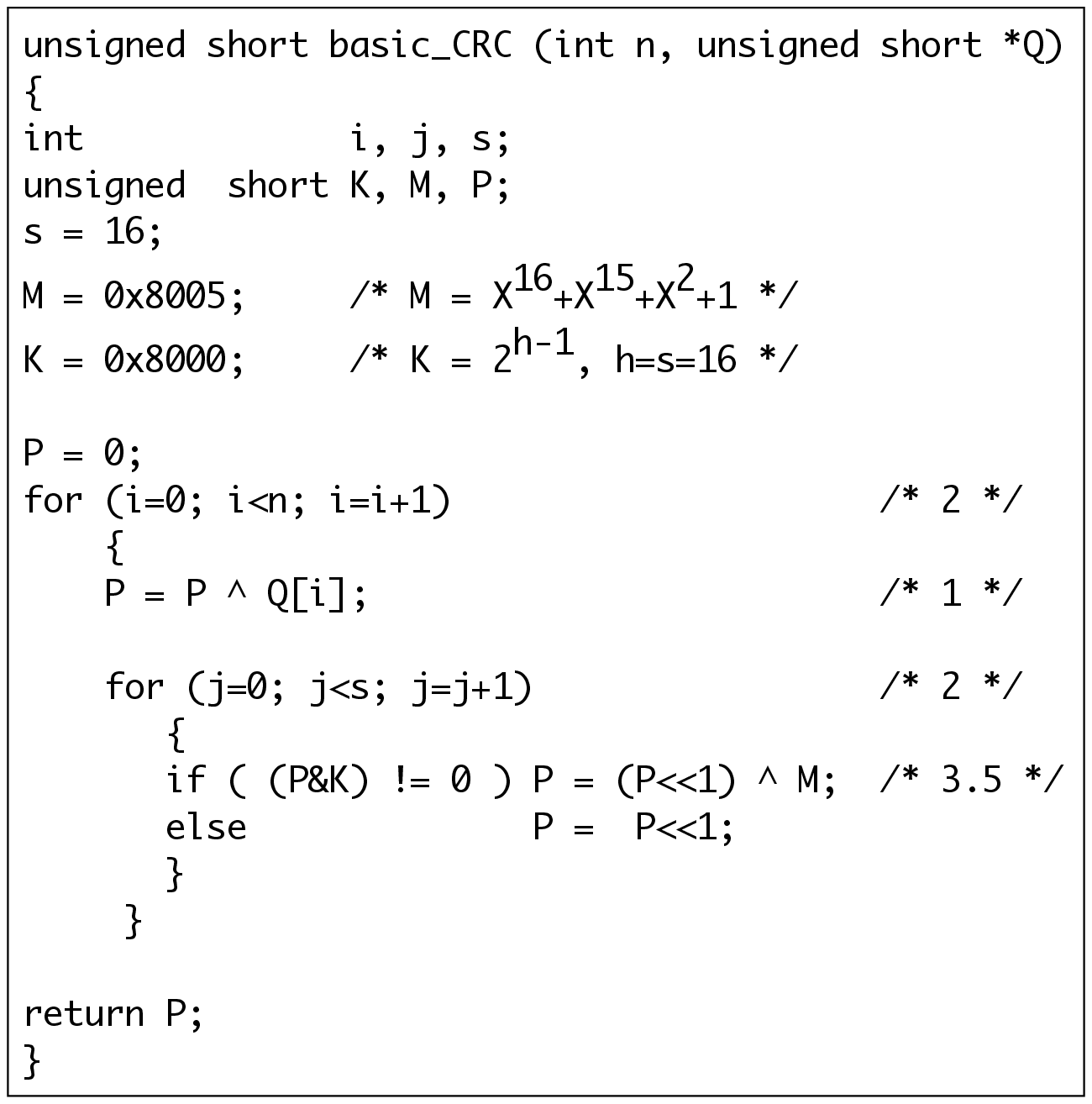}}{\fAtwo}{C program for the basic $h$-bit CRC ($s = h$).}

\figure{\includegraphics[height=7cm]{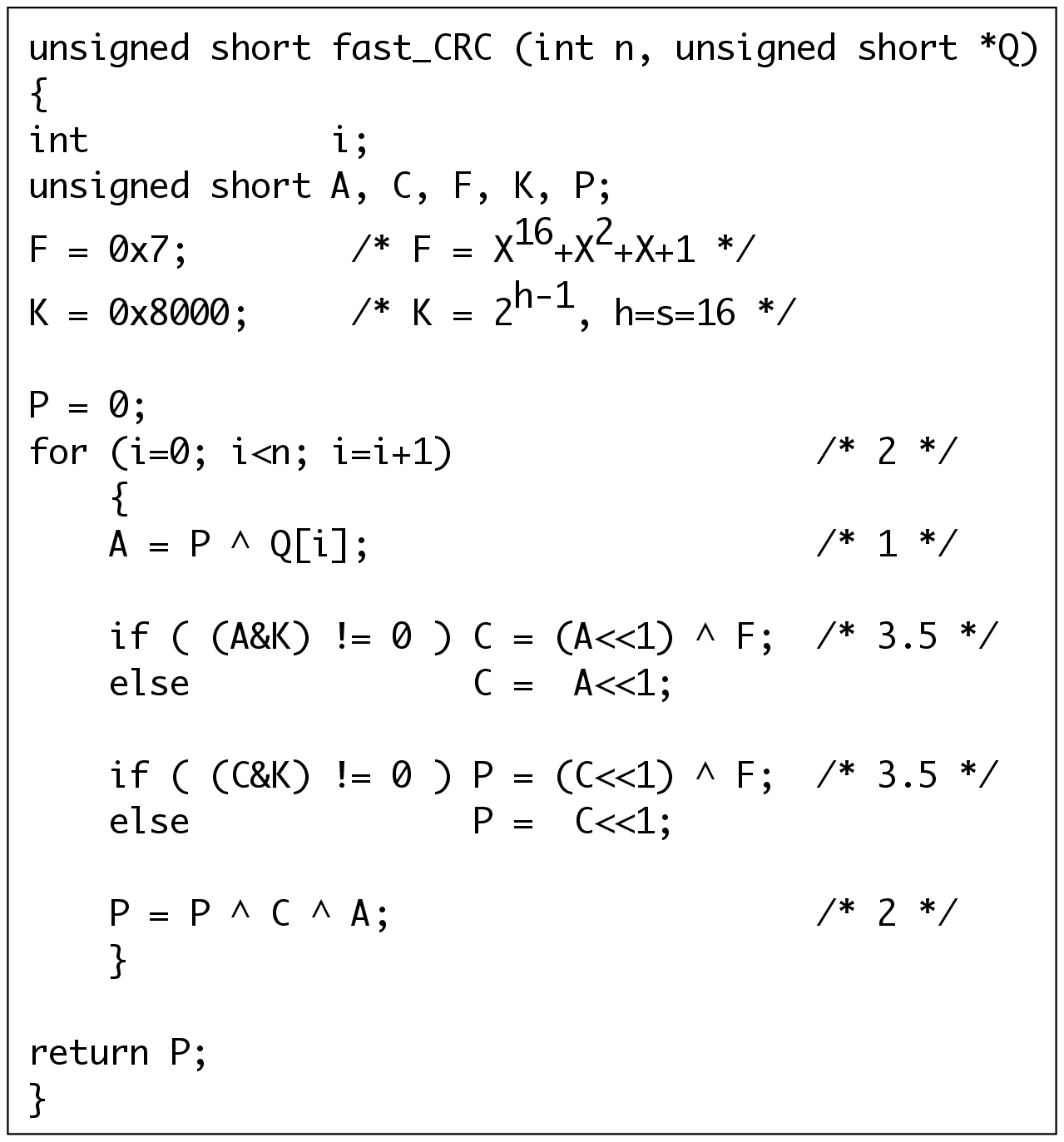} }{\fAfour}{C program for the fast $h$-bit CRC ($s = h$).}

\figure{$$
\vbox
{\ninepoint
\offinterlineskip 
\halign
{
\strut
\vrule \ #\hfill    &\vrule \ #\hfill  &\vrule \ #\hfill &\vrule \ #\hfill   &  \vrule \ #\hfill    &  \vrule  #   \cr
\noalign{\hrule} 
        &$s=8\ $  &$s=16$ &$s=32$ &$s=64\ $& \cr
        &$e_b$  &$e_b$ 	&$e_b$ 	&$e_b$& \cr
        &$e_f$  &$e_f$ 	&$e_f$ 	&$e_f$& \cr
        &$e_b/e_f$  &$e_b/e_f$ 	&$e_b/e_f$ 	&$e_b/e_f$& \cr
\noalign{\hrule}
$h=8\ $   &47.0   &45.5   &44.8   &44.4& \cr
        & {\bf 12.0}   &28.0   &36.0   &40.0& \cr
        &3.92   &1.62   &1.24   &1.11& \cr
\noalign{\hrule}
$h=16$  &48.0   &45.5   &44.8   &44.4& \cr
        &10.0   & {\bf 6.00}   &25.0   &34.5& \cr
        &4.80   &7.58   &1.79   &1.29& \cr
\noalign{\hrule}
$h=32$  &48.0   &46.0   &44.8   &44.4& \cr
        &10.0   &5.00   & {\bf 3.00}   &23.5& \cr
        &4.80   &9.20   &14.9   &1.89& \cr
\noalign{\hrule}
$h=64$  &48.0   &46.0   &45.0   &44.4& \cr
        &10.0   &5.00   &2.50   & {\bf 1.50} & \cr
        &4.80   &9.20   &18.0   &29.6& \cr
\noalign{\hrule}
}
}
$$ }{\fivtwo}{Software complexity for the basic $h$-bit CRCs ($e_b$) and the fast $h$-bit CRCs ($e_f$).}

\remark{Remark\rseven.}
There exist well-known techniques for reducing the operation counts used in CRC implementation. An example is the use of table lookup (at the cost of increased memory and cache usage), which is presented in Appendix A. Note that, to keep our C programs compact, readable, and general, we ignore software optimization techniques (such as loop unrolling) in our C programs. However, these techniques certainly can be used to reduce the operation counts in the programs. For example, if loop unrolling is used (at the cost of code size expansion) in the inner for-loop of the C program in Fig.\fAtwo, then the index increment and the end-of-loop test are eliminated, i.e., the loop overhead $l_s$ is reduced from $l_s=2$ to $l_s=0$. Using loop unrolling (i.e., $l_s=0$), it can be shown that (\eivfour) and (\eivsix) reduce to

$$
e_b = \cases {8(4+3.5s)/s     & if $s<h$  \cr   
                     8(3+3.5s)/s       & if $s \ge h$}                 
$$

$$
e_f = \cases {80/s                & if $s <h-1$  \cr
              100/s               & if $s =h-1$  \cr 
              96/s                & if $s =h$  \cr            
              8[12+3.5(s-h)]/s    & if $s > h$}   
$$ \QED

\subsection{4.3}{Other Techniques for Error-Detection Codes}
The complexity results for the basic CRC algorithm, which are rather insensitive to the input parameters $s$, $h$, and the form of the generator polynomial $M(X)$, are shown in Fig.\fivtwo. In particular, when $h=16$ and $s = 8$, we have $e_b = 48$ operations per input byte. Our CRC software implementation in C for this case is shown Fig.\fAone \ of Appendix~A, which is more efficient than the one given in [\BiR, pp.\ 555-556], which has 63 operations per input byte according to rules (R1) and (R2).

There are other CRC algorithms that are much faster than the basic algorithm. As expected, those algorithms are effective for some particular generator polynomials. For example, the clever ``add and shift" algorithm of [\Fel] is fast for the CRCs generated by $M_1(X)=X^{32}+X^{31}+X^8+1$ (for $h=32$) and  $M_2(X)=X^{64}+X^{63}+X^2+1$ (for $h=64$), which are found by computer search [\Fel]. According to rules (R1) and (R2) for determining the operation counts, these CRCs use 20 operations to process each tuple of $s=32$ input bits (see Fig.~2 in [\Fel]). Thus, these CRCs use 5 operations per input byte. In contrast, from Fig.\fivtwo, for $s=32$, our fast CRCs use only 3 and 2.5 operations per input byte for $h=32$ and $h=64$, respectively. Thus, our fast CRCs are faster than the above shift-and-add CRCs. 
Further, our fast 64-bit CRC is even much faster when $s=64$, because it uses only 1.5 operations per input byte (see Fig.\fivtwo). 

As mentioned in Section 1, alternatives to CRCs are checksums. Although checksums are weaker than CRCs, they can be substantially faster than CRCs. For example, let $s=h$ and consider the block-pariry checksum. The check tuple $P(X)$ of this checksum is simply the sum of all the input tuples, i.e., $P(X)=\sum_{i=0}^{n-1}Q_i(X)$. As shown in Section B.1, the operation count per input byte required for computing $P(X)$ of the checksum is $e=24/s$. From (\eivsix), the fast CRC has $e_f=96/s$. Thus, $e_f/e = 96/24=4$, i.e., the checksum is 4 times faster than the fast CRC.

\section {5}{SUMMARY AND EXTENSION} \secV = \pageno
Error control coding is essential for reliable transmission and storage, and CRCs are known to be effective for error detection. In software, an $h$-bit CRC is typically implemented by dividing the input message into $s$-tuples (i.e., blocks of $s$ bits). The output CRC check bits are obtained by recursively carrying the polynomial division on these tuples. 

Thus, the crucial part in CRC computation is the polynomial division on $s$-tuples. For the basic CRCs, this division requires $s$ iterations, which may be expensive for many applications. A common technique for reducing the many steps during CRC computation is to use additional memory in the form of table lookup. In this paper, we introduce the fast $h$-bit CRCs, which are generated by $F_h(X)=X^h+X^2+X+1$, as well as the new technique~(\eiiitwelve) to implement them. Using our fast CRCs, the polynomial division on $s$-tuples requires only $\max(0,s-h+2)$ iterations, which are  much less than the $s$ iterations required for the basic CRCs, as long as $s$ is chosen such that $s-h+2$ is much smaller than $s$. We study the computational complexity of the CRCs, which refers to the operation count  per input byte required for computing the CRC check tuples. Our fast CRCs have low complexity and require no table lookup. For the important case $s=h$, the fast $h$-bit CRCs are approximately $h/2$ times faster than the basic $h$-bit CRCs. 

As an illustration, we implement the CRCs in C programming language, and then study their computational complexity for the bitwise technique (i.e., without table lookup). We show that the complexity of the fast $h$-bit CRCs varies greatly with $s$, and is minimized either at $s=h-2$ or at $s=h$.  In contrast, the complexity of the  basic $h$-bit CRCs varies little with $s$. Because modern computers typically process information in bytes or words, we also present the complexity results when $s$ is restricted to multiples of byte size and word size. 

In the Appendices, we provide several extensions to the baseline ideas presented in this paper. In particular, we present the results for CRC table-lookup techniques, which illustrate tradeoffs between computational complexity and memory requirement. We show that when $s=h$, the fast CRCs can be made 20 percent faster by using tables of only 4 entries. We apply our new technique to some weaker CRCs to yield even faster CRCs, i.e., there are tradeoffs between speed and capability. 
Further, we use the new technique to construct some fast extended Hamming perfect codes. In particular, we construct $h$-bit non-CRC codes that not only have low complexity but also have the following optimal properties. They have the minimum distance $d=4$,  the burst-error-detecting capability $b=h$, and the maximum code length $2^{h-1}$.
We also apply the new technique to arbitrary CRCs, and then determine the conditions under which the new technique remains effective. In particular, the new technique is substantially faster than the basic technique for the CRC-64-ISO generated by
$ X^{64}+X^4 + X^3 + X + 1 $. {\color{red}Using computer search, we obtain CRCs that have minimum distance greater than 4 and can be efficiently implemented by the new technique. We also obtain CRC weight distributions required for estimating the undetected error probability over binary symmetric channels.}
Finally, we show how the CRCs algorithms, which are originally designed for sequential implementation on a single processor, can be adapted for parallel implementation on multiple processors.




\section{APPENDIX A}{CRC SOFTWARE IMPLEMENTATION AND COMPLEXITY EVALUATION} \secA = \pageno

\noindent The purpose of this appendix is to present software implementation for the CRC algorithms as well as to evaluate their computational complexity. {\it Software} complexity of an algorithm refers to the number of operations (i.e., operation count) used to implement the algorithm. 
Consider an $h$-bit CRC, which is generated by a polynomial $M(X)$ of degree $h$. Our goal is to compute the check $h$-tuple $P(X)$ for an input message that consists of $n$ tuples $Q_0(X), Q_1(X), \dots, Q_{n-1}(X)$. Each tuple $Q_i(X)$ has $s$ bits. 

The CRC can be implemented by any of the~4 algorithms shown in Figs.\fiitwo-4. Although the value of $h$ is fixed, we are free to choose the value of $s$. Algorithms~1 and~3 are for $s<h$, whereas Algorithms~2 and~4 are for $s \ge h$. One algorithm can be faster than another, depending the value of $s$ and the form of $M(X)$. For example, Remark\rfour \ shows that,  for bitwise implementation, Algorithm~4 is faster than Algorithm~2 when $s \ge h$. Thus, we will use Algorithm 4 for bitwise implementation when $s \ge h$, as indicated in Fig.\fvtwo.  As stated in Theorem~1, Algorithm~3 must be used for the fast CRCs when $s<h$.  Fig.\fvtwo \ lists the CRC algorithms that are used in our software implementation. 

We recognize that accurate software evaluation is complicated, and requires experiments with different processors, memory organizations, programming languages, and compilers. Other complicating factors include programming styles and the extend the CRCs must share with (or compete against) other concurrent/interupting programs.
Instead of dealing with these complex issues, which are beyond the scope of this paper, we simply use {\it software operation counts} for our complexity evaluation. Our technique of software comparison is as follows. We write a program (e.g., in C) for each CRC. We then use the operation count as the primary measure of complexity, and a CRC is said to be ``faster" than another if it has lower operation count. 

We now determine the software complexity of the CRC algorithms, which refers to the operation count per input message byte required for computing the check $h$-tuple. Let us examine Algorithms~1-4 (shown in Figs.\fiitwo-4). For each algorithm, the check tuple $P(X)$ is computed by using a loop that computes $B(X)$ for $n$ times, where $B(X)$ is given in Definition 1 for the basic CRC and in Theorem 1 for the fast CRC. In addition to $B(X)$, we also need to compute all the other terms {\it inside} the loop (which include the loop overhead). Let $r$ and $x$ be the operation counts required for computing $B(X)$ and the other terms {\it inside} the loop, respectively. Let $y$ be the operation count required for computing the terms {\it outside} the loop. Further, for each algorithm, let $t$ be the total operation count required for computing the CRC check tuple from the input message that consists of $n$ tuples. We then have $t = (x +r)n + y$. 

Let $e$ be the operation count per input {\it byte}  required for computing the check $h$-tuple. Each byte has 8 bits. Because $t$ is the operation count for computing the check $h$-tuple from the $ns$ input message bits, we have $e = 8t/(ns) = 8[(x+r)n+y]/(ns)$, i.e., 
$$
e = {8(x + r) \over s} + {8y \over ns} \eqno(\evsix)
$$
In the following, we consider $h$, $s$, and $n$ to be independent variables, and our goal is to compute $e$ in terms of $h$, $s$, and $n$ for both the basic CRCs and the fast CRCs.  That is, we can write $e=e(s,h,n)$. To compute~$e$, we need to determine $r$, $x$, and $y$, to which we add the subscripts $b$ and $f$ when they refer to the basic CRCs and the fast CRCs, respectively. That is, $r_b$, $x_b$, $y_b$, and $e_b$ refer to the basic CRCs, while $r_f$, $x_f$, $y_f$, and $e_f$ refer to the fast CRCs.

We present CRC implementation with and without table lookup. Our software programs are for $w$-bit computers that satisfy $s \le w$ and $h \le w$ (however, we allow the possibility that $h+s > w$). For example, 32-bit computers are for $s, h \le 32$ bits, while 64-bit computers are for $s, h \le 64$  bits (future 128-bit computers are for $s,h \le 128$ bits). To be specific, we implement the CRC algorithms in C, which is a highly portable general-purpose computer programming language (certainly, they can also be implemented in other computer languages). We use the following 2 simple rules to count the number of software operations [\Ngu]:

\smallskip
\remark{(R1)}  The {\it operation count} of a program statement is defined as the number of operations, other than the equal sign~(=), that appear in that statement. 

\remark{(R2)}  For an if-statement, we average the operation count of the if-statement and the operation count of its alternative (e.g., an else-statement). 
\smallskip

Let us consider examples on how to use rule (R1). The statement {\tt C = (A$<<$1)$\wedge$F} will count as 2 operations \hbox{($<<$ and $\wedge$ )}. Note that ``=" does not count as an operation. Next, consider the statement  {\tt for(i=0; i$<$n; i=i+1)$\{\ \}$}. This implements a (null) loop of $n$ iterations,  each iteration has 2 operations ($<$ and +). Thus, the total operation count for this loop statement is $2n$. The for-loop above is equivalent to the while-loop \hbox{{\tt i=0; while(i$<$n)$\{$i=i+1;$\}$}} which, of course, also has $2n$ operations. 

We now show examples about rule (R2). Suppose that {\tt K = 1}, and consider the following 2 statements:

{\obeylines
{\tt \obeyspaces{if ((A\&K) != 0) C =(A$<<$1)$\wedge$F;}}
{\tt \obeyspaces{else            C = A$<<$1;}}
}

\noindent Here, the if-statement has 4 operations (\&, !=, $<<$, $\wedge$), and the else-statement has 3 operations \hbox{(\&, !=, $<<$)}. Thus, the above 2 statements can be considered as a single statement that has 3.5 operations (i.e., the average of 4 and 3). 

The above 2 statements are equivalent to the the following 2 statements:
{\obeylines
{\tt \obeyspaces{C = A$<<$1;}}
{\tt \obeyspaces{if ((A\&K) != 0) C = C$\wedge$F;}}
}

\noindent Here, the first statement has 1 operation, and the second statement has 2.5 operations. Thus, the 2 statements together also have 3.5 operations as expected. Note that {\tt (A\&K)}$\in \{0,1\}$, because {\tt K = 1}. Here, for simplicity, we assume that {\tt (A\&K)} takes the values 0 and 1 with equal probability of 1/2. Suppose now that {\tt K~=~3}. We then have{\tt (A\&K)}$\in \{0,1,2,3\}$. By assuming that {\tt (A\&K)} takes the values 0, 1, 2, and 3 with equal probability of~1/4, the above if-statement (which has 4 operations) is executed with probability 3/4 and the else-statement (which has 3 operations) is executed with probability 1/4. Thus, these if-else statements can be considered as a single statement that has $4\times 3/4 + 3\times 1/4=3.75$ operations.

\figure{
{\figureFont
$$
\vbox
{
\offinterlineskip 
\halign
{
\strut
\vrule \quad #    &\vrule \quad #  &  \vrule \quad #    &  \vrule  #   \cr
 \noalign{\hrule} 
        &bitwise  &table lookup & \cr
\noalign{\hrule}
basic CRC   &Algo.\ 1 ($s<h$)        &Algo.\ 3 ($s<h$)& \cr
                    &Algo.\ 4 ($s \ge h$)   &Algo.\ 2 ($s \ge h$)& \cr
\noalign{\hrule}
fast CRC     &Algo.\ 3 ($s<h$)        & Algo.\ 3 ($s<h$)& \cr
                   &Algo.\ 4 ($s \ge h$)   &Algo.\ 2 ($s \ge h$)& \cr
\noalign{\hrule}
}
}
$$
}
}{\fvtwo}{CRC algorithms used in software implementation.}

\remark{Remark\reight.} 
Rules (R1) and (R2) serve as a simple technique for comparing the complexity of different CRCs, i.e., they will be used to obtain a first-order estimation of the ratio $e_b/e_f$. These rules are intended only for CRC algorithms that are implemented in C, and not for other types of algorithms or other programming languages. As seen in the following, our CRC software implementation uses only a small number  of elementary C operators (namely, +, $<<$, $>>$, =, ==, !$=$, $<$, $<$=, \&, and~$\wedge$) and C keywords (namely, {\it char, short, int, long, unsigned, if, else, for, while,} and {\it return}). Our following C programs (using the big-endian convention) for the CRCs are written in a style that is intended to be simple and straightforward. See also Remark\rnine.

Other techniques for counting operations are also possible. For example, consider rule (R1$^\prime$), which is defined as rule~(R1) but also counts the equal sign (=) as an operation. Let $e_b^\prime$ and $e_f^\prime$ denote the resulting operation counts under (R1$^\prime$). We must have $e_b^\prime > e_b$ and $e_f^\prime > e_f$. Although the difference between $e_b$ and $e_b^\prime$ (as well as between $e_f$ and $e_f^\prime$) can be significant, the difference between the ratios $e_b/e_f$ and $e_b^\prime/e_f^\prime$ are typically not significant. For example, let $s=h=32$. From Fig.\fivtwo, we have $e_b=44.8$, $e_f=3$, and $e_b/e_f=14.9$. Under rule (R1$^\prime$), it can be shown that $e_b^\prime=61.5$, $e_f^\prime=4.25$, and $e_b^\prime/e_f^\prime=14.5$, i.e., $e_b^\prime/e_f^\prime \approx e_b/e_f$. Note that rule (R1), which is used in this paper, is slightly simpler to use than rule (R1$^\prime$). Thus, our technique for counting software operations is reasonable for the purpose of {\it complexity  comparison}, i.e., we are more interested in the ratio $e_b/e_f$, rather than in $e_b$ and $e_f$.

Here, for simplicity, we assign the same unit cost to each operation. A more elaborate technique would assign different costs to different operations. However, this assignment depends on many factors (such as computer hardware, operating system, processor architecture, and memory organization), which are outside the focus of this paper.
\QED

Let us now compute $x$ and $y$ in (\evsix). The computation of $r$ is deferred to later subsections.
First, consider Fig.\fAone, which shows the C program for bitwise implementation of the basic CRC for the case $s<h$. As indicated in Fig.\fvtwo, this program is based on Algorithm 1. In this program, we assume that $h \in \{8, 16, 32, 64 \}$, i.e., $h$ is the size (in bits) of one of the natural {\it unsigned types} of C: unsigned char, unsigned short int, unsigned int, or unsigned long int. The input is the $n$ message $s$-tuples $Q[0], Q[1], \dots, Q[n-1]$,  and the output is the CRC check $h$-tuple $P$.

We then apply rules (R1) and (R2) to the program shown in Fig.\fAone \ to obtain the desired operation counts. The non-zero operation count for each program statement is recorded between the comment quotes (/* */). Recall that the total operation count for computing the check tuple $P$ from the $n$ input tuples is $t_b=(x_b+r_b)n+y_b$. Here, $r_b$ is the operation count required for computing $B(X)$, which is inside the loop indexed by $i$, $x_b$ is the operation count required for computing all the other terms in the loop besides $B(X)$, and  $y_b$ is the operation count required for computing all the terms outside the loop. From Fig.\fAone, we have $x_b=4$ and $y_b=0$.

To summarize, for $h \in \{8, 16, 32, 64 \}$ and $s < h$, we have $x_b = 4$ and $y_b = 0$, which are recorded in Fig.\fvfour. For illustration, we let $h=16$, $s=8$ and $M(X) = X^{16}+X^{15}+X^2+1$ (which generates the well-known CRC-16) in Fig.\fAone. When $h \notin \{8, 16, 32, 64 \}$ and $s < h$, the computational complexity is slightly higher, namely,  $x_b = 4$ and $y_b = 1$. 

Fig.\fAtwo \ shows the C program for the basic CRC when $s=h$. Next, consider Fig.\fAthree,  which shows the C program for the fast CRC when $h \in \{8, 16, 32, 64 \}$ and $s<h-1$. It can be shown that $x_f = 6$ and $y_f = 0$  for this case. Again, for illustration, we let $h=16$ and $s=8$ in Fig\fAthree. Similarly, we can compute the values of $x$ and $y$ for all the cases for both the basic CRCs and the fast CRCs. The results are summarized in Fig.\fvfour. 

Using Fig.\fvfour, the expression (\evsix) can be simplified as follows. Let $z$ be the ratio of the 2 terms on the right-hand side of (\evsix), i.e., 
$$
\eqalign{z &= \frac{{8(x+r)}/{s}}{{(8y)}/{(ns)}} \cr
			&=\frac{(x+r)n}{y} }
$$	 
From Fig.\fvfour, we have $0 \le y \le 1$ and $x \ge 3$, which implies that $z \ge (x+r)n \ge (3+r)n \ge 3n$.
Thus, $8y/(ns)$ is much smaller than $8(x + r)/s$, because we assume in this paper that $n$ is not too small (i.e., we assume that $n>4$). Thus, the term $8y/(ns)$ can be dropped from (\evsix). The operation count per input byte required for computing the CRC check $h$-tuple then simplifies to    
$$
e =  {8(x + r) \over s}  \eqno(\evten)
$$
where $x$ is determined from Fig.\fvfour, which depends only on $s$ and $h$, i.e., $x=x(s,h)$. Recall that $r$ denotes the operation count required for computing $B(X)$, which also depends only on $s$ and $h$ [see (\eiitwentyeight)], i.e., $r=r(s,h)$. It follows from (\evten) that $e$ now also depends only on $s$ and $h$, i.e., $e=e(s,h)$. From (\evten), we also have
$$
\frac{e_b}{e_f} = \frac{x_b+r_b}{x_f+r_f}
$$
where $x_b$ and $x_f$ are given in Fig.\fvfour.  In the following, using rules~(R1) and (R2), we compute~$r_b$ and $r_f$ for both the bitwise and the table-lookup techniques.

\figure{ \includegraphics[height=7.5cm]{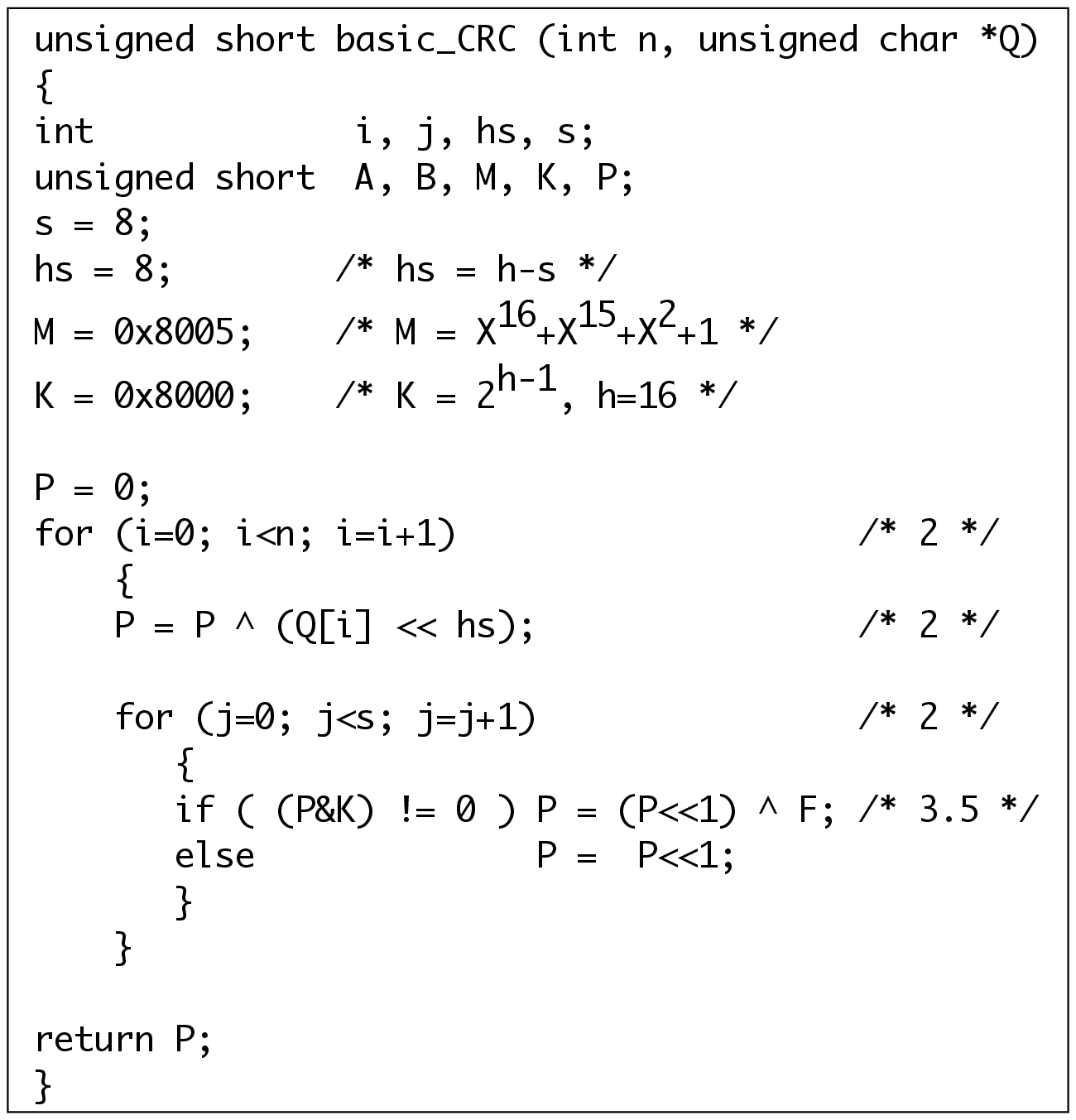} } 
{\fAone}{C program for the basic $h$-bit CRC ($s < h$).}

\figure{ \includegraphics[height=6cm]{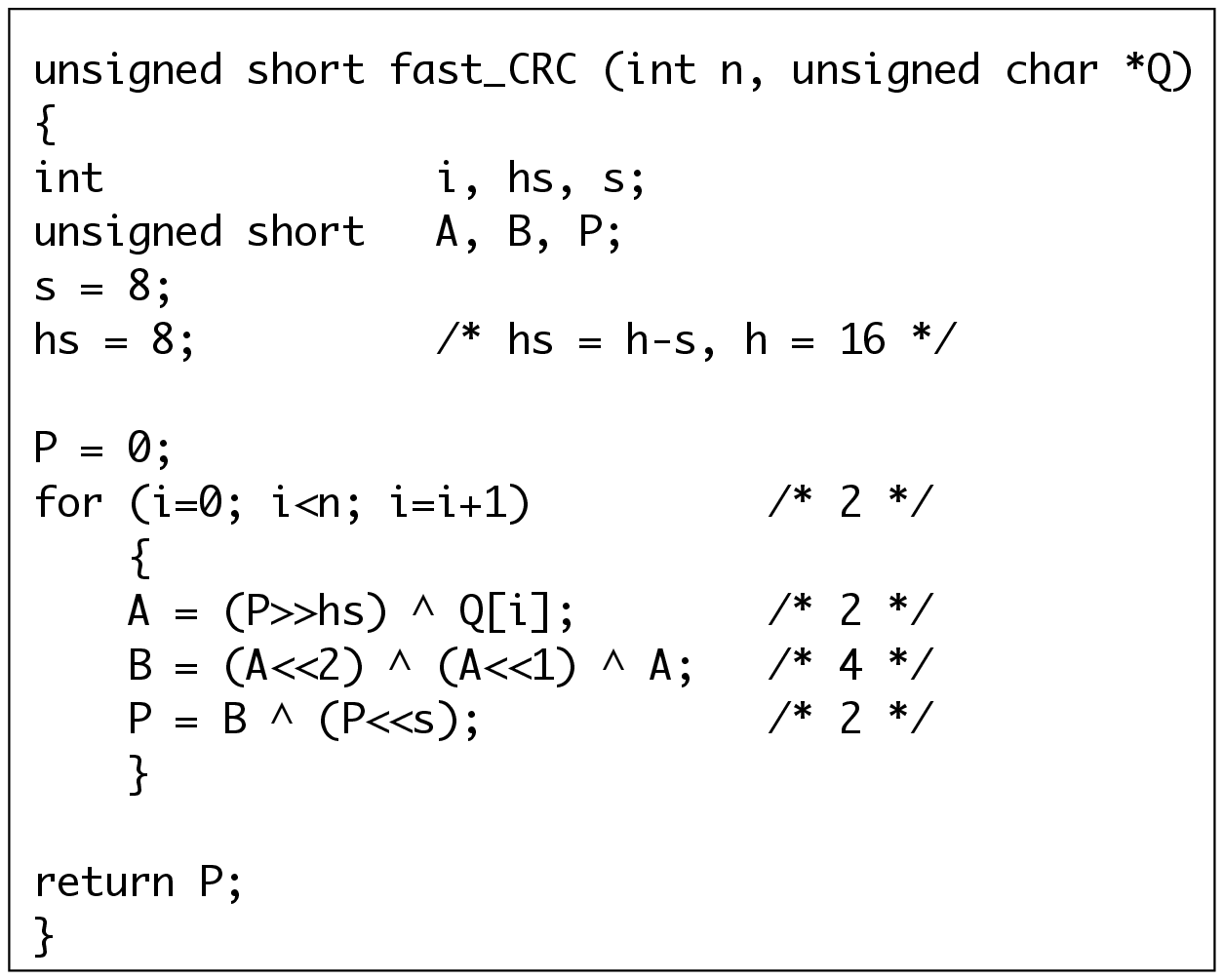} } {\fAthree}{C program for the fast $h$-bit CRC ($s < h-1$).}

\figure{
{\figureFont 
$$
\vbox
{
\offinterlineskip 
\halign
{
\strut
\vrule \  #    &\vrule \  #   &# \hfill &\vrule \ #  \hfill  &  \vrule  # \cr
 \noalign{\hrule} 
        &$x$  & &$y$ &\cr
\noalign{\hrule}
Algo.\ 1      &4 & &0 if $h = 8,16,32,64$   &\cr
$(s<h)$      &  & &1 if $h \ne 8,16,32,64$   &\cr
\noalign{\hrule}
Algo.\ 2         &3 &if $s=h$ &0   &\cr
$(s \ge h)$     &4 &if $s>h$ &0  &\cr
\noalign{\hrule}
Algo.\ 3         &6 & if $h = 8,16,32,64$     &0 if $h = 8,16,32,64$   &\cr
$(s<h)$      &7  & if $h \ne 8,16,32,64$  &1 if $h \ne 8,16,32,64$   &\cr
\noalign{\hrule}
Algo.\ 4         &3 &   &0 if $s=h$&\cr
$(s \ge h)$     &  &   &1 if $s>h$&\cr
\noalign{\hrule}
}
}
$$
}
}{\fvfour}{Values of $x$ and $y$.}

\subsection {A.1}{CRC Software Implementation: Bitwise Technique (Without Table Lookup)}
According to Fig.\fvtwo, the bitwise implementation of the the basic CRCs uses Algorithm 1 for $s<h$, and Algorithm 4 for $s \ge h$. From Fig.\fvfour, we then have 
$$
x_b = \cases {4     & if $s<h$  \cr
              3      & if $s \ge h$}   
$$
Substituting $x_b$ into (\evten), we have   
$$
e_b = \cases {8(4+r_b)/s     & if $s<h$  \cr
                     8(3+r_b)/s      & if $s \ge h$}   \eqno(\evtwelve)                 
$$
where $r_b$ denotes the operation count required for computing $B(X)$ of the basic CRCs.

According to Fig.\fvtwo, the bitwise implementation of the the fast CRCs uses Algorithm~3 for $s<h$, and Algorithm 4 for $s \ge h$. From Fig.\fvfour, we then have
$$
x_f = \cases {6             & if $s < h$ and $h=8,16,32,64$ \cr
                      7             & if $s <h$ and $h \ne 8,16,32,64$ \cr
                      3             & if $s \ge h$} 
$$
Substituting $x_f$ into (\evten), we have  
$$
e_f = \cases {8(6+r_f)/s             & if $s < h$ and $h=8,16,32,64$ \cr
                      8(7+r_f)/s             & if $s <h$ and $h \ne 8,16,32,64$ \cr
                      8(3+r_f)/s             & if $s \ge h$} \eqno(\evfourteen)  
$$
where $r_f$ denotes the operation count required for computing $B(X)$ of the fast CRCs. Both $r_b$ and $r_f$ are computed in the following subsections.
\subsubsection{A.1.1}{Basic CRCs} 
Recall that $B(X)$ for the basic CRCs is given in Definition 1. First, consider the case $s<h$, and let us revisit Fig.\fAone. This figure contains the loop (indexed by $j$) for computing $B(X)$, which is based on Remark\rsix. The figure shows that the operation count required for computing $B(X)$ is $r_b = 5.5s$. Next,  for the case $s \ge h$, it can also be shown that $r_b = 5.5s$ (see Fig.\fAtwo). To summarize, we have   
$$
r_b = 5.5s \eqno(\evtwenty)
$$
Substitute (\evtwenty) into (\evtwelve) we have   
$$
e_b = \cases {8(4+5.5s)/s     & if $s<h$  \cr
              8(3+5.5s)/s     & if $s \ge h$}   \eqno(\evtwentytwo)                
$$  
Note that (\evtwenty) is derived from the C programs that do not use loop unrolling (which is also the case for the C programs presented in [\BiR]). If loop unrolling is used, (\evtwenty) reduces to $r_b=3.5s$.  

Here, our software implementations of the basic CRCs are general, i.e., they are applicable to all generator polynomials $M(X)$ and to a wide range of processor architectures. For some specific generator polynomials that have some desirable properties, alternative implementations (such as shift and add [\Fel], and on the fly [\Per, \RaG]) may have lower complexity. 
Thus, we concentrate on the general nature of the algorithms rather than attempting to deal with specific types of generator polynomials. Also, for our C programs, we are more concerned with their readability and less concerned with optimization techniques such as loop unrolling and use of register variables (see Remark\rseven). 

\subsubsection{A.1.2}{Fast CRCs} 
Recall that $B(X)$ of the fast CRCs is given in Theorem 1. First, assume that $s<h-1$. The C program for this case is shown in Fig.\fAthree, which contains the procedure for computing $B(X)$. Applying rules (R1) and (R2) to Fig.\fAthree, we observe that the operation count required for computing $B(X)$ is $r_f = 4$. Next, assume that $s=h$. The C program for this case is shown in Fig.\fAfour, which yields $r_f=9$. The C programs for all the other cases can also be written, and the resulting software complexity can also be determined. Following is the list of the operation counts for all the cases:   
$$
r_f = \cases {4                &if $s < h-1$  \cr
              6.5             &if $s=h-1$   \cr
              9               &if $s=h$   \cr
              9+5.5(s-h)  &if $s > h$} \eqno(\evtwentyfour) 
$$
Substituting (\evtwentyfour) into (\evfourteen), we have
   
$$
e_f = \cases {80/s             & if $s <h-1$ and $h=8,16,32,64$ \cr
                     88/s             & if $s <h-1$ and $h \ne 8,16,32,64$ \cr
                     100/s           & if $s =h-1$ and $h=8,16,32,64$ \cr
                     108/s           & if $s =h-1$ and $h \ne 8,16,32,64$ \cr
                     96/s            & if $s=h$ \cr
                     8[12+5.5(s-h)]/s    & if $s > h$} \eqno(\evtwentysix)   
$$ 

The operation count per input byte $e_f$ for the fast $h$-bit CRC given in (\evtwentysix) is a function of $s$, which is the size of each input tuple $Q_i(X)$. We now determine the value of $s$ that minimizes $e_f$. These optimal values are denoted by $s^*$ and $e^*_f$. 

First, assume that $h \in \{8,16,32, 64\}$. For each   $h \in \{8,16,32, 64\}$, we can search for an $s \in \{1, 2, \dots, 64\}$ such that $e_f$ in (\evtwentysix) is minimized. Our search shows that
$$
e^*_f = \cases {80/(h-2) & if $h=16,32,64$ \cr
                96/h     & if $h=8$} \eqno(\evtwentysixa)
$$
which is achieved when
$$
s^* = \cases {h-2  & if $h=16,32,64$ \cr
              h    & if $h=8$} \eqno(\evtwentysixb) 
$$

Next, assume that $h \notin \{ 8,16,32, 64\}$. For each   $h \in \{8,16,32, 64\}$, $4 \le h \le 64$, we can search for an $s \in \{1, 2, \dots, 64\}$ such that $e_f$ in (\evtwentysix) is minimized. Our search shows that

$$
e^*_f = \cases {88/(h-2) & if $h > 24$ \cr
                96/h     & if $4 \le h \le 24$} \eqno(\evtwentysixc) 
$$
which is achieved when
$$
s^* = \cases {h-2  & if $h > 24$ \cr
              h    & if $4 \le h \le 24$} \eqno(\evtwentysixd) 
$$

Thus, (\evtwentysixb) and (\evtwentysixd) show that the complexity of the fast $h$-bit CRCs is minimized (i.e., $e_f$ is minimized) at either $s=h$ or $s=h-2$, where $s$ is the number of bits  in each input tuple $Q_i(X)$.

For example, by letting $h=16$, the optimal size for each input tuple $Q_i(X)$ is $s^* = h-2=14$ [by (\evtwentysixb)], and the  corresponding minimum operation count is $e^*_f = 80/(h-2) = 80/14 = 5.71$ [by (\evtwentysixa)].  Information on computers is typically organized in bytes or words. Thus, it is of interest to determine the optimal value of $e_f$ when $s$ is restricted to a multiple of byte size and word size, i.e., when $s$ is a multiple of 8, 16, 32, 64. These optimal values, which are obtained from (\evtwentysix), are shown in Fig.\fvfive. 

Recall that $e_f = e_f(s,h)$, i.e., $e_f$ is a function of $s$ and $h$. In Fig.\fvfive, for a given $h$, $s^{\rm (opt)}$ denotes the value of  $s$, $1 \le s \le 64$, that minimizes $e_f(s,h)$, and the corresponding minimum $e_f(s,h)$ is denoted by $e_f^{\rm (opt)}$. Thus, we have $e_f^{\rm (opt)}=e_f(s^{\rm (opt)},h) \le e_f(s,h)$ for all $1 \le s \le 64$. Similarly, $s^{\rm (byte)}$ denotes the value of  $ s \in \{8,16,24,32,40,48,56,64\}$ that minimizes $e_f(s,h)$, and the corresponding minimum $e_f(s,h)$ is denoted by $e_f^{\rm (byte)}$. Finally, $s^{\rm (word)}$ denotes the value of $ s \in \{8,16,32,64\}$ that minimizes $e_f(s,h)$, and the corresponding minimum $e_f(s,h)$ is denoted by $e_f^{\rm (word)}$. For example, by letting $h=64$, we have $s^{\rm (opt)}=62$, $e_f^{\rm (opt)}= 1.29$, $s^{\rm (byte)}=56$, $e_f^{\rm (byte)}= 1.43$, $s^{\rm (word)}=64$, $e_f^{\rm (word)}= 1.50$.  In general, we must have $e_f^{\rm (opt)} \le e_f^{\rm (byte)} \le e_f^{\rm (word)}$. 


\figure{$$
\vbox
{\eightpoint
\offinterlineskip 
\halign
{
\strut
\vrule \quad #      &\vrule \quad # &\vrule \quad #   &  \vrule \quad #    &  \vrule  #   \cr
\noalign{\hrule} 
$h$ & $s^{\rm (opt)}$   & $s^{\rm (byte)}$   & $s^{\rm (word)}$   & \cr
    & $e^{\rm (opt)}_f$ & $e^{\rm (byte)}_f$ & $e^{\rm (word)}_f$ & \cr
\noalign{\hrule}
4 & 4 & 8 & 8& \cr
  & 24.0   & 34.0   & 34.0 & \cr
\noalign{\hrule}
6 & 6 & 8 & 8& \cr
  & 16.0   & 23.0   & 23.0 & \cr
\noalign{\hrule}
8 & 8 & 8 & 8& \cr
  & 12.0   & 12.0   & 12.0 & \cr
\noalign{\hrule}
10 & 10 & 8 & 8& \cr
  & 9.60   & 11.0   & 11.0 & \cr
\noalign{\hrule}
12 & 12 & 8 & 8& \cr
  & 8.00   & 11.0   & 11.0 & \cr
\noalign{\hrule}
16 & 14 & 16 & 16& \cr
  & 5.71   & 6.00   & 6.00 & \cr
\noalign{\hrule}
20 & 20 & 16 & 16& \cr
  & 4.80   & 5.50   & 5.50 & \cr
\noalign{\hrule}
24 & 22 & 24 & 16& \cr
  & 4.00   & 4.00   & 5.50 & \cr
\noalign{\hrule}
32 & 30 & 32 & 32& \cr
  & 2.67   & 3.00   & 3.00 & \cr
\noalign{\hrule}
40 & 38 & 40 & 32& \cr
  & 2.32   & 2.40   & 2.75 & \cr
\noalign{\hrule}
48 & 46 & 48 & 32& \cr
  & 1.91   & 2.00   & 2.75 & \cr
\noalign{\hrule}
56 & 54 & 56 & 32& \cr
  & 1.63   & 1.71   & 2.75 & \cr
\noalign{\hrule}
64 & 62 & 56 & 64& \cr
  & 1.29   & 1.43   & 1.50 & \cr
\noalign{\hrule}
}
}
$$
}{\fvfive}{The optimal values of $s$ and $e_f$ for the $h$-bit fast CRCs \hbox{({$s^{\rm (opt)}$ = best of $s \in \{1,2, \dots, 63, 64\}$}, $s^{\rm (byte)}$ = best of $s \in \{8,16,24,32,40,48,56,64\}$}, \hbox{$s^{\rm (word)}$ = best of $s \in \{8,16,32,64\}$)}.}

\remark{Remark\rnine.} Our C programs for the CRCs, which follow directly from the pseudocodes in Figs.\fiitwo-4, are written in a style that is intended to be simple and straightforward. For readability, we use  an array (instead of a pointer) for the input $s$-tuples $Q_i$. We also avoid using any C syntax that obscures the operation counts. For example, the more explicit syntax \ {\tt if((P\&K)!=0)} \ is used instead of the shorthand \ {\tt if(P\&K)}. Although these 2 expressions are equivalent, the former shows 2 operations more clearly. If desired, these C programs can be rewritten in pointer and shorthand style, for example, as shown in Figs.\fAtwop \ and\fAfourp, which are equivalent to Figs.\fAtwo \ and\fAfour, respectively. \QED

\figure{ \includegraphics[height=8cm]{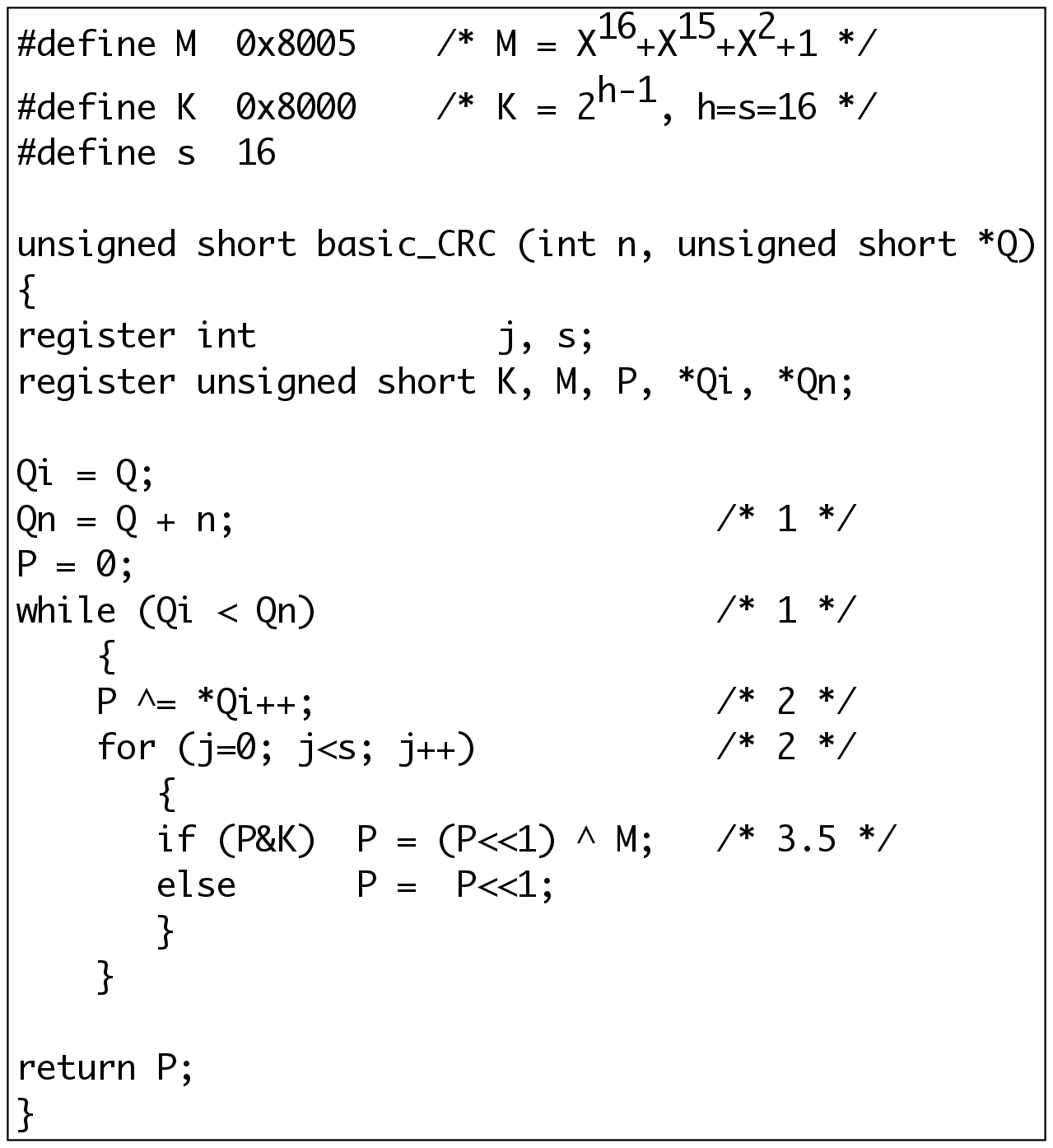} } 
{\fAtwop}{C program for the basic $h$-bit CRC  in pointer style ($s = h$).}

\figure{ \includegraphics[height=8cm]{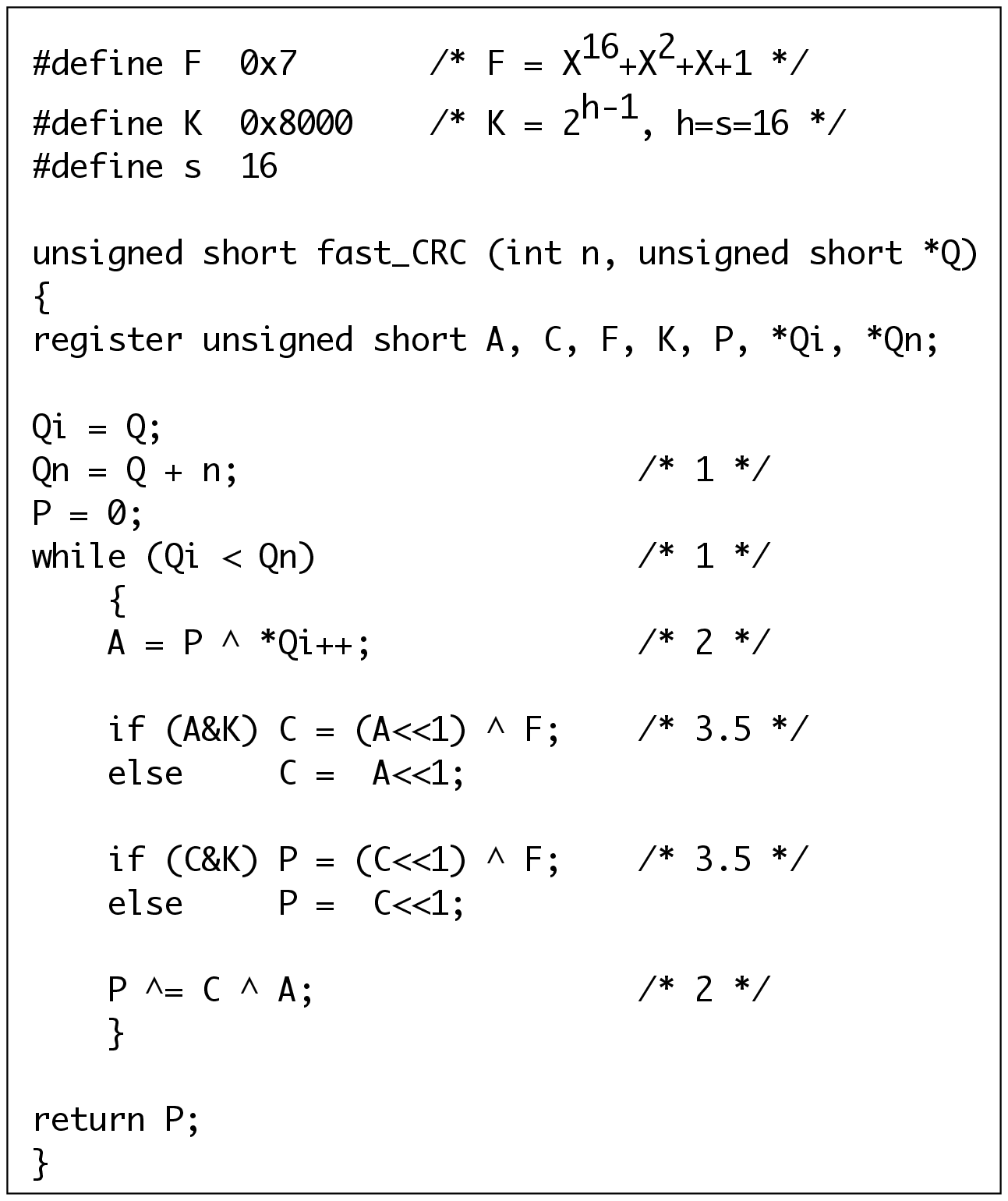} }
{\fAfourp}{C program for the fast $h$-bit CRC  in pointer style ($s = h$). } 

\remark{Remark\rten.} When $s$ is small, we can compute $B(X) = \MOD{A(X)X^h}{M(X)}$, where degree$(A(X))<s$, using a series of if-else statements as follows. For example, suppose that $s=2$ and $M(X) = F_h(X) = X^h+X^2+X+1$. Then $A(X) \in \{0, 1, X, X+1 \}$, and it can be shown that
 $$
 B(X) = \cases {0                   & if $A(X) = 0$      \cr
                         X^2+X+1      & if $A(X) = 1$      \cr
                         X^3+X^2+X  & if $A(X) = X $     \cr
                         X^3+1          & if $A(X) = X+1$}
 $$                        
 Note that polynomials can also be represented as integer numbers, e.g., the polynomial $X^3+X^2+X$ is equivalent to the decimal number 14. Thus, $B(X)$ can be computed using the C program segment shown in Fig.\fAfive. Applying rules (R1) and (R2) to this C program segment, the operation count for computing $B(X)$ is 1, 2, 3, or 3 if $A(X)$ is 0, 1, 2, or 3 (in integer representation), respectively. We now assume that  the bits 0 and 1 of the input message occur equally likely. Thus, $A(X)$ assumes one of the values 0, 1, 2, 3 with equal probability of 1/4. Then, on the average, the operation count for computing $B(X)$ is $(1+2+3+3)/4$ = 2.25. In general, this technique for computing $B(X)$ can be applied to any generator polynomial $M(X)$.

Let $k$ denote the operation count required for computing $B(X)= \MOD{A(X)X^h}{M(X)}$ using this if-else technique, where degree$(A(X))<s$. Note that $k$ depends on $s$, i.e., $k=k(s)$. As shown above, we then have $k(2)=2.25$, which is smaller than both $r_b$ and $r_f$ given in (\evtwenty) and (\evtwentyfour). In general, it can be shown that $k(s) = 2^{s-1} + 2^{-1} - 2^{-s}$ for $s \ge 1$. In particular, $k(1)=1$. Thus, this if-else technique is effective for small $s$, such as $s=1, 2$, or 3.  However, in this paper, we are mainly  concerned with the case $s \ge 8$, which is more commonly used in practice. For this case, $k(s)$ is much greater than both $r_b$ and $r_f$. Thus, when $s \ge 8$, the if-else technique is much more expensive than the basic and the new techniques, and it will not be discussed further in this paper. Note also that this if-else technique is different from the table-lookup technique (which will be discussed later).

\figure{ \includegraphics[height=3.25cm]{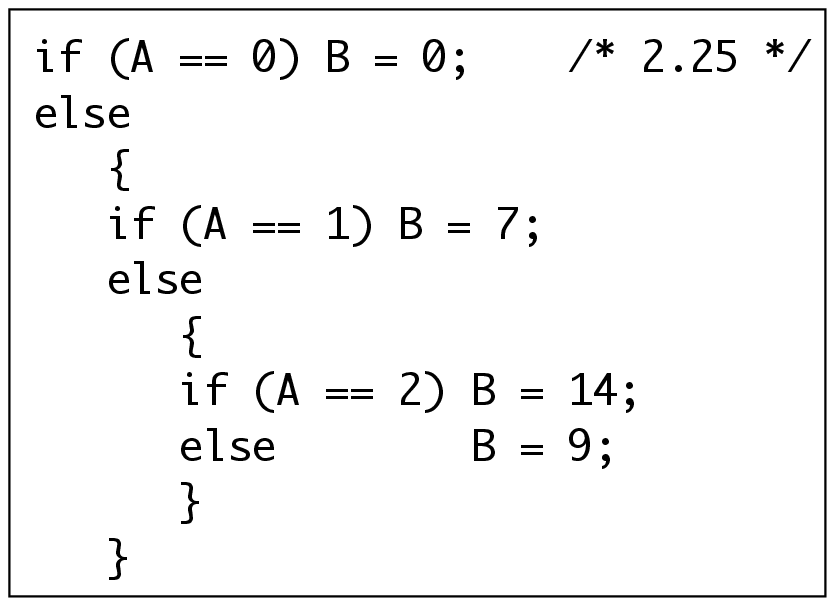} }{\fAfive}{C program segment for computing $B(X) = \MOD{A(X)X^h}{F_h(X)}$ when  $s=2$.} \QED

\remark{Remark\releven.} 
Consider an input message $U(X)$, which is protected by an $h$-bit CRC. Recall that we implement this CRC by first dividing the input message into $n$ $s$-tuples $Q_i(X)$, i.e., we have $U(X)=(Q_0(X), Q_1(X),\dots,Q_{n-1}(X))$. These $s$-tuples $Q_i(X)$ then become the input to one of the CRC algorithms. Fig.\fivtwo \ shows that the complexity of the basic CRCs is rather insensitive to the values of $s$, whereas the complexity of the fast CRCs is very sensitive to the values of $s$. 
Recall that the operation count per input byte $e_f$ in (\evtwentysix) is a function of $s$ and $h$, i.e., $e_f=e_f(s,h)$. For example, Fig.\fivtwo \ shows that $e_f(8,16)=10$ and $e_f(16,16)=6$, i.e., $e_f(16,16) < e_f(8,16)$. 

So far, we do not address the cost of obtaining the tuples $Q_i(X)$. We now address the impact of this cost by considering the fast 16-bit CRC, i.e., $h=16$. Suppose that the input message $U(X)$ originally  consists of $m$ bytes, $m \ge 4$, denoted by $I_0(X), I_1(X), \dots , I_{m-1}(X)$. Each $I_i(X)$ is an 8-tuple. Thus, we need to organize the bytes $I_j(X)$ into the $s$-tuples $Q_i(X)$. One technique is to simply set $Q_i(X) = I_i(X)$, i.e., each $Q_i(X)$ is an 8-tuple. Let $e$ be the operation count  per input byte required for CRC encoding. We then have $s=8$, and hence $e = e_f(8,16)=10$.

An alternative technique is first to pair 2 adjacent  input bytes to form 16-bit tuples from which the check bits are then computed. More precisely, we now let $s=16$ and define the new 16-tuples $Q_i(X)$ by
$$
Q_0(X)= \cases{(I_0(X),I_1(X)) & if $m$ is even \cr
               \quad \quad (0,I_0(X)) & if $m$ is odd}
$$ 
and               
$$
Q_i(X)= \cases{(I_{2i}(X),I_{2i+1}(X)) & if $m$ is even \cr
               (I_{2i-1}(X),I_{2i}(X)) & if $m$ is odd}
$$ 
for
$$
0<i \leq \cases{(m-2)/2 & if $m$ is even \cr
                (m-1)/2 & if $m$ is odd}
$$               
The algorithm for pairing the bytes and then computing the fast 16-bit CRC is shown in Fig.\fvsix. Using this algorithm, it can be shown that the operation count per input byte is $e=7.5$, which is lower than $e_f(8,16)=10$ of the non-pairing technique. Note that $e=7.5 > e_f(16,16)=6$, because of the additional cost for pairing the input bytes to form  the new 16-bit tuples to be used for the CRC computation.            
\figure{
{
\figureFont
$$
\vbox
{ 
\offinterlineskip 
\halign
{ %
\strut 
\vrule \ \hss #  \vrule & \ # \hss \vrule \cr
\noalign{\hrule}
1&	if ($m$ is even)	\cr
2&	\kern 4ex $\{Q = I_0X^8 + I_1; \ i=2;\}$	\cr
3&	else	\cr
4&	\kern 4ex $\{Q = I_0; \ i=1;\}$	\cr
5&	$B = \MOD{Q(X^2 + X +1)}{F_{16}}$;\cr
6&	while $(i < m-1)$ \cr
7& \kern 4ex $\{$		\cr
8&	\kern 4ex $Q = I_iX^8; \ i=i+1$;	\cr
9&	\kern 4ex $Q = Q + I_i; \ i=i+1$;	\cr
10&	\kern 4ex $A = B + Q$;	\cr
11&	\kern 4ex $B = \MOD{A(X^2 + X +1)}{F_{16}}$;\cr
12& \kern 4ex $\}$				\cr
13& 	$P = B$;	\cr
14&	return $P$;	\cr
\noalign{\hrule}
} %
} 
$$
} 
}{\fvsix}{Algorithm for computing the fast 16-bit CRC directly from the $m$ input bytes $I_i$.}
\QED

\subsection {A.2}{CRC Software Implementation: Table-Lookup Technique}
Recall that the complexity of the fast CRCs is low even without using table lookup. With table lookup, the operation count is reduced at the cost of additional memory resource. Although our focus in this paper is on bitwise algorithms, we now also present table-lookup algorithms to illustrate tradeoffs between operation count and table size. Our formulation and results here are straightforward generalizations or variations of well-known results, which are available in [\BiR, \Fel, \KoB, \Per, \RaG, \Sar].  Note that, with  table lookup, speed directly correlates with operation count under ideal conditions (e.g., the table is stored in the fastest cache, and there is no cache miss). Otherwise, speed may not correlate directly with operation count (e.g., when the impact of cache miss is not negligible [\Fel]). 

For table-lookup implementation, according to Fig.\fvtwo, we use Algorithm~3 (when $s<h$) and Algorithm~2 (when $s \ge h$) for both the basic CRCs and the fast CRCs. From (\eiitwentyeight), we then have
$$
B(X) = \MOD{A(X)X^h}{M(X)}
$$
where degree$(A(X))<s$. In the following, $B(X)$ is computed by table lookup. Let $g_b$ and $g_f$ be the total number of table entries for the basic CRCs and the fast CRCs, respectively.

\subsubsection {A.2.1}{Basic CRCs} 
According to Fig.\fvtwo, Algorithms 2 and 3 are used for the basic CRCs. Substituting the values of $x$ from Fig.\fvfour \ for Algorithms~2 and~3 into (\evten), we have

$$
e_b = \cases {8(6+r_b)/s     & if $s<h$ and $h=8,16,32,64$ \cr
             8(7+r_b)/s   & if $s<h$ and $h \ne 8,16,32,64$ \cr
              8(3+r_b)/s     & if $s=h$  \cr
              8(4+r_b)/s     & if $s > h$}   \eqno(\evthirty)                 
$$
where $r_b$ is the operation count required for computing $B(X)$ via table lookup. The required tables are defined below. First, we write 
$$
s = t_1+t_2+ \cdots +t_m
$$ 
for some $m$ and $t_i$ such that $1 \le m \le s$ and  $1 \le t_i \le s$ ($i=1,2, \dots, m$). Next, we decompose $A(X)$ into $m$ polynomials $A_1(X), A_2(X), \dots, A_{m-1}(X), A_m(X)$ such that 
$$
\eqalign{A(X) &= A_1(X)X^{(t_2+s_3+\cdots+t_m)}+A_2(X)X^{(t_3+s_4+ \cdots+t_m)}+ \cdots +A_{m-1}(X)X^{t_m}+A_m(X) \cr
                      &=\sum_{i=1}^{m-1}A_i(X)X^{(t_{i+1}+ \cdots +t_m)}+A_m(X)}
$$
with degree$(A_i(X)) < t_i$, $i=1, 2, \dots, m$. We then have  
$$
\eqalign{B(X) &= \MOD{A(X)X^h}{M(X)}   \cr
                      &= \MOD{\sum_{i=1}^{m-1}A_i(X)X^{(t_{i+1}+ \cdots +t_m+h)} + A_m(X)X^h}{M(X)} \cr
                      &= \sum_{i=1}^{m-1} \MOD{A_i(X)X^{(t_{i+1}+ \cdots +t_m+h)}}{M(X)} + \MOD{A_m(X)X^h}{M(X)} \cr 
                      &= \sum_{i=1}^{m-1}T_i[A_i] + T_m[A_m]}   \eqno(\evthirtytwo)                  
$$
where the tables $T_i[\ ]$ are defined by  
$$
T_i[A_i] = \cases {\MOD{A_i(X)X^{(t_{i+1}+ \cdots +t_m+h)}}{M(X)} & if $1 \le i < m$ \cr
                            \MOD{A_m(X)X^h}{M(X)}       & if $i=m$} \eqno(\evthirtyfour) 
$$
where $A_i$ denotes the $t_i$-tuple that is composed of the binary coefficients of $A_i(X)$. For example, if $t_i=4$ and $A_i(X) = X^2+1$, then $A_i = (0101)$, which is equivalent to the decimal integer 5. 

Thus, regardless of whether $s<h$ or $s \ge h$, the table $T_i[\ ]$ has $2^{t_i}$ entries, each entry is an $h$-tuple. (For example, let $h=16$ and $t_i=8$. The table $T_i[\ ]$ then has $2^8$ entries, 16 bits each, i.e., the total memory storage for this particular table is $2^8 \times 16$ bits = 512 bytes). Finally, the total number of entries for the $m$ tables, denoted by $g_b$, is
$$
g_b =  \sum_{i=1}^{m-1}2^{t_i} + 2^{t_m} =  \sum_{i=1}^m 2^{t_i} \eqno(\evthirtyeight) 
$$

To summarize, for a given polynomial $A(X)$ of degree less than $s$, let $m, t_1, t_2, ..., t_m$ be such that $ s = \sum_{i=1}^m t_i $. The term
$$
B(X) = \MOD{A(X)X^h}{M(X)} 
$$ 
can then be computed using the $m$ tables defined by (\evthirtyfour). The total number of entries for these tables is $g_b =  \sum_{i=1}^m 2^{t_i}$. Further, regardless of whether $s<h$ or $s \ge h$, it can be shown that, using the $m$ tables, the number of operations required for computing $B(X)$ is 
$$
r_b = 3(m-1) \eqno(\evfourtytwo) 
$$

We now consider the special case $t_1 = t_2 = \cdots = t_m = s/m$. The $m$ tables defined in (\evthirtyfour) then becomes
$$
T_i[A_i] = \MOD{A_i(X)X^{h+s(m-i)/m}}{M(X)} 
$$
$i=1, 2, \dots, m$. Each of the $m$ tables has $2^{s/m}$ entries. From (\evthirtyeight), the total number of table entries is 
$$
g_b =  \sum_{i=1}^m 2^{s/m} = m 2^{s/m}  \eqno(\evfourtyfour) 
$$
Equations (\evfourtytwo) and (\evfourtyfour) show tradeoffs between the operation count $r_b$ and the table size $g_b$. That is, to decrease the table size, we must increase $m$ in $g_b = m 2^{s/m}$, and this in turn will increase the operation count $r_b = 3(m-1)$. Thus, smaller (larger) table size $g_b$ will yield larger (smaller) operation count $r_b$. In particular, when $m=1$, we have $r_b = 0$ and $g_b = 2^s$. When $m=s$, we have $r_b = 3(s-1)$ and $g_b = 2s$.

Substituting $r_b = 3(m-1)$ into (\evthirty), we have 
$$
e_b = \cases {(24m+24)/s     & if $s<h$ and $h=8,16,32,64$ \cr
             (24m+32)/s   & if $s<h$ and $h \ne 8,16,32,64$ \cr
              24m/s     & if $s=h$  \cr
              (24m+8)/s     & if $s > h$}   \eqno(\evfourtysix)                 
$$
Note that our formulation is a straightforward generalization of [\RaG], which contains the results for the special cases $h=16$, $s \in \{8,16\}$, and $m \in \{1,s\}$. Our results [e.g., (\evfourtyfour)] also resemble those of [\KoB], which presents in-depth studies of the case $h=32$.

Note that the function $f(x)=2^x$ is convex. Given $s=  \sum_{i=1}^m t_i$, from Jensen's inequality, it can then be shown that $m^{-1}\sum_{i=1}^m 2^{t_i}  \ge 2^{s/m}$, i.e., $\sum_{i=1}^m 2^{t_i}  \ge m2^{s/m}$. This implies that, for a given $m$, the table size $g_b$ in (\evthirtyeight) is minimized when $t_1 = t_2 = \cdots = t_m$. Thus, we focus on only this special case ($t_i=s/m$) in this paper. 

For example, let $h=16$, $s=8$, and $m=1$. That is, we use a basic 16-bit CRC to protect a message consisting of input bytes ($s=8$). This CRC is implemented using one lookup table ($m=1$), which has $g_b=m2^{s/m}=2^8$ entries. Using (\evfourtysix) with $s<h$, we have $e_b=(24m+24)/s=6$. That is, 6 operations are required for computing the check tuple per input byte. These results are recorded in the first row of Fig.\fAonni. Note that, because each table entry has $h=16$ bits, the total storage is $hg_b=16\times2^8$ bits = 512 bytes (which is not shown in the figure). The results for other values of $h, s$, and $m$ are shown in Fig.\fAonni.

From Fig.\fAonni, we observe the followings.  First, the results for the cases $h=8$ and $h=16$ are identical for $s=32, 64$, i.e., they differs only for $s=8, 16$. This follows directly from (\evfourtysix). Similarly, the results for the cases $h=16$ and $h=32$ are identical for $s=8,64$, i.e., they differs only for $s=16,32$. Although the number of table entries $g_b=m2^{s/m}$ depends on only $s$ and $m$, the total storage is $hg_b = hm2^{s/m}$, which also depends on $h$. 

Recall from Fig.\fivtwo \ that the complexity results for bitwise implementation of the basic CRCs vary little over a wide range of $s$ values. In contrast, as seen in Fig.\fAonni, those for table-lookup implementation vary greatly with $s$. These results can also be used to optimize CRC table-lookup implementation. For example, suppose that $h=16$. Let us compare the 2 cases: ($s=8, m=1$) and ($s=16, m=4$) in Fig.\fAonni. In both cases, the required operation count per input byte is $e_b=6$. However, the first case requires one table of $2^8$ entries ($=16 \times 2^8 = 512$ bytes), while the second case requires 4 tables totaling only $2^6$ entries ($=16 \times 2^6 = 128$ bytes), which is 75\% less than the first case. More generally, Fig.\fAonni \ shows that, for a given $e_b$, the total number of table entries $g_b$ is minimized when $s=h$. 

\figure{$$
\vbox
{\eightpoint
\offinterlineskip
\halign
{
\strut
\vrule \  #   &\vrule \ #\hfill &\vrule \ #\hfill &\vrule \ #\hfill &\vrule \ #\hfill &\vrule \ # \hfill  &  \vrule \ # \hfill   &  \vrule #   \cr
\noalign{\hrule}
&       	&$h=8$  &$h=16$ &$h=32$	&$h=64$&     & \cr
\noalign{\hrule}
& $m$		&$e_b$ 	&$e_b$  &$e_b$	&$e_b$	&$g_b$& \cr
\noalign{\hrule}
$s=8$& 1	&3		&6		&6		&6		&$2^{8}$& \cr
     & 2	&6		&9		&9		&9		&$2^{5}$& \cr
     & 4	&12		&15		&15		&15		&$2^{4}$& \cr
\noalign{\hrule}
$s=16$& 1	&2		&1.5	&3		&3		&$2^{16}$& \cr
     & 2	&3.5	&3	 	&4.5	&4.5	&$2^{9}$& \cr
     & 4	&6.5	&6		&7.5	&7.5	&$2^{6}$& \cr
     & 8	&12.5	&12		&13.5	&13.5	&$2^{5}$& \cr
\noalign{\hrule}
$s=32$& 1	&1		&1 		&0.75	&1.5	&$2^{32}$& \cr
     & 2	&1.75	&1.75	&1.5	&2.25	&$2^{17}$& \cr
     & 4	&3.25	&3.25	&3		&3.75	&$2^{10}$& \cr
     & 8	&6.25	&6.25	&6		&6.75	&$2^{7}$& \cr
     & 16	&12.25	&12.25	&12		&12.75	&$2^{6}$& \cr
\noalign{\hrule}
$s=64$& 1	&0.5	&0.5	&0.5	&0.375	&$2^{64}$& \cr
     & 2	&0.875	&0.875	&0.875	&0.75	&$2^{33}$& \cr
     & 4	&1.625	&1.625	&1.625	&1.5	&$2^{18}$& \cr
     & 8	&3.125	&3.125	&3.125	&3		&$2^{11}$& \cr
     & 16	&6.125	&6.125	&6.125	&6		&$2^{8}$& \cr
     & 32	&12.125	&12.125	&12.125	&12		&$2^{7}$& \cr
\noalign{\hrule}
}
}
$$
}
{\fAonni}{Complexity results for table-lookup technique for the basic $h$-bit CRCs \hbox{($e_b$ = operation count per input byte, $g_b$ = total number of entries from $m$ tables).}}

\remark{Remark\rtwelve.} Both $g_b$ and $e_b$ depend on $s$, $h$, and $m$, i.e., we can write $g_b=g_b(s,h,m)$ and $e_b=e_b(s,h,m)$. Consider the 2 special cases: $m=s/2$ and $m=s$. From (\evfourtyfour) and (\evfourtysix), it can be shown that $g_b(s,h,s/2) = g_b(s,h,s) = 2s$ and $e_b(s,h,s/2) < e_b(s,h,s)$. That is, these 2 cases yield the same table size, but the case $m=s/2$ always yields lower operation count than the case $m=s$. Thus, the case $m=s$ can be eliminated from our discussion. \QED

 \remark{Remark\rfourteen.} So far, $B(X)$ is computed by either the bitwise technique or the table-lookup technique. However, $B(X)$ can also be computed using both techniques as follows. Recall from (\evthirtytwo) that $B(X)$ is the sum of $m$ terms. Suppose that we now use tables to compute the first $m-1$ terms, and use no tables to compute the last term. More precisely, from (\evthirtytwo), we have 
$$
\eqalign {B(X) &=  \sum_{i=1}^{m-1} \MOD{A_i(X)X^{(t_{i+1}+ \cdots +t_m+h)}}{M(X)} + \MOD{A_m(X)X^h}{M(X)} \cr
                       &= \sum_{i=1}^{m-1}T_i[A_i] +  \MOD{A_m(X)X^h}{M(X)} }
$$                    
where the $m-1$ tables $T_i[\ ]$ are defined by $T_i[A_i] = \MOD{A_i(X)X^{(t_{i+1}+ \cdots +t_m+h)}}{M(X)}$, $1 \le i < m$.  Assume now that $\MOD{A_m(X)X^h}{M(X)}$ is computed without using tables. Thus, $B(X)$ can be computed using the~2 techniques at the same time: the table lookup technique (with the $m-1$ tables  $T_i[\ ]$) and the bitwise technique (for computing $\MOD{A_m(X)X^h}{M(X)}$ without using tables). In the following, this mixed technique is applied to the fast CRCs to yield small table size when $s \approx h$. \QED

\subsubsection{A.2.2}{Fast CRCs} 
Recall from Section 2.1 that, when implementing an $h$-bit CRC, we are free to choose the value of $s$, which is the size of each input tuple $Q_i(X)$. That is, we can choose $s<h$, $s=h$, or $s>h$. Fig.\fivtwo \ shows that, under bitwise implementation, the fast CRCs are much faster than the basic CRCs for $s \le h$, in the sense that $e_f$ is much smaller than $e_b$. 
Further, by comparing Fig.\fivtwo \ with Fig.\fAonni, we see that the bitwise implementation of the fast CRCs (i.e., $g_f=0$) is even faster than the table-lookup implementation of the basic CRCs (i.e., $g_b>0$) in many cases. For example, consider the case $s=h=32$. Fig.\fAonni \ shows that $e_b=6$ when $g_b=2^7$ (at $m=8$), and $e_b=12$ when $g_b=2^6$ (at $m=16$). On the other hand, Fig.\fivtwo \ shows that $e_f=3$ when $g_f=0$. 
The same figures also show that, although the fast CRC requires no table lookup (i.e., $g_f=0$) and the basic CRC requires a table of $g_b=2^{10}$ entries (at $m=4$), both CRCs have the same operation count $e_f=e_b=3$. 

Recall from (\evtwentysixb) and (\evtwentysixd) that, under bitwise implementation, $e_f$ is minimized either at $s=h-2$ or at $s=h$. Thus, by choosing $s$ to be at (or near) these optimal values, the fast CRCs require no table lookup (i.e., $g_f=0$) and still have low operation count (i.e., $e_f$ is small).

We now discuss table-lookup techniques for the fast CRCs generated by $F_h(X) = X^h+X^2+X+1$. An obvious technique is to apply the table-lookup technique in Section A.2.1 for the basic CRCs to the fast CRCs by simply letting $M(X) = F_h(X)$. The required total number of table entries is then given by (\evfourtyfour), i.e., $g_f=g_b = m2^{s/m}$.
In the following, for the case $s \ge h$, we present another table-lookup technique (which is similar to the mixed technique in Remark\rfourteen\  with $m=2$) that exploits the special structure of $F_h(X)$ to yield $g_f = 2^{s-h+2}$, which is small when $s\approx h$ , e.g., $g_f=4$ when $s=h$.

Recall that $r_f$ denotes the operation count required for computing $B(X)$. Without using tables (i.e., when $g_f=0$), $r_f$ is given by (\evtwentyfour), i.e.,  $r_f = 9+5.5(s-h)$ for $s \ge h$. We show below that $r_f$ is slightly reduced by using a small lookup table.

Assume that $s \ge h$. According to Fig.\fvtwo, we use Algorithm~2 to implement the table-lookup technique for the fast CRCs when $s \ge h$. From Fig.\fvfour, we then have 
$$
x = \cases {3  & if $s=h$ \cr
            4  & if $s>h$}
$$
which is inserted in (\evten) to yield

$$
e_f = \cases {8(3+r_f)/s  & if $s=h$ \cr
              8(4+r_f)/s  & if $s>h$} \eqno(\evfifty)
$$ 
where $r_f$, which denotes the operation count required for computing $B(X)$ via table lookup, is determined in the following.

First, we decompose $A(X)$ into $A_1(X)$ and $A_2(X)$ such that
$$
A(X) = A_1(X)X^{h-2} + A_2(X)
$$
where degree$(A_1(X)) < s-h+2$ and degree$(A_2(X)) < h-2$.  Using (\eiitwentyeight) with $M(X)=F_h(X)$, we have
$$
\eqalignno {B(X) 	&= \MOD{A(X)X^h}{F_h(X)} \cr
				&= \MOD{A(X)(X^h+F_h(X))}{F_h(X)} \cr
				&= \MOD{A(X)(X^2+X+1)}{F_h(X)} \cr
               	&= \MOD{A_1(X)X^{h-2}(X^2+X+1)}{F_h(X)} + A_2(X)(X^2+X+1) \cr
               	&=T_f[A_1] + A_2(X)(X^2+X+1) & (\evfiftyA)}                       
$$
where $T_f[\ ]$ is the table defined by
$$
T_f[A_1] =  \MOD{A_1(X)X^{h-2}(X^2+X+1)}{F_h(X)} \eqno(\evfiftyone) 
$$
where $A_1$ denotes the $(s-h+2)$-tuple that is composed of the binary coefficients of the polynomial $A_1(X)$ of degree less than $s-h+2$. The table $T_f[\ ]$ has $g_f = 2^{s-h+2}$ entries, and each entry contains $h$ bits. Using this table, it can be shown that the operation count required for computing $B(X)$ is $r_f = 7$. To summarize, when $s \ge h$, we have
$$
r_f = 7 ,  \quad g_f=2^{s-h+2}  \eqno(\evfourtyeight) 
$$                   
             
Substituting (\evfourtyeight) into (\evfifty), we then have the following operation count per input byte and the table size for the case $s \ge h$:

$$
\cases {e_f=80/s, \quad g_f=4 & if $s=h$ \cr
        e_f=88/s, \quad g_f = 2^{s-h+2}& if $s>h$} \eqno(\evfiftytwo)
$$
i.e., the table size $g_f$ grows exponentially with the difference $s-h$. Thus, this table-lookup technique is not recommended for large $s-h$. To have a small table, we must choose $s$ that is sufficiently close to $h$. The table size is minimized when $s=h$, which yields $g_f=4$, i.e., the fast $h$-bit CRCs now require $e_f=80/h$ operations per input byte and a small table of only $g_f=4$ entries. This is 20\% lower than the bitwise technique (i.e., $g_f=0$) that requires $e_f=96/h$ operations per input byte.
Fig.\fAtwze \ shows the numerical values of (\evfiftytwo) for $s, h \in \{8, 16, 32, 64\}$, which vary greatly with $h$ and $s$. In particular, the table size is large ($g_f\ge 2^{10}$) when $s > h$, but is very small ($g_f=4$) when $s=h$. 

For the special case $s=h$ (i.e., $g_f = 4$), it can be shown that the 4 entries of the table (\evfiftyone) are given by
$$
\eqalign {T_f[0] &= 0 \cr
               T_f[1] &= X^{h-1} + X^{h-2} + X^2 + X + 1 \cr
               T_f[2] &= X^{h-1} + X^3 + 1 \cr
               T_f[3] &= X^{h-2} + X^3 + X^2 + X} \eqno(\evfiftyfour)
$$
These entries in hexadecimal are shown in Fig.\fvtwelve.

The table-lookup algorithm for the fast CRC (when $s \ge h$) is given in Fig.\fvfourteen, where the $2^{s-h+2}$ entries of the table $T_f[\ ]$ defined by (\evfiftyone) are stored in the top part of the algorithm. The C program for the special case $s=h=16$, which is based on Fig.\fvfourteen, is given in Fig.\fAsix.

\figure{
{\figureFont    
$$ 
\vbox
{
\offinterlineskip 
\halign
{
\strut
\vrule \quad #   &\vrule \quad # &\vrule \quad #   &  \vrule \quad #    &  \vrule  #   \cr
\noalign{\hrule}
& $s$	&$e_f$&	$g_f$& \cr
\noalign{\hrule}
$h=8$ & 8	&10	  	&$2^{2}$&  \cr
      & 16	&5.5	  	&$2^{10}$& \cr
      & 32	&2.75	&$2^{26}$& \cr
      & 64	&1.375	&$2^{58}$& \cr
\noalign{\hrule}
$h=16$& 16	&5	 	&$2^{2}$&  \cr
      & 32	&2.75	&$2^{18}$& \cr
	  & 64	&1.375	&$2^{50}$& \cr
\noalign{\hrule}
$h=32$&32	&2.5	&$2^{2}$&  \cr
	  &64	&1.375	&$2^{34}$& \cr
\noalign{\hrule}
$h=64$&64	&1.25	&$2^{2}$&  \cr
\noalign{\hrule}
}
}
$$
} 
}{\fAtwze}{Complexity results for the fast $h$-bit CRCs with table lookup, $s \ge h$ 
\hbox{($e_f$ = operation count per input byte, $g_f$ = total number of table entries).}}

\figure{ $$
\vbox
{\eightpoint
\offinterlineskip 
\halign
{
\strut
\vrule \ #  \hfill\vrule    & \ # \hfill\vrule    &\ # \hfill\vrule    &\ # \hfill\vrule &  \ # \hfill&   \vrule  #  \cr
\noalign{\hrule} 
$h$ & 	$T_f[0]$	& $T_f[1]$ & $T_f[2]$	& $T_f[3]$ &\cr
\noalign{\hrule} 	
8	&0	&c7	     &89	&4e	&\cr
16	&0	&c007	 & 8009	&400e	&\cr
32	&0	&c0000007	 & 80000009	&4000000e &\cr
64  &0 &  c000000000000007  &8000000000000009 &  400000000000000e &\cr          
\noalign{\hrule} 	
}
}
$$
 }
{\fvtwelve}{Four-entry tables for the fast $h$-bit CRCs generated by $F_h(X) = X^h+X^2+X+1$ ($s = h$).}

%
\figure{
{\figureFont
$$
\vbox
{
\offinterlineskip 
\halign
{
\strut 
\vrule \ \hss #  \vrule & \ # \hss \vrule \cr
\noalign{\hrule}
1&	store $T_f[0], \dots, T_f[2^{s-h+2} -1]$;	\cr
2&	$B=0$;	\cr
3&	for $(0 \le i < n)$ \cr
4& 	\kern 4ex $\{$		\cr
5&	\kern 4ex $A = BX^{s-h} + Q_i$;	\cr
6&	\kern 4ex $A_1=(s-h+2)$ left-hand bits of $A$;	\cr
7&	\kern 4ex $A_2=(h-2)$ right-hand bits of $A$;	\cr
8&	\kern 4ex $B = T_f[A_1] + A_2(X^2 + X +1)$;	\cr
9& \kern 4ex $\}$				\cr
10& 	$P = B$;	\cr
11&	return $P$;	\cr
\noalign{\hrule}
}
}
$$
}
}{\fvfourteen}{Table-lookup algorithm for the fast $h$-bit CRC generated by $F_h(X) = X^h+X^2+X+1$ ($s \ge h$).} 

\figure{ \includegraphics[height=6cm]{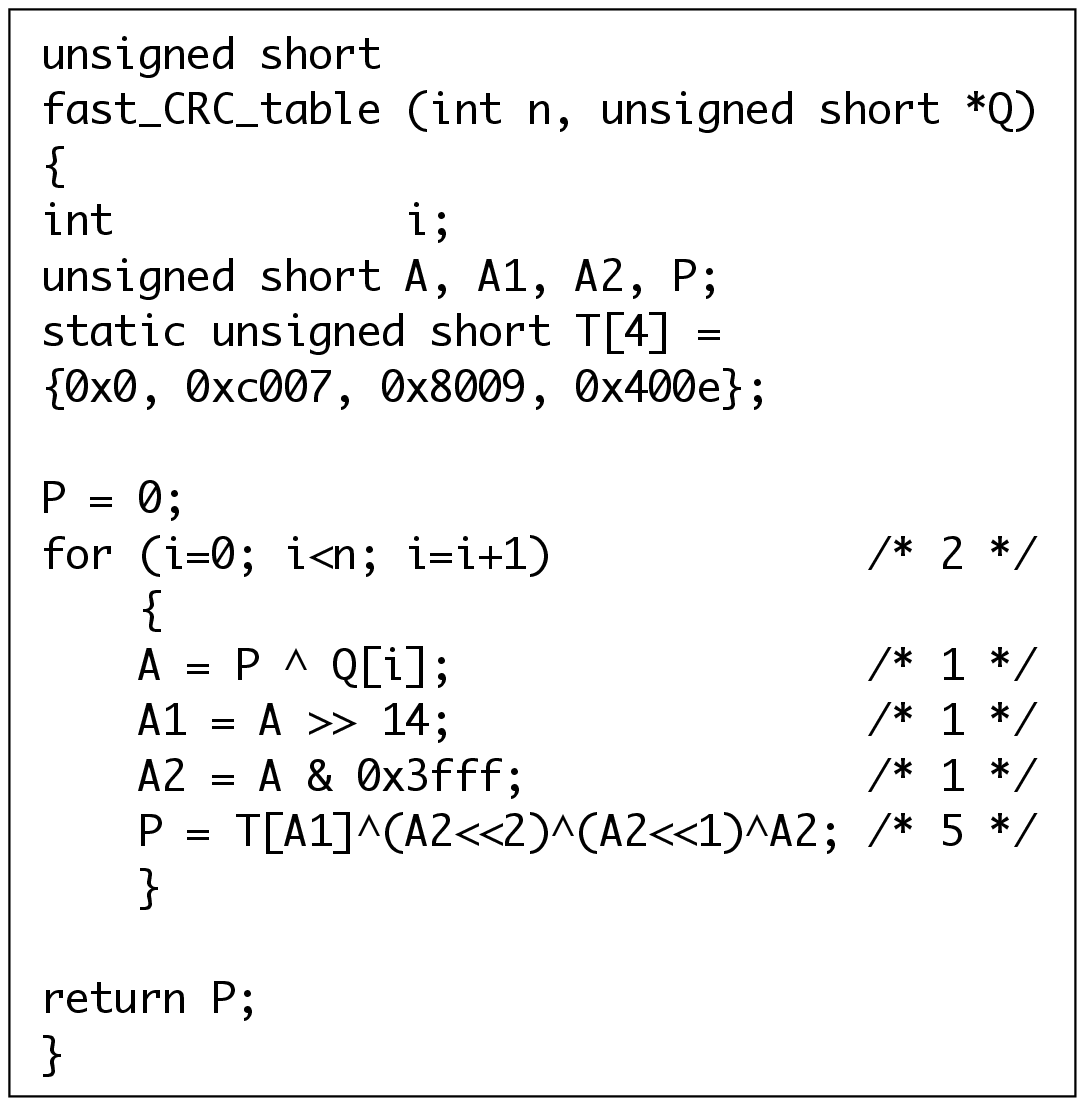} }
{\fAsix}{C program with table lookup for the fast 16-bit CRC generated by $F_{16}(X) = X^{16}+X^2+X+1$  ($s=h=16$).}

{\color{red}
\remark{Remark\rSawadaTable.} Based on the suggestion by Y.~Sawada, the C program in Fig.\fAsix \ can be improved to yield the C program shown in Fig.\fSawadaTable. This improvement follows from the observation that {\tt A1} and {\tt A2} are the left and right parts of {\tt A}, respectively, i.e., {\tt A2} can be determined from {\tt A1} and {\tt A}. Thus, by modifying the table {\tt T[A1]} as shown in Fig.\fSawadaTable, we can replace {\tt A2} by {\tt A}, i.e., {\tt A2} is now no longer needed. 

Using the above improvement, it can be shown in general that the table given in (\evfiftyfour) can now be simplified to become the new table defined by

$$
\eqalign {T_f[0] &= 0 \cr
               T_f[1] &= X^2 + X + 1 \cr
               T_f[2] &= X^3 + 1 \cr
               T_f[3] &= X^3 + X^2 + X} \eqno(\evfiftysix)
$$
i.e., $T_f[0]=0, \ T_f[1]={0x7}, \ T_f[2]={0x9}, \ T_f[3]={0xe}$ in hexadecimal.
\figure{ \includegraphics[height=5.5cm]{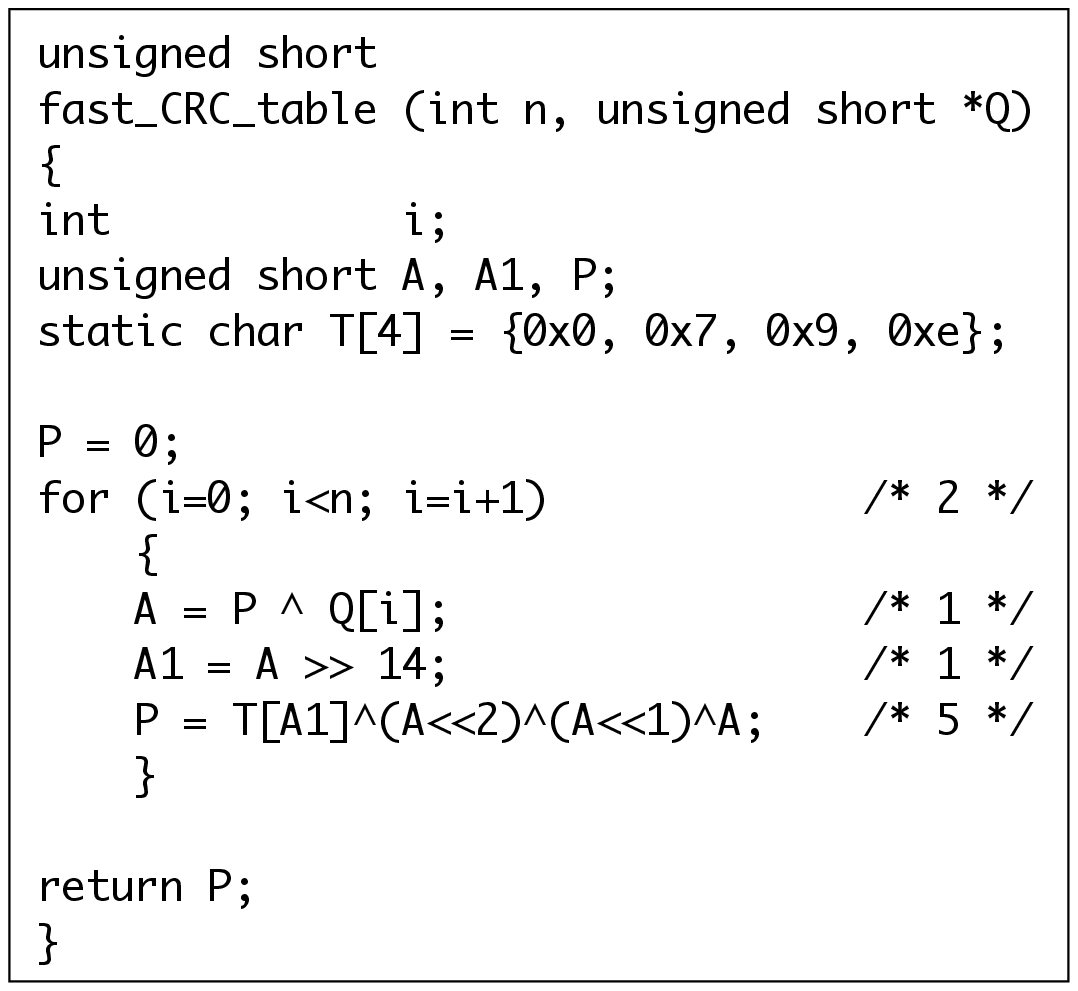} }
{\fSawadaTable}{{Improved C program with table lookup for the fast 16-bit CRC} generated \hbox{by $F_{16}(X) = X^{16}+X^2+X+1$  ($s=h=16$)}.} 
} \QED

\section {APPENDIX B}{OTHER FAST ERROR-DETECTION CODES} \secB = \pageno
So far, we apply the new technique (\eiiitwelve) to $F_h(X)=X^h+X^2+X+1$ to yield the fast $h$-bit CRCs.
{\color{blue}We now apply this same technique to binomials and trinomials to yield even faster (but weaker) CRCs. We then construct some non-CRC error-detection codes, which are not only fast but also have optimal guaranteed error-detecting capability.} 

Recall from Section 3.3 that the {\sl maximum length} of an error-detection code is defined to be the total bit length at or below which its minimum distance is $d \ge 3$, i.e., beyond which its minimum distance will reduce to $d=2$. Theorem~2 shows that the maximum length of a CRC is the {\sl period} of its generator polynomial. In the following, $n_{\rm b}$ denotes the total bit length of a code. 

\subsection {B.1}{Fast CRCs Generated by Binomials} Consider the $h$-bit CRC generated by the {\it binomial} $M(X)=X^h+1$, which has period $h$. To avoid triviality, we assume that this CRC includes at least one input bit, i.e., $n_{\rm b} > h$. From Theorem~2, this CRC then has the minimum distance $d=2$, i.e., it is a weak code for error detection. This CRC can be implemented via Fig.\fiifour \ (for $s<h$) or Fig.\fiifive \ (for $s \ge h$). Applying the new technique (\eiiitwelve) to $M(X)=X^h+1$, the term $B(X)$ in these figures is given by
$$
B(X) = \cases{A(X)                  & if $s<h$ \cr
            \MOD{A(X)X^{s-h}}{N(X)} & if $s \ge h$}
$$
where $N(X)=(X^h+1)X^{s-h}$. Note that by choosing $s \le h$, we have $B(X)= A(X)$, i.e., the polynomial division is eliminated. 

Suppose now that $s=h$. The CRC generated by $X^h+1$ can then be implemented by Fig.\fivone \ with $B(X)= A(X)$. Fig.\fivone \ can be further simplified to yield the following pseudocode for computing the check $h$-tuple $P(X)$: 

{\figureFont
$$
\vbox
{
\offinterlineskip 
\halign
{
\strut 
\vrule \ \hss #  \vrule & \ # \hss \vrule \cr
\noalign{\hrule}
1&	$P=0$;	\cr
2&	for $(0 \le i < n)$ \cr 
3&	\kern 4ex $P=P+Q_i$;\cr
4&	return $P$;	\cr
\noalign{\hrule}
}
}
$$
}
which yields 
$$
P(X) = \sum_{i=0}^{n-1}Q_i(X)
$$
i.e., the CRC generated by $M(X)=X^h+1$ is identical to the bock-parity checksum [\Fel]. From the above pseudocode, it can be shown that computing the check tuple $P(X)$ for this checksum requires $e=24/s$ operations per input byte. Recall from Section 4 that $e_b = 8(3+5.5s)/s$ and $e_f = 96/s$. We then have $e_f/e = 96/24=4$ and $e_b/e = 8(3+5.5s)/24 = (3+5.5s)/3$.

For example, if $s=h=16$, then computing the check tuple $P(X)$ for the 16-bit bock-parity checksum requires $e=24/16=1.5$ operations per input byte. We then have $e_f/e = 4$ and $e_b/e = (3+5.5\times 16)/3 = 30.33$. Thus, as expected, the bock-parity checksum (which has minimum distance $d=2$) is substantially faster than the fast and basic CRCs (both of which have minimum distance $d=4$). 

\subsection {B.2}{Fast CRCs Generated by Trimomials} 
Let $C$ be the CRC generated by the {\it trinomial} $T_h(X)=X^h+X+1$. The periods $t$ of the trinomials are given in Fig.\fvitwo \ for $h \ge 3$. Note that the periods $t$ for the important cases $h$ = 8, 16, 32, 64, 128, are unusually small. In fact, \ Fig.\fvitwo \ shows  that the period is $t=h^2 -1$ when $h$ is a power of 2.
Because $T_h(X)$ is a codeword of weight 3, the minimum distance $d$ of this CRC must satisfy $d \le 3$. From Theorem 2, we then have $d=3$ if $n_b \le t$, and $d=2$ if $n_b > t$. This CRC can be implemented via Fig.\fiifour \ (for $s<h$) or Fig.\fiifive \ (for $s \ge h$). Applying the new technique (\eiiitwelve) to $M(X)=T_h(X)$, the term $B(X)$ in these figures is given by
$$
B(X) = \cases{\MOD{A(X)(X+1)}{T_h(X)}      & if $s<h$ \cr
            \MOD{A(X)X^{s-h}(X+1)}{N(X)} & if $s \ge h$} \eqno(\eviten)
$$
where $N(X)=(X^h+X+1)X^{s-h}$. Remark\rtwo \ implies that it is simpler to compute the $B(X)$ in (\eviten) than the $B(X)$ in (\eiiifourteen). Thus, the CRC generated by the trinomial $T_h(X)=X^h+X+1$ is faster than the fast CRC generated by $F_h(X)=X^h+X^2+X+1$. However, the former has minimum distance only $d=3$, whereas the latter has minimum distance $d=4$. Further, for the important cases of $h=8, 16, 32, 64, 128$, the maximum length of the faster CRC generated by the trinomial $T_h(X)$ is much shorter than that of the fast CRC generated by $F_h(X)$. For example, the faster 16-bit CRC generated by $T_{16}(X)$ has $d=3$ and the maximum length of only 255 bits (see Fig.\fvitwo), whereas the fast CRC generated by $F_{16}(X)$ has $d=4$ and the maximum length of $2^{15} -1 = 32767$ bits (see Section 3.2). Thus, these 2 types of CRCs illustrate tradeoffs between code capability and complexity. 
\figure{
{\figureFont
$$
\vbox
{
\offinterlineskip 
\halign
{
\strut
\vrule \quad #  \vrule    &\quad # \vrule &  \quad# &  \vrule  #  \cr
\noalign{\hrule} 
$h$ & 	period	& $2^h-1\over{\rm period}$ & \cr
\noalign{\hrule} 	
3	&7				&1						& \cr
4	&15				&1						& \cr
7	&127			&1						& \cr
8	&63				&4.05					& \cr
15	&32767			&1						& \cr
16	&255			&257					& \cr
23	&2088705		&4.02					& \cr
24	&2097151		&8						& \cr
31	&2097151		&1024					& \cr 
32	&1023			&$4.2 \times 10^6$		& \cr
63  &$2^{63}-1$ 	&1  					& \cr               
64  &4095 			&  $4.5 \times 10^{15}$	& \cr
127 &$2^{127}-1$	&1  					& \cr
128 &16383			&$2.1 \times 10^{34}$ 	&\cr          
\noalign{\hrule} 	
}
}
$$
}
}{\fvitwo}{The period of trinomial $T_h(X)=X^h+X+1$.}

\subsection {B.3}{Fast and Optimal Error-Detection Codes}
In the following, we construct codes that are not only fast but also have optimal error-detecting capability. The $h$-bit CRC in Section B.2, which is denoted by $C$ and has minimum distance $d=3$, can be extended to yield a code that has $d=4$ by adding an overall parity bit to the $h$-bit CRC. Note that this extended code, denoted by $C^*$, has $h^*=h+1$ check bits and is not a CRC. The $h$-bit CRC has burst-error-detecting capability $b=h$. The following theorem shows that the extended code $C^*$ has burst-error-detecting capability $b=h^*=h+1$.

\remark{Theorem 4.} Let $C$ be an $h$-bit CRC generated by a polynomial $M(X)$ of degree $h$. Assume that $M(X)$ is not a multiple of $X$, i.e., gcd$(X,M(X))=1$, and that $M(X)$ has odd weight, i.e., it has an odd number of terms. Let $C^*$ be the non-CRC code that is obtained by adding an overall parity check bit to $C$, i.e., $C^*$ has $h^*=h+1$ check bits. Then $C^*$ detects all error bursts of length $h+1$ or less, i.e., its burst-error-detecting capability is $b=h+1$.
 
\remark{Proof.} Let $V^*(X)$ be a codeword of $C^*$. By definition of $C^*$, we have $V^*(X)=V(X)X+{\rm parity}(V(X))$, where $V(X)$ is a codeword of $C$. Because $V(X)$ is a codeword of the CRC generated by $M(X)$, we have $V(X)=K(X)M(X)$ for some polynomial $K(X)$ (see the proof of Theorem 2). We then have
$$
V^*(X) = K(X)M(X)X + {\rm parity}(K(X)M(X))
$$

Let $E^*(X)$ be an error burst of length $h+1$ or less, which has the form
$$
E^*(X) = X^i(E(X) + 1) 
$$
where $i \ge 0$, and $E(X)$ is a polynomial such that $E(X) \ne 1$ and degree$(E(X)) \le h$. Using proof by contradiction, we now show that $E^*(X)$ cannot be a codeword of $C^*$. Thus, assume that $E^*(X)$ is a nonzero codeword of $C^*$, i.e.,
$$
X^i(E(X) + 1) = K(X)M(X)X + {\rm parity}(K(X)M(X))
$$
We consider 2 cases: $i=0$ and $i>0$.

Case 1: $i=0$. We then have
$$
E(X) + 1 = K(X)M(X)X + {\rm parity}(K(X)M(X))
$$
This implies that ${\rm parity}(K(X)M(X))=1$. Thus, $K(X) \ne 0$, which implies that degree$(K(X)M(X)X) > h$. But we also have $E(X) = K(X)M(X)X$, which implies that degree$(K(X)M(X)X) \le h$, which is a contradiction to the previous statement. 

Case 2: $i>0$. We then have ${\rm parity}(K(X)M(X))= 0$. Thus, $X^i(E(X) + 1) = K(X)M(X)X$. Because gcd$(X,M(X))=1$, we must have $X^i=K(X)X$. Thus, $K(X) = X^{i-1}$. We then have ${\rm parity}(K(X)M(X))={\rm parity}(M(X))= 1$, which is a contradiction to the previous statement that ${\rm parity}(K(X)M(X))= 0$. \QED

Let $t$ be the period of the polynomial $M(X)$ in Theorem 4. The extended code $C^*$ in Theorem 4 then has $h^*=h+1$ check bits, the burst-error-detecting capability $b=h^*=h+1$, the minimum distance $d=4$, and the maximum length of $t+1$ bits. In the following, we show that $C^*$ becomes fast by choosing $M(X)=T_h(X)=X^h+X+1$, i.e., $M(X)$ is a trinomial.

Thus, let $M(X)=X^h+X+1$, and $P_{\rm CRC}(X)$ be the check $h$-tuple for the CRC generated by this particular $M(X)$. Suppose that $s=h+1$. Because $s>h$, the check tuple $P_{\rm CRC}(X)$ can be computed by Algorithm~4 (see Fig.\fiifive), in which the term $B(X)$ is computed by (\eviten), i.e., 
$$
\eqalign{B(X) &= \MOD{A(X)X(X+1)}{N(X)} \cr
              &= \MOD{A(X)(X^2+X)}{N(X)}}
$$
where $N(X)=(X^h+X+1)X$, and degree$(A(X))<s=h+1$.

Recall from Theorem 4 that the non-CRC code $C^*$ is obtained by adding an overall parity check bit to the above CRC. The overall parity bit of $C^*$ is computed as follows. First, we define 
$$
W(X)=\sum_{i=0}^{n-1}Q_i(X) + P_{\rm CRC}(X)X
$$
where $Q_0(X),\dots,Q_{n-1}(X)$ are the input $s$-tuples. The overall parity bit of $C^{*}$ is also the parity bit of $W(X)$. The check polynomial of $C^*$ is then 
$$
P(X)=P_{\rm CRC}(X)X + {\rm parity}(W(X))
$$
which is a polynomial of degree $<h+1$. 

Fig.\fvifour \ shows an implementation of $C^*$, which is based on Fig.\fiifive \ (with $s=h^*=h+1$ and $M(X)=X^h+X+1$) and includes the calculation of the overall parity bit of $C^*$.  
Let $e^*$ be the operation count per input byte required for computing the check tuple $P(X)$ for the code $C^*$. By ignoring the negligible complexity due to the terms outside the loop indexed by $i$ in Fig.\fvifour, it can be shown that $e^* = 96/h^*$. It is shown in (\evtwentysix) of Appendix A that the complexity for the fast $h^*$-bit CRC is also given by $e_f = 96/h^*$ (when $s=h^*$). Thus, $e^*=e_f$, i.e., the (non-CRC) $h^*$-bit code $C^*$ is as fast as the fast $h^*$-bit CRC.

Let $M(X)$ be a primitive polynomial of degree $h$, i.e., the period of $M(X)$ is $t=2^h-1$. Let us now compare the capability and complexity for the  following 2 codes, each of which has $h^*=h+1$ check bits. The first code is the familiar basic CRC generated by $(X+1)M(X)$, which has $d=4$, $b=h+1$, and the maximum length of $2^h - 1$ bits. An example is the well-known CRC-16, which is generated by $M(X)=(X+1)(X^{15}+X+1)=X^{16}+X^{15}+X^2+1$. Under bitwise implementation, this basic CRC requires $e_b=45.5$ operations per input byte for computing its check tuple, provided that the input message is composed of 16-tuples, i.e., $s=h=16$ (see Fig.\fivtwo).

The second code is the non-CRC code $C^*$ as described in Theorem 4, which has $d=4$, $b=h+1$, and the maximum length of $t+1=2^h$ bits (which is 1 bit longer than the basic $(h+1)$-bit CRC above). 
It is well-known that any code that has $h+1$ check bits and  the minimum distance $d=4$ must satisfy the following 2 constraints: (1) the burst-error detecting capability $b \le h+1$ and (2) the maximum length $\le 2^h$. Thus, the non-CRC code $C^*$ is optimal for error detection in the sense that, with $h^*=h+1$ check bits and $d=4$, it has the optimal $b=h^*$ and the optimal maximum length $2^h$. In fact, at the maximum length of $2^h$ bits, the code  $C^*$ is a $(2^h, 2^h-h-1, 4)$ extended Hamming perfect code with the optimal burst-error-detecting capability $b=h+1$. Also, it is well-known that the undetected error probability of this perfect code is bounded above by $2^{-(h+1)}$.

As shown in Fig.\fvifour, the code $C^*$ is fast when  $M(X)=T_h(X)=X^h+X+1$, i.e., $M(X)$ is a trinomial. It is known that $T_h(X)$ is primitive for some values of $h$ [\Zie], including $h=3$, 7, 15, 63, and 127 (i.e., $h+1=4$, 8, 16, 64, and 128).  
For example, let $h=15$ and $s=h^*=h+1=16$. Using Fig.\fvifour, it can be shown that the operation count per input bye required for computing the check tuple for the (non-CRC) 16-bit code $C^*$ is $e^* = 96/16=6$ (which is much smaller than $e_b = 45.5$ of the basic CRC-16 above). To summarize, the (non-CRC) $h^*$-bit code $C^*$ (e.g., with $h^*=4$, 8, 16, 64, or 128 check bits) constructed from a primitive trinomial and an overall parity check bit has (a) the optimal error-detection capability and (b) a fast bitwise implementation. Note that, as discussed later in Section C.2, other fast and optimal codes can also be constructed from polynomials different from trinomials.

\figure{
{\figureFont
$$
\vbox
{
\offinterlineskip 
\halign
{
\strut 
\vrule \ \hss #  \vrule & \ # \hss \vrule \cr
\noalign{\hrule}
1&	$W=0$;	\cr
2&	$B=0$;	\cr
3&	for $(0 \le i < n)$ \cr
4& \kern 4ex $\{$		\cr
5& \kern 4ex $A = B+ Q_i$;	\cr
6& \kern 4ex $B = \MOD{A(X^2 + X)}{N}$;\cr
7& \kern 4ex $W = W+ Q_i$;	\cr
8& \kern 4ex $\}$				\cr
9& 	$W = W+ B$;	\cr
10&  $P = B+{\rm parity}(W)$;	\cr
11&	return $P$;	\cr
\noalign{\hrule}
}
}
$$
} 
}{\fvifour}{Algorithm for computing the fast $(h+1)$-bit non-CRC code from \hbox{the $h$-bit CRC (generated by $X^h+X+1$) and an overall parity bit}.} 

\section{APPENDIX C}{APPLICATION OF THE NEW TECHNIQUE TO GENERAL CRC GENERATOR POLYNOMIALS} \secC = \pageno
So far, we apply the new technique (\eiiitwelve) to the polynomials $X^h+X^2+X+1$, $X^h+X+1$, and $X^h+1$ to yield fast CRCs. In this appendix, we apply the same technique to more general generator polynomials,  and then determine the conditions under which the new technique is faster than the basic technique. In particular, we show later in Section C.1.2.1 that, when applied to the CRC-64-ISO generated by
$ X^{64}+X^4 + X^3 + X + 1$, the new technique is~15 times faster than the basic technique.  This appendix presents only bitwise algorithms.

Consider an $h$-bit CRC that is generated by a general polynomial
$$
\eqalign{F(X) &= X^h + X^{i_k} + X^{i_{k-1}} + \cdots + X^{i_1} + 1	\cr
			  &= X^h + H(X) }	\eqno(\eDtwo)
$$
where $k>0$, $h>i_{k}>i_{k-1} > \cdots > i_1 > 0$, and 
$$
H(X) = X^{i_k} + X^{i_{k-1}} + \cdots + X^{i_1} + 1	\eqno(\eDthree)
$$
Note that $i_k \ge k > 0$, $i_k = {\rm degree}(H(X))$, and $k=$ weight of $(H(X)+1)$. Here, we have $F(X) \ne X^h+1$ because $i_1>0$, i.e., $H(X) \ne 1$. The case $F(X)=X^h+1$ is already discussed in Section B.1, where it is shown that the CRC reduces to the block-parity checksum.  

{\color{red}
For example, let $F(X)=X^{32} + X^{7} + X^{6} + X^{2} + 1$. We then have $h=32$, $k=3$, $i_3=7$, $i_2=6$, $i_1=2$, $H(X)=X^{7} + X^{6} + X^{2} + 1$, and ${\rm degree}(H(X))=7$.
}

The $h$-bit CRC generated by (\eDtwo) can be computed either by the basic technique (see Definition 1) or by the new technique (\eiiitwelve). Recall that CRC complexity refers to the operation count per input byte (denoted by $e_b$ and $e_f$ for the basic and the fast CRCs, respectively) required for computing the CRC check tuple. Again, we assume that the CRCs are implemented in C, and the operations are counted according to rules (R1) and (R2) stated in Appendix~A.
\subsection{C.1}{General CRC Generator Polynomials}
First, suppose that the basic technique is used to compute the check tuple of the CRC generated by (\eDtwo), i.e., $B(X)$ is computed as in Definition 1 with $M(X)=F(X)$ in (\eiitwentyeight). From (\evtwentytwo), we have	
$$
e_b = \cases {8(4+5.5s)/s             & if $s <   h$ \cr
              8(3+5.5s)/s             & if $s \ge h$ } \eqno(\eDfour)
$$

Next, suppose that the new technique is used to compute the check tuple of the CRC generated by (\eDtwo). By letting $M(X)=F(X)$ in (\eiiitwelve), we have

$$
B(X) = \cases {\MOD{A(X)(X^h + F(X))}{F(X)}  &if $s < h$  \cr
               \MOD{A(X)(X^s + N(X))}{N(X)}   &if $s \ge h$}  \eqno(\eDfourA)     
$$
where $N(X) = F(X)X^{s-h}$ and degree$(A(X))<s$. Substituting (\eDtwo) into (\eDfourA), we have
$$
B(X) = \cases {\MOD{A(X)H(X)}{F(X)}  &if $s < h$  \cr
               \MOD{A(X)H(X)X^{s-h}}{N(X)}   &if $s \ge h$}       \eqno(\eDfive) %
$$

To briefly illustrate the main idea, consider the special case $s=h$. Then $B(X)=\MOD{A(X)X^h}{F(X)}$ under the basic technique, and  $B(X)=\MOD{A(X)(X^{i_k}+X^{i_{k-1}}+\cdots+X^{i_1}+1)}{F(X)}$ under the new technique. Intuition suggests that computing $B(X)$ via the new technique is faster than the basic technique if $i_k$ is sufficiently small. More precise conditions on $i_k$ are given in the following.

Let $e$ be the operation count  per input byte required for computing the CRC check tuple under the new technique (\eDfive). We have $e=e_f$ for the special case $F(X)=F_h(X)=X^h+X^2+X+1$. Although Fig.\fivtwo \ shows that $e_f < e_b$, it may not be the case that $e < e_b$ for the more general polynomial $F(X)$. Thus, in the following, we determine the conditions on $F(X)$ so that $e < e_b$ or $e_b/e > 1$ (i.e., the conditions under which the new technique is faster than the basic technique). Thus, the new technique serves as a faster alternative to the basic technique when $e_b/e > 1$.
{\color {blue}The computation of $e_b/e$ for many CRCs are given later in Sections~C.2-C.4.}
Before continuing, we present the following remarks, which contain some results that will be used later to determine the operation count required for computing $B(X)$.

{\remark{Remark\rfourteenA.} Let $r^\prime$ be the number of operations required for computing 
$$
\eqalign{B^\prime(X) &= A(X)(X^{j_n}+\cdots+X^{j_1}+1) \cr
	&= A(X)X^{j_n}+\cdots+A(X)X^{j_1}+A(X)}
$$ 
where $n \ge 1$, and we assume that the tuple $A(X)X^{j_i}$ can be stored in a single computer word. Computing $A(X)X^{j_i}$ is then equivalent to shifting $A(X)$ to the left by $j_i$ bits, which can be done by a single operation on most computers. Thus, for a given $A(X)$, we can compute $B^\prime(X)$ by using $n$ left-shift operations and $n$ addition operations. We then have $r^\prime=2n$. \QED

{\remark{Remark\rfourteenB.} Let $M^*(X)$ and $A^*(X)$ be 2 polynomials. Let $n$ and $q$ be such that $n \ge q$.
Let $r^*$ be the number of operations required for computing
$$
\eqalign{B^*(X)&=\MOD{A^*(X)(X^{j_n}+\cdots+X^{j_q}+1)}{M^*(X)} \cr
&=\MOD{A^*(X)X^{j_n}}{M^*(X)}+\cdots+\MOD{A^*(X)X^{j_q}}{M^*(X)}+\MOD{A^*(X)}{M^*(X)} }
$$

We assume in the following that $\MOD{A^*(X)X^{j_i}}{M^*(X)}\not\equiv A^*(X)X^{j_i}$, $i=n,\dots, q$, i.e., the polynomial division is needed. Define 
$$
\eqalign{C_{q-1}(X) 		&=  \MOD{A^*(X)}{M^*(X)} 				\cr
		C_q(X)  &=	\MOD{C_{q-1}(X)X^{j_q}}{M^*(X)}	\cr
		C_{q+1}(X)  &=	\MOD{C_q(X)X^{j_{q+1}-j_q}}{M^*(X)}	\cr
					& \cdots									\cr
        C_{m+q}(X)  &=	\MOD{C_{m+q-1}(X)X^{j_{m+q}-j_{m+q-1}}}{M^*(X)}	\cr
        				& \cdots									\cr
        C_{n}(X)		&=	\MOD{C_{n-1}(X)X^{j_{n}-j_{n-1}}}{M^*(X)} }
$$

Let $r_0$ be the operation count required for computing $C_{q-1}(X)$. Given $C_{q-1}(X)$, the term $C_q(X) = \MOD{C_{q-1}(X)X^{j_q}}{M^*(X)}$ can be computed with $5.5j_q$ operations, and so on. Given $C_{n-1}(X)$, the final term $C_n(X) = \MOD{C_{n-1}(X)X^{j_{n}-j_{n-1}}}{M^*(X)}$ can be computed in $5.5(j_n - j_{n-1})$ operations. Thus, computing $C_{q-1}(X),C_q(X),\dots,C_n(X)$ altogether requires 
$$
r_0 + 5.5j_q + 5.5(j_{q+1}-j_q) + \cdots + 5.5(j_n-j_{n-1}) = r_0 + 5.5j_n
$$
operations. Because $B^*(X) = C_{q-1}(X)+C_q(X)+\cdots+C_n(X)$, the tuple $B^*(X)$ is computed by using $(n-q+1)$ addition operations. Overall, $B^*(X)$ can be computed with 
$$
r^* = 5.5j_n + n-q+1 + r_0
$$ 
operations, where $r_0$ is the operation count required for computing $\MOD{A^*(X)}{M^*(X)}$. \QED 
{\color{red}
Recall that, given the CRC generated by (\eDtwo), our goal here is to determine the complexity for computing the check $h$-tuple $P(X)$ for an input message that consists of $n$ tuples $Q_0(X), Q_1(X), \dots, Q_{n-1}(X)$. Each tuple $Q_i(X)$ has $s$ bits. 
As shown in Figs.\fiitwo-4, the check tuple $P(X)$ is computed by using a loop that computes $B(X)$ for $n$ times, where $B(X)$ can be computed by  the basic technique (see Definition 1) or by the new technique (\eiiitwelve). In the following, we compare the complexity between these 2 techniques.
We consider 2 cases: $s \ge h$ and $s < h$.
} 
  
\subsubsection{C.1.1}{Case: $s \ge h$} 
In this case, according to Fig.\fvtwo, the new technique uses Algorithm 4 (shown in Fig.\fiifive), which contains the computation of $B(X)$. 
From (\eDthree) and (\eDfive), we have 
$$
\eqalign{B(X) &= \MOD{A(X)H(X)X^{s-h}}{N(X)} \cr
			&=\MOD{A^*(X)X^{i_k}}{N(X)}+\MOD{A^*(X)X^{i_{k-1}}}{N(X)}+\cdots+\MOD{A^*(X)X^{i_1}}{N(X)}+\MOD{A^*(X)}{N(X)}    }
$$
where $A^*(X)=A(X)X^{s-h}$. 
Using Remark\rsix, it can be shown that $\MOD{A^*(X)}{N(X)}=\MOD{A(X)X^{s-h}}{N(X)}$ can be computed with $r_0=5.5(s-h)$ operations (see Appendix~C).  Applying Remark\rfourteenB \ with $M^*(X)=N(X)$, $n=k$, $j_n=i_k$, $q=1$, and $r_0=5.5(s-h)$, the tuple $B(X)$ can be computed with $r = 5.5(s-h+i_k) + k$ operations. 

Fig.\fvfour \ shows that $x=3$ for $s \ge h$ under Algorithm 4. By substituting the values of $x$ and $r$ into (\evten), the operation count per input byte required for computing the check tuple under the new technique is
$$
e = \frac{8[3 + 5.5(s-h + i_k) + k]}{s} \eqno(\eDfiveA)
$$
\noindent Using (\eDfour) and (\eDfiveA), we have 
$$
\frac{e_b}{e} = \frac{3+5.5s}{3 + 5.5(s-h + i_k) + k} \eqno(\eDfiveB)
$$
\noindent Thus, the new technique is faster than the basic technique when $e_b/e > 1$, i.e., $3+5.5s > 3 + 5.5(s-h + i_k) + k$, which is equivalent to
$$
i_k < h- \frac{k}{5.5}
$$
where $i_k = {\rm degree}(H(X))$ and $k=$ weight of $(H(X)+1)$.  

\remark{Remark\rfifteen.} Suppose that $F(X)$ is either $X^h + X^2 + X + 1$ or $X^h + X + 1$. Then $i_k \le 2$, i.e., $i_k$ is a very small value. Thus, it is appropriate to use loop unrolling in the calculation of $C_m(X)$ above. Then (\eDfiveA) reduces to $e = 8[3 + 5.5(s-h) +3.5i_k + k]/s$, and then  
$$
\frac{e_b}{e} = \frac{3+5.5s}{3 + 5.5(s-h) +3.5i_k + k} \eqno(\eDfiveD)
$$
Note that it is common to choose $s, h \in \{8,16, 32, 64 \}$, i.e., the typical values of $s$ and $h$ are not very small, even when $i_k$ is very small. 

For example, suppose now that $s=h$ and $F(X)=F_h(X)=X^h+X^2+X+1$, i.e., $k=i_k=2$ and $e=e_f$. Substituting $s=h$ and $k=i_k=2$ into (\eDfiveD), we have $$
\frac{e_b}{e} = \frac{e_b}{e_f} =\frac{3+5.5h}{3 +3.5\times 2 + 2} = 0.25+0.458h
$$
as previously shown in (\eivthreeI). \QED

\subsubsection{C.1.2}{Case: $s < h$} 
In this case, according to Fig.\fvtwo, the new technique uses Algorithm 3 (shown in Fig.\fiifour), which contains the computation of $B(X)$. From (\eDfive), we have 
$ B(X)= \MOD{A(X)H(X)}{F(X)} $. 
From Fig.\fvfour, we have 
$$
x = \cases {6             & if $h=8,16,32,64$ \cr
              7             & if $h \ne 8,16,32,64$ } 
$$
By substituting the values of $x$ into (\evten), the operation count per input byte required for computing the CRC check tuple under the new technique is
$$
e = \cases {8(6+r)/s             & if $h=8,16,32,64$ \cr
            8(7+r)/s             & if $h \ne 8,16,32,64$ } \eqno(\eDfiveE)
$$
where $r$ is the number of operations required for computing $ B(X)= \MOD{A(X)H(X)}{F(X)} $. From (\eDfour), we have $e_b = 8(4+5.5s)/s$ for $s<h$, which is used with (\eDfiveE) to yield
$$
\frac{e_b}{e} = \cases{ (4+5.5s)/(6+r) &if $h=8,16,32,64$ \cr	
(4+5.5s)/(7+r) & if $h \ne 8,16,32,64$}\eqno(\eDfiveF)
$$
where $r$, which depends on whether $i_k \le h-s$ or $i_k < h-s$, is computed in the following subsections. As seen below, the condition $i_k \le h-s$ implies that $B(X) = \MOD{A(X)H(X)}{F(X)} = A(X)H(X)$, i.e., the polynomial division is eliminated.

\subsubsection{C.1.2.1}{Case: $s < h$ \ and \ $i_k \le h-s$} 
Because degree$(A(X)) < s$ and degree$(H(X))=i_k$, we have degree$(A(X)H(X)) < s+i_k$. The assumption $i_k \le h-s$ then implies that degree$(A(X)H(X)) < h$. Thus, $B(X) = \MOD{A(X)H(X)}{F(X)} = A(X)H(X)$, i.e., the polynomial division is eliminated.
Let $r$ be the number of operations required for computing $B(X) = A(X)H(X)$. Using (\eDthree), we have 
$$
B(X) = A(X)X^{i_k} + A(X)X^{i_{k-1}} + \cdots + A(X)X^{i_1} + A(X)
$$
Applying Remark\rfourteenA, we then have $r=2k$, which is substituted into (\eDfiveE) and (\eDfiveF) to yield
$$
e = \cases {8(6+2k)/s             & if $h=8,16,32,64$ \cr
            8(7+2k)/s             & if $h \ne 8,16,32,64$ } \eqno(\eDsix)
$$
$$
\frac{e_b}{e} = \cases{ (4+5.5s)/(6+2k) &if $h=8,16,32,64$ \cr	
(4+5.5s)/(7+2k) & if $h \ne 8,16,32,64$}\eqno(\eDseven)
$$

Thus, the new technique is faster than the basic technique if $e_b/e > 1$, i.e.,
$$
{4+5.5s}> \cases{6+2k & if $h=8,16,32,64$ \cr
				 7+2k & if $h \ne 8,16,32,64$}
$$
which is equivalent to 
$$
k < \cases{2.75s -1 &if $h=8,16,32,64$\cr
2.75s -1.5& if $h \ne 8,16,32,64$}\eqno(\eDeight)
$$
where $i_k = {\rm degree}(H(X))$ and $k=$ weight of $(H(X)+1)$.  	

For example, consider the CRC-64-ISO generated by the primitive polynomial
$$
F(X) = X^{64}+X^4 + X^3 + X + 1
$$
Here, we have $h=64$, $k=3$, and $i_k=4$. Assume that $s \le h-i_k = 60$. Under the new technique, we then have 
$$
\eqalign{B(X) &= \MOD{A(X)(X^4 + X^3 + X + 1)}{F(X)} \cr
			&= A(X)X^4 + A(X)X^3 + A(X)X + A(X)}
$$
i.e., the polynomial division is eliminated. Substituting $k=3$ into (\eDseven), we have
$$
\frac{e_b}{e} = \frac{4+5.5s}{12}
$$
For the special case $s=32$, we have $e_b/e = 15$, i.e., the new technique is 15 times faster than the basic technique for the CRC-64-ISO. We also have $e_b/e = 15$ when $s=32$ for a 64-bit CRC generated by a polynomial that has the following more general form
$$
F(X) = X^{64} + X^{i_3} + X^{i_2} + X^{i_1} + 1
$$
where $32 \ge i_3 > i_2 > i_1>0$.

{\color{red}
\remark{Remark\rSawadaISO.} For the CRC-64-ISO, the value $e_b/e$ shown above is based on the implementation for which $B(X)$ is computed directly by the single statement $B(X)=A(X)X^4 + A(X)X^3 + A(X)X + A(X)$. An alternative implementation, suggested by Y.~Sawada, is to compute $B(X)$ via 2 statements: first, compute $B'(X)=A(X)X+A(X)$, and then compute $B(X)=B'(X)X^3+B'(X)$. \QED 
}
\subsubsection{C.1.2.2}{Case: $s < h$ \ and \ $i_k > h-s$} 
As seen below, the computation of $B(X)$ requires the polynomial division in this case. The assumption $i_k > h-s$ implies that there exists $m^*$ such that $1 \le m^* \le k$ , $i_{m^*} > h-s$, and $i_j \le h-s$ for all $j < m^*$. There are 3 subcases to consider.

Case 1: $1<m^* < k$. By letting $m=m^* - 1$, we have 
$$
F(X) = X^h + X^{i_k} +\cdots+X^{i_{m+1}}+X^{i_m}+\cdots+X^{i_1}+1
$$
where $h>i_k> i_{m+1} > i_m \ge i_1 > 0$, $ i_{m+1}>h-s$, and $i_m \le h-s $.

Because $i_m \le h-s$, we have degree$(A(X)X^{i_n})<h$, for $1\le n \le m$. Thus, 
$$
\MOD{A(X)X^{i_n}}{F(X)} = A(X)X^{i_n}
$$
for $1\le n \le m$. From (\eDfive), we then have
$$
B(X) = \MOD{A(X)X^{i_k}}{F(X)}+\cdots+\MOD{A(X)X^{i_{m+1}}}{F(X)}+A(X)X^{i_m}+\cdots+A(X)X^{i_1}+A(X)
$$
By letting $A^*(X) = A(X)X^{i_{m+1}}$, we can write 
$$
B(X) = B_1(X) + B_2(X)
$$
where
$$
B_1(X)=\MOD{A^*(X)X^{i_k-i_{m+1}}}{F(X)}+\cdots+\MOD{A^*(X)X^{i_{m+2}-i_{m+1}}}{F(X)} + \MOD{A^*(X)}{F(X)}
$$
and
$$
B_2(X) = A(X)X^{i_m}+\cdots+A(X)X^{i_1}+A(X)
$$

Because $\MOD{A^*(X)}{F(X)} = \MOD{A(X)X^{i_{m+1}}}{F(X)} = \MOD{(A(X)X^{h-s})X^{i_{m+1}-(h-s)}}{F(X)}$, the term $\MOD{A^*(X)}{F(X)}$ can be computed with $r_0 = 1 + 5.5[i_{m+1} - (h-s)]$ operations for a given $A(X)$. Using Remark\rfourteenB, $B_1(X)$ can be computed with 
$$
\eqalign{r_1 &= 5.5(i_k - i_{m+1}) + k-(m+2)+1 + r_0 \cr
			&= 5.5[i_k -(h-s)] + k-m}
$$
operations. Using Remark\rfourteenA, $B_2(X)$ can be computed with $r_2 = 2m$ operations. Overall, the number of operations required for computing $B(X)$ is 
$$
\eqalign{r 	&= r_1+r_2 +1 \cr
			&= 5.5[i_k -(h-s)]+k+m + 1 } \eqno(\eFastB)
$$
which is substituted into (\eDfiveF) to yield
$$
\frac{e_b}{e} = \cases{ (4+5.5s)/(6+ 5.5[i_k -(h-s)]+k+m +1) & if $h=8,16,32, 64$\cr
(4+5.5s)/(7+ 5.5[i_k -(h-s)]+k+m+ 1) & if $h \ne 8,16,32, 64$} \eqno(\eDnine)
$$
Thus, the new technique is faster than the basic technique when
$$
4+5.5s > \cases {6+5.5[i_k -(h-s)]+k+m + 1 & if $h=8,16,32, 64$\cr
7+5.5[i_k -(h-s)]+k+m + 1 & if $h \ne 8,16,32, 64$}
$$
which is equivalent to 
$$
i_k < \cases{h - (3+k+m)/{5.5} & if $h=8,16,32, 64$\cr
h - (4+k+m)/{5.5} & if $h \ne 8,16,32, 64$}
$$
where $i_k = {\rm degree}(H(X))$ and $k=$ weight of $(H(X)+1)$. Recall that we also assume that $h>i_k> i_{m+1} > i_m \ge i_1 > 0$, $ i_{m+1}>h-s$, and $i_m \le h-s $.

For example, consider the CRC-32-IEEE 802.3 generated by the following primitive polynomial:
$$
F(X)=X^{32} + X^{26} + X^{23} + X^{22} + X^{16} + X^{12} + X^{11} + X^{10} + X^8 + X^7 + X^5 + X^4 + X^2 + X + 1 \eqno(\eDnineA)
$$
i.e., $h=32$, $k=13$, $i_k=26$. Assume that $s=16$. We then have $m=10$. Substituting these values into~(\eDnine) yields
$e_b/e = 92/85$, i.e., the new technique is slightly faster than the basic technique.

Case 2: $m^* = 1$. We then have 
$$
F(X) = X^h + X^{i_k} + \cdots + X^{i_1} + 1
$$
where $i_1 > h-s$. We have 
$$
\eqalign{B(X) &= \MOD{A(X)X^{i_k}}{F(X)}+\cdots+\MOD{A(X)X^{i_2}}{F(X)}+\MOD{A(X)X^{i_1}}{F(X)}+A(X) \cr
			&= \MOD{A^*(X)X^{i_k-i_1}}{F(X)}+\cdots+\MOD{A^*(X)X^{i_2-i_1}}{F(X)}+\MOD{A^*(X)}{F(X)}+A(X) \cr 
			&= B_1(X) + A(X) }
$$
where $A^*(X) = A(X)X^{i_1}$ and 
$$
B_1(X) = \MOD{A^*(X)X^{i_k-i_1}}{F(X)}+\cdots+\MOD{A^*(X)X^{i_2-i_1}}{F(X)}+\MOD{A^*(X)}{F(X)}
$$
Because $\MOD{A^*(X)}{F(X)} = \MOD{A(X)X^{i_1}}{F(X)} = \MOD{(A(X)X^{h-s})X^{i_1-(h-s)}}{F(X)}$, the term $\MOD{A^*(X)}{F(X)}$ can be computed with $r_0 = 1 + 5.5[i_1 - (h-s)]$ operations for a given $A(X)$. Using Remark\rfourteenB, $B_1(X)$ can be computed with 
$$
\eqalign{r_1 &= 5.5(i_k - i_1) + k-2+1 + r_0 \cr
			&= 5.5[i_k -(h-s)] + k}
$$
operations. Thus, the number of operations required for computing $B(X)$ is 
$$
\eqalign{r 	&= r_1+1 \cr
			&= 5.5[i_k -(h-s)]+k+ 1 }
$$
which is substituted into (\eDfiveF) to yield
$$
\frac{e_b}{e} = \cases{ (4+5.5s)/(6+ 5.5[i_k -(h-s)]+k +1) & if $h=8,16,32, 64$\cr
(4+5.5s)/(7+ 5.5[i_k -(h-s)]+k+ 1) & if $h \ne 8,16,32, 64$} \eqno(\eDten)
$$

Case 3: $m^* = k$. We then have 
$$
F(X) = X^h + X^{i_k} + \cdots + X^{i_1} + 1
$$
where $i_k > h-s$, and $i_n \le h-s$ for all $n<k$.  We have 
$$
\eqalign{B(X) &= \MOD{A(X)X^{i_k}}{F(X)}+A(X)X^{i_{k-1}}+\cdots + A(X)X^{i_1}+A(X) \cr
			&= \MOD{A(X)X^{i_k}}{F(X)}+B_2(X) }
$$
where 
$$
B_2(X) = A(X)X^{i_{k-1}}+\cdots + A(X)X^{i_1}+A(X)
$$

Because $\MOD{A(X)X^{i_k}}{F(X)} = \MOD{(A(X)X^{h-s})X^{i_k-(h-s)}}{F(X)}$, the term $\MOD{A(X)X^{i_k}}{F(X)}$ can be computed with $r_0 = 1 + 5.5[i_k - (h-s)]$ operations for a given $A(X)$. Using Remark\rfourteenA, $B_2(X)$ can be computed with 
$r_2 = 2(k-1)$
operations. Thus, the number of operations required for computing $B(X)$ is 
$$
\eqalign{r 	&= r_0+r_2+1 \cr
			&= 5.5[i_k -(h-s)]+2k }
$$
which is substituted into (\eDfiveF) to yield
$$
\frac{e_b}{e} = \cases{ (4+5.5s)/(6+ 5.5[i_k -(h-s)]+2k) & if $h=8,16,32, 64$\cr
(4+5.5s)/(7+ 5.5[i_k -(h-s)]+2k) & if $h \ne 8,16,32, 64$} \eqno(\eDtenA)
$$ 

For example, consider the CRC-32-IEEE 802.3 generated by (\eDnineA), 
i.e., $h=32$, $k=13$, $i_k=26$. Assume that $s=8$. Substituting these values into (\eDtenA) yields
$e_b/e = 48/43$, i.e., the new technique is slightly faster than the basic technique.

\subsection{C.2}{CRC Generator Polynomials of Weight 3}
We now consider the special case $k=1$, i.e., $F(X)$ is a polynomial of weight 3:
$F(X) = X^h + X^{i_1} + 1	$.
By defining $i=i_1$, we have
$$
F(X) = X^h + X^i + 1	\eqno(\eDsixteen)
$$
where $h>i>0$. Note that $F(X) = T_h(X)=X^h+X+1$ for the special case $i=1$. Fig.\fDfour \ lists some weight-3 polynomials along with their periods, for $h \le 32$. In Section B.2, the fast $h^*$-bit perfect codes are constructed from the CRCs generated by $T_h(X)$ for $h^* = 4,8, 16, 64, 128$, where $h^*=h+1$. In the following, we show that other fast perfect codes can also be constructed from CRCs generated by weight-3 polynomials.

Let $C$ be the $h$-bit CRC generated by $F(X)$  in (\eDsixteen). Recall from Theorem 2 that the maximum length of $C$ equals the period of $F(X)$. Assume that $s \le h-i$. Using the new technique, we have $B(X)=\MOD{A(X)(X^i+1)}{F(X)}=A(X)(X^i+1)$, i.e., the polynomial division is eliminated. Let $e$ be the operation count per input byte required for computing the check tuple $P(X)$ of the $h$-bit CRC $C$. Then $e$ is given by~(\eDsix). 

Let $C^*$ be the non-CRC code that is constructed by adding an overall parity bit to the $h$-bit CRC $C$, and $P^*(X)$ be the check tuple  of $C^*$. Let $e^*$ be the operation count per input byte required for computing $P^*(X)$. Note that $P^*(X)$ has $h+1$ bits, which can be computed by an algorithm that is similar to Fig.\fvifour. Using (\eDsix) and Fig.\fvifour, it can then be shown that 
$$
e^* = \cases {8(7+2k)/s             & if $h=8,16,32,64$ \cr
            8(8+2k)/s             & if $h \ne 8,16,32,64$ }	\eqno(\eDeighteen)
$$
Substituting $k=1$ into (\eDeighteen), we have
$$
e^* = \cases {72/s             & if $h=8,16,32,64$ \cr
            80/s             & if $h \ne 8,16,32,64$ }	\eqno(\eDtwenty)
$$

Let us now compare the speed of the $(h+1)$-bit code $C^*$ with that of the fast $(h+1)$-bit CRC generated by $F_{h+1}(X) = X^{h+1}+X^2+X+1$. From (\evtwentysix), we have

$$
e_f = \cases {80/s             & if $h+1=8,16,32,64$ \cr
            88/s             & if $h+1 \ne 8,16,32,64$ }	\eqno(\eDtwentytwo)
$$
for $s < h$. By comparing (\eDtwenty) with (\eDtwentytwo), we have 
$$
e^* \le e_f 		\eqno(\eDtwentyfour)
$$
for $s \le h-i$. Thus, (\eDtwentyfour) shows that the $(h+1)$-bit non-CRC code $C^*$ is at least as fast as the fast $(h+1)$-bit CRC generated by $F_{h+1}(X)$, i.e., $C^*$ is also a fast code for $s \le h-i$. Further, at its maximum length, the code $C^*$ is the $(2^h, 2^h-h-1, 4)$ extended Hamming perfect code, provided that $F(X)=X^h + X^i + 1$ is a primitive polynomial (i.e., its period is $2^h-1$).  

For example, let $F(X)=X^{11}+X^2+1$, i.e., $h=11$ and $i=2$. Fig.\fDfour \ shows that $F(X)$ is  primitive. Let $C$ be the 11-bit CRC generated by $F(X)$. The non-CRC code $C^*$, which is constructed by adding an overall parity bit to $C$, is the $(2048, 2036, 4)$ extended Hamming perfect code. Note that both $C$ and $C^*$ are fast if we choose $s \le h-i = 9$. Suppose that we choose $s=8$. From (\eDtwenty) and (\eDtwentytwo), we then have $e^* = 80/8 = 10$ and $e_f = 88/8 = 11$, i.e., $e^*<e_f$.  Thus, for $s=8$, the non-CRC 12-bit code $C^*$ is faster than the fast 12-bit CRC generated by $F_{12}(X)=X^{12}+X^2+X+1$. Further, the maximum length of the non-CRC code (which is 2,048 bits) is also much longer than that of the fast CRC generated by $F_{12}(X)$ (which is 595 bits), and 2 bits longer than that of the CRC generated by $F(X) = X^{12}+X^3+X+1$ (which is 2,046 bits, as discussed later in Section C.3). 

Similarly, we can construct a 32-bit extended Hamming perfect code $C^*$ by adding an overall parity bit to the CRC $C$ generated by $F(X)=X^{31}+X^3+1$ (see Fig.\fDfour). We have $h=31$ and $i=3$. Both $C$ and $C^*$ are fast if we choose $s \le h-i=28$. 

\figure{
{\eightpoint
$$
\vbox
{
\offinterlineskip 
\halign
{
\strut
\vrule \quad #  \vrule    &\quad # \vrule &  \quad# &  \vrule  #  \cr
\noalign{\hrule} 
$X^h+X^i+1$ & 	period	& $2^h-1\over{\rm period}$ & \cr
\noalign{\hrule} 	
$ X^{3}+X+1 $ &    7 &      1 &     \cr
 $ X^{4}+X+1 $ &    15 &     1 &     \cr
 $ X^{5}+X+1 $ &    21 &     1.47619 &       \cr
 $ X^{5}+X^{2}+1 $ &    31 &     1 &     \cr
 $ X^{6}+X+1 $ &    63 &     1 &     \cr
 $ X^{7}+X+1 $ &    127 &    1 &     \cr
 $ X^{8}+X+1 $ &    63 &     4.04762 &       \cr
 $ X^{8}+X^{3}+1 $ &    217 &    1.17512 &       \cr
 $ X^{9}+X+1 $ &    73 &     7 &     \cr
 $ X^{9}+X^{2}+1 $ &    465 &    1.09892 &       \cr
 $ X^{9}+X^{4}+1 $ &    511 &    1 &     \cr
 $ X^{10}+X+1 $ &   889 &    1.15073 &       \cr
 $ X^{10}+X^{3}+1 $ &   1023 &   1 &     \cr
 $ X^{11}+X+1 $ &   1533 &   1.33529 &       \cr
 $ X^{11}+X^{2}+1 $ &   2047 &   1 &     \cr
 $ X^{12}+X+1 $ &   3255 &   1.25806 &       \cr
 $ X^{13}+X+1 $ &   7905 &   1.03618 &       \cr
 $ X^{13}+X^{3}+1 $ &   8001 &   1.02375 &       \cr
 $ X^{14}+X+1 $ &   11811 &          1.3871 &        \cr
 $ X^{15}+X+1 $ &   32767 &          1 &     \cr
 $ X^{16}+X+1 $ &   255 &    257 &   \cr
 $ X^{16}+X^{3}+1 $ &   57337 &          1.14298 &       \cr
 $ X^{16}+X^{7}+1 $ &   63457 &          1.03275 &       \cr
 $ X^{17}+X+1 $ &   273 &    480.114 &       \cr
 $ X^{17}+X^{2}+1 $ &   114681 &         1.14292 &       \cr
 $ X^{17}+X^{3}+1 $ &   131071 &         1 &     \cr
 $ X^{18}+X+1 $ &   253921 &         1.03238 &       \cr
 $ X^{18}+X^{7}+1 $ &   262143 &         1 &     \cr
 $ X^{19}+X+1 $ &   413385 &         1.26828 &       \cr
 $ X^{19}+X^{3}+1 $ &   491505 &         1.0667 &        \cr
 $ X^{19}+X^{6}+1 $ &   520065 &         1.00812 &       \cr
 $ X^{20}+X+1 $ &   761763 &         1.37651 &       \cr
 $ X^{20}+X^{3}+1 $ &   1048575 &        1 &     \cr
 $ X^{21}+X+1 $ &   5461 &   384.023 &       \cr
 $ X^{21}+X^{2}+1 $ &   2097151 &        1 &     \cr
 $ X^{22}+X+1 $ &   4194303 &        1 &     \cr
 $ X^{23}+X+1 $ &   2088705 &        4.01618 &       \cr
 $ X^{23}+X^{2}+1 $ &   7864305 &        1.06667 &       \cr
 $ X^{23}+X^{5}+1 $ &   8388607 &        1 &     \cr
 $ X^{24}+X+1 $ &   2097151 &        8 &     \cr
 $ X^{24}+X^{5}+1 $ &   16766977 &       1.00061 &       \cr
 $ X^{25}+X+1 $ &   10961685 &       3.06107 &       \cr
 $ X^{25}+X^{2}+1 $ &   25165821 &       1.33333 &       \cr
 $ X^{25}+X^{3}+1 $ &   33554431 &       1 &     \cr
 $ X^{26}+X+1 $ &   298935 &         224.493 &       \cr
 $ X^{26}+X^{3}+1 $ &   2094081 &        32.0469 &       \cr
 $ X^{26}+X^{5}+1 $ &   67074049 &       1.00052 &       \cr
 $ X^{27}+X+1 $ &   125829105 &      1.06667 &       \cr
 $ X^{27}+X^{8}+1 $ &   133693185 &      1.00392 &       \cr
 $ X^{28}+X+1 $ &   17895697 &       15 &    \cr
 $ X^{28}+X^{3}+1 $ &   268435455 &      1 &     \cr
 $ X^{29}+X+1 $ &   402653181 &      1.33333 &       \cr
 $ X^{29}+X^{2}+1 $ &   536870911 &      1 &     \cr
 $ X^{30}+X+1 $ &   10845877 &       99 &    \cr
 $ X^{30}+X^{7}+1 $ &   1073215489 &     1.00049 &       \cr
 $ X^{31}+X+1 $ &   2097151 &        1024 &          \cr
 $ X^{31}+X^{2}+1 $ &   22362795 &       96.0293 &       \cr
 $ X^{31}+X^{3}+1 $ &   2147483647 &     1 &     \cr
 $ X^{32}+X+1 $ &   1023 &   $4.2\times10^6$ &    \cr
 $ X^{32}+X^{3}+1 $ &   1409286123 &     3.04762 &       \cr
 $ X^{32}+X^{5}+1 $ &   3758096377 &     1.14286 &       \cr
 $ X^{32}+X^{15}+1 $ &  4292868097 &     1.00049 &       \cr   
\noalign{\hrule} 	
}
}
$$
}
}{\fDfour}{The period of $X^h+X^i+1$.}

\subsection{C.3}{CRC Generator Polynomials of Weight 4}
We now consider the special case $k=2$, i.e., $F(X)$ is a polynomial of weight 4:
$$
F(X) = X^h + X^{i_2} + X^{i_1} + 1	
$$
where $h>i_2>i_1>0$. In particular, $F(X)=F_h(X)=X^h+X^2+X+1$ when $i_2=2$ and $i_1=1$. Fig.\fvisix \ lists some weight-4 polynomials $F(X)=X^h + X^{i_2} + X^{i_1} + 1$, which have periods that are greater than those of $F_h(X)$, for $h \le 32$.  Recall from Theorem 2 that the maximum length of a CRC equals the period of its generator polynomial. In the following, we consider the application of the new technique to weight-4 generator polynomials for CRCs such as CRC-16 and CRC-CCITT. For brevity, we only present the results for $s<h$ (the case $s \ge h$ can be handled similarly). There are 3 cases to consider.

Case 1: $i_2 \le h-s$ (i.e., $s \le h-i_2$). Using the new technique, we have $B(X)=\MOD{A(X)(X^{i_2}+X^{i_1}+1)}{F(X)}=A(X)(X^{i_2}+X^{i_1}+1)$, i.e., the polynomial division is eliminated. Substituting $k=2$ into (\eDsix), we have
$$
e = \cases {80/s             & if $h=8,16,32,64$ \cr
            88/s             & if $h \ne 8,16,32,64$ }	\eqno(\eDeleven)
$$
By comparing (\eDeleven) with (\evtwentysix), we have 
$$
e = e_f 		\eqno(\eDelevenA)
$$
for $s \le h-i_2$. Using $k=2$ and $s \le h-i_2$ in (\eDeight), it can be shown that the new technique is faster than the basic technique when 
$$
2 \le s \le h-i_2 \eqno(\eDtwelve)
$$

For example, let $F(X)=X^{32}+X^4+X+1$, i.e., $h=32$ and $i_2=4$. It follows from (\eDtwelve) that the new technique is faster the basic technique when $2 \le s \le 28$. Under this condition, we have  
$$
B(X)=A(X)(X^4+X+1) \eqno(\eDfourteen)
$$
i.e., the polynomial division is eliminated. Fig.\fvisix \ shows that the 32-bit CRC generated by $F(X)$ has the maximum length of 2,147,483,647 $=2^{31}-1$ bits ($\approx$~268,435,456 bytes). Recall from Fig.\fiiitwo \ that the original fast 32-bit CRC, generated by $F_{32}(X)=X^{32}+X^2+X+1$, has the maximum length of 2,097,151 $\approx (2^{31}-1)/1024$ bits ($\approx$~262,143 bytes). Thus, the maximum length of the CRC generated by $F(X)$ is substantially larger than that of the fast CRC generated by $F_{32}(X)$. However, (\eDelevenA) shows that these 2 CRCs have identical complexity when $s \le 28$.

Consider the 12-bit CRC generated by $F(X)=X^{12}+X^3+X+1$. Fig.\fvisix \ shows that this CRC has the maximum length of 2,046 bits, which is much larger than that of the fast CRC generated by $F_{12}(X)=X^{12}+X^2+X+1$, which has the maximum length of only 595 bits (see Fig.\fiiitwo). However, (\eDelevenA) shows that these 2 CRCs have identical complexity when $s \le 9$.

Case 2: $i_2 > h-s$ and $i_1 \le h-s$.
Using the new technique, we have $B(X)=\MOD{A(X)(X^{i_2}+X^{i_1}+1)}{F(X)} = \MOD{A(X)X^{i_2}}{F(X)}+A(X)X^{i_1}+A(X)$. Substituting $k=2$ into (\eDseven) yields
$$
\frac{e_b}{e} = \cases{ (4+5.5s)/(10 + 5.5[i_2 -(h-s)]) & if $h=8,16,32, 64$\cr
(4+5.5s)/(11 + 5.5[i_2 -(h-s)]) & if $h \ne 8,16,32, 64$}
$$

For example, consider the CRC-CCITT generated by $F(X)=X^{16}+X^{12}+X^5+1$, i.e., $h=16$, $i_2=12$, and $i_1=5$. Assume that $s=8$. We then have $e_b/e = (4+5.5 \times 8)/(10+5.5 \times 4) = 48/32=1.5$. Thus, for the 16-bit CRC-CCITT, the new technique is 50\% faster than the basic technique.

Next, consider the CRC-16 generated by $F(X)=X^{16}+X^{15}+X^2+1$, i.e., $h=16$, $i_2=15$, and $i_1=2$. Assume also that $s=8$. We then have $e_b/e = (4+5.5 \times 8)/(10+5.5 \times 7) = 48/48.5$. Thus, for the CRC-16, the new technique is slightly slower than the basic technique. 

Case 3: $i_1>h-s$. Using the new technique, we have $B(X)=\MOD{A(X)(X^{i_2}+X^{i_1}+1)}{F(X)} = \MOD{A(X)X^{i_2}}{F(X)}+\MOD{A(X)X^{i_1}}{F(X)}+A(X)$. Substituting $k=2$ into (\eDten) yields
$$
\frac{e_b}{e} = \cases{ (4+5.5s)/(9 + 5.5[i_2 -(h-s)]) & if $h=8,16,32, 64$\cr
(4+5.5s)/(10 + 5.5[i_2 -(h-s)]) & if $h \ne 8,16,32, 64$}
$$ 
\figure{
{\eightpoint
$$
\vbox
{
\offinterlineskip 
\halign
{
\strut
\vrule \quad #  \vrule    &\quad # \vrule &  \quad# &  \vrule  #  \cr
\noalign{\hrule} 
$X^h+X^{i_2}+X^{i_1}+1$ & 	period	& $2^{h-1}-1\over{\rm period}$ & \cr
\noalign{\hrule} 	
$ X^{5}+X^{3}+X+1 $ &      15 &     1 &     \cr
 $ X^{7}+X^{3}+X^{2}+1 $ &      62 &     1.01613 &       \cr
 $ X^{7}+X^{4}+X^{2}+1 $ &      63 &     1 &     \cr
 $ X^{9}+X^{5}+X^{3}+1 $ &      255 &    1 &     \cr
 $ X^{10}+X^{3}+X^{2}+1 $ &     511 &    1 &     \cr
 $ X^{11}+X^{3}+X+1 $ &     1023 &   1 &     \cr
 $ X^{12}+X^{3}+X+1 $ &     2046 &   1.00049 &       \cr
 $ X^{12}+X^{7}+X^{2}+1 $ &     2047 &   1 &     \cr
 $ X^{15}+X^{3}+X^{2}+1 $ &     16382 &          1.00006 &       \cr
 $ X^{15}+X^{5}+X^{3}+1 $ &     16383 &          1 &     \cr
 $ X^{17}+X^{3}+X+1 $ &     63457 &          1.03275 &       \cr
 $ X^{17}+X^{4}+X^{3}+1 $ &     65534 &          1.00002 &       \cr
 $ X^{17}+X^{10}+X^{4}+1 $ &    65535 &          1 &     \cr
 $ X^{18}+X^{5}+X^{2}+1 $ &     98301 &          1.33336 &       \cr
 $ X^{18}+X^{5}+X^{4}+1 $ &     131071 &         1 &     \cr
 $ X^{19}+X^{3}+X^{2}+1 $ &     262142 &         1 &     \cr
 $ X^{19}+X^{5}+X^{3}+1 $ &     262143 &         1 &     \cr
 $ X^{20}+X^{4}+X^{3}+1 $ &     521985 &         1.00441 &       \cr
 $ X^{20}+X^{7}+X^{5}+1 $ &     524286 &         1 &     \cr
 $ X^{20}+X^{11}+X^{2}+1 $ &    524287 &         1 &     \cr
 $ X^{21}+X^{3}+X+1 $ &     1048575 &        1 &     \cr
 $ X^{22}+X^{3}+X+1 $ &     491460 &         4.26719 &       \cr
 $ X^{22}+X^{3}+X^{2}+1 $ &     2094081 &        1.00147 &       \cr
 $ X^{22}+X^{7}+X^{4}+1 $ &     2097151 &        1 &     \cr
 $ X^{23}+X^{3}+X+1 $ &     4161409 &        1.0079 &        \cr
 $ X^{23}+X^{6}+X+1 $ &     4194300 &        1 &     \cr
 $ X^{23}+X^{7}+X^{6}+1 $ &     4194302 &        1 &     \cr
 $ X^{23}+X^{8}+X^{2}+1 $ &     4194303 &        1 &     \cr
 $ X^{25}+X^{3}+X+1 $ &     4194303 &        4 &     \cr
 $ X^{25}+X^{4}+X+1 $ &     7864260 &        2.13335 &       \cr
 $ X^{25}+X^{4}+X^{3}+1 $ &     12070842 &       1.3899 &        \cr
 $ X^{25}+X^{5}+X+1 $ &     16766977 &       1.00061 &       \cr
 $ X^{25}+X^{6}+X^{3}+1 $ &     16777212 &       1 &     \cr
 $ X^{25}+X^{9}+X^{2}+1 $ &     16777214 &       1 &     \cr
 $ X^{25}+X^{14}+X^{2}+1 $ &    16777215 &       1 &     \cr
 $ X^{26}+X^{3}+X+1 $ &     32505732 &       1.03226 &       \cr
 $ X^{26}+X^{4}+X+1 $ &     33554431 &       1 &     \cr
 $ X^{27}+X^{3}+X^{2}+1 $ &     67108862 &       1 &     \cr
 $ X^{27}+X^{5}+X+1 $ &     67108863 &       1 &     \cr
 $ X^{28}+X^{3}+X+1 $ &     97517382 &       1.37635 &       \cr
 $ X^{28}+X^{5}+X^{2}+1 $ &     134217727 &      1 &     \cr
 $ X^{29}+X^{11}+X+1 $ &    268435455 &      1 &     \cr
 $ X^{30}+X^{3}+X+1 $ &     536870908 &      1 &     \cr
 $ X^{30}+X^{7}+X^{6}+1 $ &     536870911 &      1 &     \cr
 $ X^{31}+X^{3}+X^{2}+1 $ &     50133510 &       21.4176 &       \cr
 $ X^{31}+X^{4}+X+1 $ &     1073213442 &     1.00049 &       \cr
 $ X^{31}+X^{6}+X^{2}+1 $ &     1073602561 &     1.00013 &       \cr
 $ X^{31}+X^{6}+X^{3}+1 $ &     1073741822 &     1 &     \cr
 $ X^{31}+X^{12}+X^{2}+1 $ &    1073741823 &     1 &     \cr
 $ X^{32}+X^{3}+X+1 $ &     21691754 &       99 &    \cr
 $ X^{32}+X^{3}+X^{2}+1 $ &     22362795 &       96.0293 &       \cr
 $ X^{32}+X^{4}+X+1 $ &     2147483647 &     1 &     \cr     
\noalign{\hrule} 	
}
}
$$
}
}{\fvisix}{The period of $X^h+X^{i_2}+X^{i_1}+1$.}

\break
\color{blue}
\subsection{C.4}{CRC Generator Polynomials of Weights Greater Than 4} 
We now consider the case $k\ge 3$, i.e., the CRC generator polynomial
$$
F(X) = X^h + X^{i_k} + X^{i_{k-1}} + \cdots + X^{i_1} + 1	\eqno(\eVIIIii)
$$
has weight greater than 4, i.e., it contains more than 4 terms. Our goal here is to find generator polynomials for CRCs that (a) have minimum distance $d_{\rm min} > 4$ and (b) can be efficiently implemented by the new technique~(\eiiitwelve), i.e., they have low complexity. Codes with minimum distance $d_{\rm min}$ detect all patterns of less than $d_{\rm min}$ errors. For example, the fast CRCs generated by $F_h(X)=X^h+X^2+X+1$, which have $d_{\rm min}=4$, detect all patterns of 1, 2, and 3 errors.

An error pattern $E(X)$ is detected by the CRC generated by $F(X)$ if $E(X)$ is not a multiple of $F(X)$, i.e., $\MOD{E(X)}{F(X)} \not = 0$  (see the proof of Theorem 2). This fact can be used to search for CRCs that can detect specified sets of error patterns.

\color{blue}
In the following, for a given $m>0$, we search for $h$-bit CRCs of length $l$ that can detect all patterns of $m$ errors:
$$
E(X) = X^{a_{m-1}} + X^{a_{m-2}} + \cdots + X^{a_1}+ X^{a_0} 
$$
where $l>a_{m-1}>a_{m-1}> \cdots > a_1 > a_0 \ge 0$. There are $l \choose m$ such $m$-error patterns. Let $l_m$ be the {\it maximum} length of a CRC that can detect all patterns of $m$ errors. A CRC with $d_{\rm min}=m+1$ detects all patterns of $m$ errors or less, and fails to detect some patterns of $m+1$ errors. Thus, a CRC will have $d_{\rm min}=m+1$ if its total code length $ = \min \{l_1, l_2, \dots, l_m \}$ bits, and $d_{\rm min}\ge m+1$ if its total code length $ \le \min \{l_1, l_2, \dots, l_m \}$ bits. Note that $l_2$ of a CRC is also its period.  If $F(X)$ in (\eVIIIii) has {\it even} weight, then all patterns of odd number of errors are detected, i.e., $l_m = \infty$ for odd~$m$. Thus, for odd $m$, the CRC has $d_{\rm min} = m+1$ if its total code length $ = \min \{l_2, l_4,\dots, l_{m-1} \}$ bits. For example, suppose that $k$ is even, i.e., $F(X)$ in (\eVIIIii) has even weight. The CRC generated by $F(X)$ then has $d_{\rm min}=6$ if its total code length $= \min \{l_2, l_4 \}$ bits. Further, this CRC has $d_{\rm min}=8$ if its total code length $= \min \{l_2, l_4, l_6 \}$ bits.

\color{blue}
A straightforward technique to show that a CRC of length $l$ will detect all $m$-error patterns is to verify that each of such $l \choose m$ $m$-error patterns is not a multiple of the CRC generator polynomial. This brute-force technique has computational complexity $\hbox{O}(l^m)$  [\Koo]. A faster technique, which has computational complexity $\hbox{O}(l^{m-1})$, is presented in Remark\rIIiv. 
As an example, Fig.\fweightEqualSix \ shows some $h$-bit CRC generator polynomials of weight 6, for $h=16, 24, 32$, and 64. These CRCs and their $l_2$ and $l_4$ are found by computer search. For each value of $h$, these CRC generator polynomials are arranged according to increasing $l_4$ (see Remark\rIIvi).
As discussed above, these CRCs detect all odd numbers of errors because their generator polynomials have even weights.
Further, these CRCs have $d_{\rm min}=6$ when their total code lengths $l = \min \{l_2, l_4 \}$.   

In the following examples, we discuss the performance and implementation of some CRCs that are generated by the polynomials shown in Fig.\fweightEqualSix. Here, we have $k=4$ and $ F(X)=X^{h} + X^{i_4} + X^{i_3} + X^{i_2} + X^{i_1} + 1 $.  
We assume that the input message is divided into $n$ tuples $Q_0(X), Q_1(X), \dots, Q_{n-1}(X)$. Each tuple $Q_i(X)$ has $s$ bits (e.g., $s=8$, 16, 32 and 64).  
Recall that the $h$-bit CRC generated by (\eVIIIii) can be computed either by the basic technique (see Definition 1) or by the new technique (\eiiitwelve). 
Let $e$ and $e_b$ denote the operation count per input byte required for computing the CRC check tuple under the new technique and under the basic technique, respectively. Thus, the new technique is $e_b/e$ times faster than the basic technique. 
The calculation of $e$, $e_b$, and $e_b/e$ for general CRC generator polynomials is given in Section~C.1.
As shown in the following examples, with proper choice of $s$ for the CRCs, the new technique can be much faster than the basic technique. 

\figure{
{\eightpoint
\vbox
{
\offinterlineskip 
\halign
{
\strut
\vrule \quad # \hss \vrule    &\quad # \vrule &\quad # \vrule &  \quad# &  \vrule  #  \cr
\noalign{\hrule} 
CRC generator polynomial & 	$l_4$ & $l_2=\hbox{period}$	& $2^{h-1}-1\over{\rm period}$ & \cr
\noalign{\hrule}
$ X^{16} + X^{4} + X^{3} + X^{2} + X + 1 $ &   17 &    31620 &         1.03627 &        \cr
 $ X^{16} + X^{5} + X^{3} + X^{2} + X + 1 $ &   67 &    534 &   61.3614 &        \cr
 $ X^{16} + X^{6} + X^{5} + X^{2} + X + 1 $ &   74 &    12264 &         2.6718 &         \cr
 $ X^{16} + X^{7} + X^{6} + X^{4} + X^{3} + 1 $ &       77 &    28658 &         1.14338 &        \cr
 $ X^{16} + X^{8} + X^{4} + X^{3} + X + 1 $ &   115 &   28658 &         1.14338 &        \cr
 $ X^{16} + X^{10} + X^{8} + X^{7} + X^{3} + 1 $ &      128 &   254 &   129.004 &        \cr
 $ X^{16} + X^{14} + X^{11} + X^{5} + X^{2} + 1 $ &     130 &   258 &   127.004 &        \cr
\noalign{\hrule}	
 $ X^{24} + X^{4} + X^{3} + X^{2} + X + 1 $ &   25 &    1048572 &       8.00003 &        \cr
 $ X^{24} + X^{5} + X^{3} + X^{2} + X + 1 $ &   461 &   2446675 &       3.42857 &        \cr
 $ X^{24} + X^{7} + X^{6} + X^{4} + X^{2} + 1 $ &       530 &   344043 &        24.3824 &        \cr
 $ X^{24} + X^{7} + X^{6} + X^{4} + X^{3} + 1 $ &       561 &   2046 &  4100 &   \cr
 $ X^{24} + X^{8} + X^{5} + X^{4} + X^{2} + 1 $ &       691 &   8388607 &       1 &      \cr
 $ X^{24} + X^{14} + X^{13} + X^{11} + X^{10} + 1 $ &   1024 &  2046 &  4100 &   \cr
 $ X^{24} + X^{16} + X^{12} + X^{10} + X + 1 $ &        1030 &  7161 &  1171.43 &        \cr
 $ X^{24} + X^{16} + X^{15} + X^{9} + X^{8} + 1 $ &     2048 &  4094 &  2049 &   \cr
 $ X^{24} + X^{18} + X^{13} + X^{11} + X^{6} + 1 $ &    2050 &  4098 &  2047 &   \cr
\noalign{\hrule} 	
 $ X^{32} + X^{4} + X^{3} + X^{2} + X + 1 $ &   33 &    1610612724 &    1.33333 &        \cr
 $ X^{32} + X^{5} + X^{3} + X^{2} + X + 1 $ &   2948 &  133693185 &     16.0628 &        \cr
 $ X^{32} + X^{5} + X^{4} + X^{3} + X + 1 $ &   3258 &  805306362 &     2.66667 &        \cr
 $ X^{32} + X^{6} + X^{4} + X^{3} + X^{2} + 1 $ &       3501 &  2139094785 &    1.00392 &        \cr
 $ X^{32} + X^{7} + X^{6} + X^{5} + X^{2} + 1 $ &       4145 &  1761607470 &    1.21905 &        \cr
 $ X^{32} + X^{11} + X^{10} + X^{9} + X^{4} + 1 $ &     4198 &  1408426068 &    1.52474 &        \cr
 $ X^{32} + X^{11} + X^{10} + X^{9} + X^{5} + 1 $ &     4480 &  2013265905 &    1.06667 &        \cr
 $ X^{32} + X^{12} + X^{8} + X^{4} + X^{3} + 1 $ &      4856 &  2147483647 &    1 &      \cr
 $ X^{32} + X^{17} + X^{15} + X^{13} + X^{2} + 1 $ &    4989 &  2147483647 &    1 &      \cr
 $ X^{32} + X^{18} + X^{17} + X^{15} + X^{14} + 1 $ &   32770 &         65538 &         32767 &          \cr
\noalign{\hrule}
$ X^{64} + X^{4} + X^{3} + X^{2} + X + 1 $ &    65 &    $2.69\times 10^{18}$ \hss &        3.42857 &    \cr
 $ X^{64} + X^{5} + X^{3} + X^{2} + X + 1 $ &   $>10^5$\hss &        $3.46\times 10^{18}$ \hss &    2.66797 &    \cr
  \noalign{\hrule}	
}
}
}
}{\fweightEqualSix}{CRC generator polynomials of weight 6 \hbox{($d_{\rm min} \ge 4$ if total lengths $\le l_2$, and $d_{\rm min} = 6$ if total lengths $\le \min\{l_2, l_4\}$)}.}

\remark{Example 1.} Consider the 16-bit CRC generated by $ X^{16} + X^{8} + X^{4} + X^{3} + X + 1 $. From Fig.\fweightEqualSix, we have $l_4=115$ and $l_2=28658$. Thus, this CRC detects (a) up to 5 errors if its total length $\le$ 115 bits, and (b) up to 3 errors if its total length $\le$ 28658 bits. In other words, this CRC has $d_{\rm min}=6$ if its total length $\le$ 115 bits, and $d_{\rm min}=4$ if its total length $\le$ 28658 bits.
Here, we have $h=16$, $k=4$, $i_k = i_4 = 8$. For implementation, we consider 2 cases: $s=8$ and 16.

Case 1: $s=8 < h$. Then $8=i_4 = i_k \le h-s=8$. Thus, the results of Section C.1.2.1 can be used in this case. From (\eDseven), we have 
$$\frac{e_b}{e} = \frac{4+5.5s}{6+2k}=\frac{4+5.5\times8}{6+2\times4} = \frac{48}{14}=3.43 
$$
i.e., the new technique is 3.43 times faster than the basic technique when $s=8$.

Case 2: $s=16$. Because $s\ge h$, the results of Section C.1.1 can be used in this case. From (\eDfiveB), we have 

$$
\frac{e_b}{e} = \frac{3+5.5s}{3 + 5.5(s-h + i_k) + k} = \frac{3+5.5\times16}{3+5.5\times8+4}=\frac{91}{51} = 1.78
$$
i.e., the new technique is only 1.78 times faster than the basic technique when $s=16$.

Thus, under the new technique, using $s=8$ is much faster than using $s=16$ for this CRC. The C program for computing the CRC check bits using $s=8$ is shown in Fig.\fCCodeCRCxvi.
\figure{ \includegraphics[height=6cm]{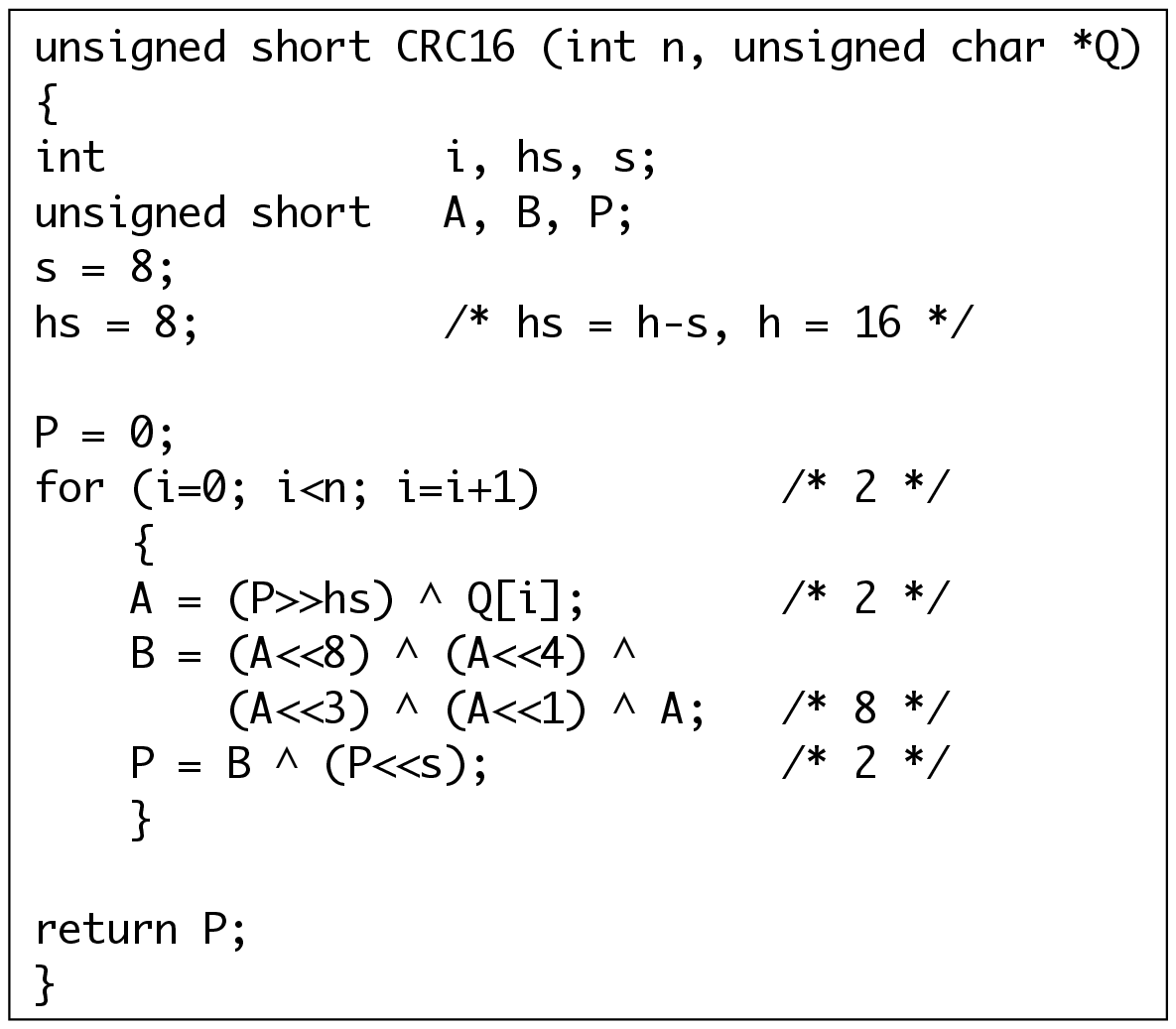} }
{\fCCodeCRCxvi}{C program with $s=8$ for 16-bit CRC generated by $ X^{16} + X^{8} + X^{4} + X^{3} + X + 1 $.}
\QED

\remark{Example 2.} Consider the 24-bit CRC generated by $ X^{24} + X^{8} + X^{5} + X^{4} + X^{2} + 1 $. From Fig.\fweightEqualSix, we have $l_4= 691 $ and $l_2= 8388607 $. Thus, this CRC detects (a) up to 5 errors if its total length $\le$ 691 bits, and (b) up to 3 errors if its total length $\le$ 8388607 bits.
Here, we have $k=4$, $i_k = i_4 = 8$, and $h=24\not = 8,16, 32, 64$. For implementation, we consider 2 cases: $s=8$ and 16.

Case 1: $s=8 < h$. We have $8=i_4 = i_k \le h-s=16$. Thus, the results of Section C.1.2.1 can be used in this case. From (\eDseven), we have 
$$\frac{e_b}{e} = \frac{4+5.5s}{7+2k}=\frac{4+5.5\times8}{7+2\times4} = \frac{48}{15}=3.20 
$$

Case 2: $s=16 < h$.
We have $8=i_4 = i_k \le h-s=8$. Thus, the results of Section C.1.2.1 can be used in this case. From (\eDseven), we have 
$$\frac{e_b}{e} = \frac{4+5.5s}{7+2k}=\frac{4+5.5\times16}{7+2\times4} = \frac{92}{15}=6.13 
$$

Thus, under the new technique, using $s=16$ is much faster than using $s=8$ for this CRC. \QED

\remark{Example 3.} Consider the 24-bit CRC generated by $ X^{24} + X^{16} + X^{15} + X^{9} + X^{8} + 1 $. From Fig.\fweightEqualSix, we have $l_4= 2048 $ and $l_2= 4094 $. Thus, this CRC detects (a) up to 5 errors if its total length $\le$ 2048 bits, and (b) up to 3 errors if its total length $\le$ 4094 bits.
Here, we have $k=4$, $i_k = i_4 = 16$, and $h=24\not = 8,16, 32, 64$. For implementation, we consider 2 cases: $s=8$ and 16.

Case 1: $s=8 < h$.
We have $16=i_4 = i_k \le h-s=16$. Thus, the results of Section C.1.2.1 can be used in this case. From (\eDseven), we have 
$$\frac{e_b}{e} = \frac{4+5.5s}{7+2k}=\frac{4+5.5\times8}{7+2\times4} = \frac{48}{15}=3.20 
$$

Case 2: $s=16 < h$.
We have $16=i_4 = i_k > h-s=8$. Thus, the results of Section C.1.2.2 can be used in this case. Further, consider Case 1 of Section C.1.2.2, which requires the existence of an $m$ such that $i_{m+1} > h-s \ge i_m$. We have $9=i_2 > h-s = 8 \ge i_1=8$. Thus, $m=1$. Using (\eDnine), we have
$$
\frac{e_b}{e} = \frac{ 4+5.5s}{7+ 5.5[i_k -(h-s)]+k+m +1} = \frac{4+5.5\times16}{7+5.5\times(16-8)+4+1+1} = \frac{92}{57} = 1.61 
$$

Thus, under the new technique, using $s=8$ is much faster than using $s=16$ for this CRC.
\QED


\remark{Example 4.} Consider the 32-bit CRC generated by $ X^{32} + X^{12} + X^{8} + X^{4} + X^{3} + 1 $. From Fig.\fweightEqualSix, we have $l_4= 4856 $ and $l_2= 2147483647 $. Thus, this CRC detects (a) up to 5 errors if its total  length $\le$ 4856 bits, and (b) up to 3 errors if its total length $\le$ 2147483647 bits.
Here, we have $h=32=\hbox{degree}(F(X))$, $k=4$, $i_k = i_4 = 12$, $i_3=8$, $i_2=4$, and $i_1=3$. For implementation, we consider 3 cases: $s=8, 16$, and 32.

Case 1: $s=8 < h$. 
Then $12=i_4=i_k \le h-s=24$. Thus, the results of Section C.1.2.1 can be used in this case. From (\eDseven), we have 
$$\frac{e_b}{e} = \frac{4+5.5s}{6+2k}=\frac{4+5.5\times8}{6+2\times4} = \frac{48}{14}=3.43
$$

Case 2: $s=16 < h$. Then $12=i_4=i_k \le h-s=16$. Thus, the results of Section C.1.2.1 can be used in this case. From (\eDseven), we have 
$$\frac{e_b}{e} = \frac{4+5.5s}{6+2k}=\frac{4+5.5\times16}{6+2\times4} = \frac{92}{14}=6.57
$$

Case 3: $s=32 \ge h$. 
Then $12=i_4=i_k > h-s = 0$. Thus, the results of Subsection C.1.2.2 can be used in this case. Further, we have $3=i_1 > h-s = 0$. Thus, Case 2 of Section C.1.2.2 is applicable here. Using~(\eDten), we have

$$
\frac{e_b}{e} = \frac{ 4+5.5s}{6+ 5.5[i_k -(h-s)]+k +1} = \frac{4+5.5\times32}{6+5.5\times12+4+1} = \frac{180}{77} = 2.34 
$$

Thus, under the new technique, using $s=16$ is much faster than using $s=8$ and 32 for this CRC. The C program for computing the CRC check bits using $s=16$ is shown in Fig.\fCCodeCRCThirtytwo.
\figure{ \includegraphics[height=6cm]{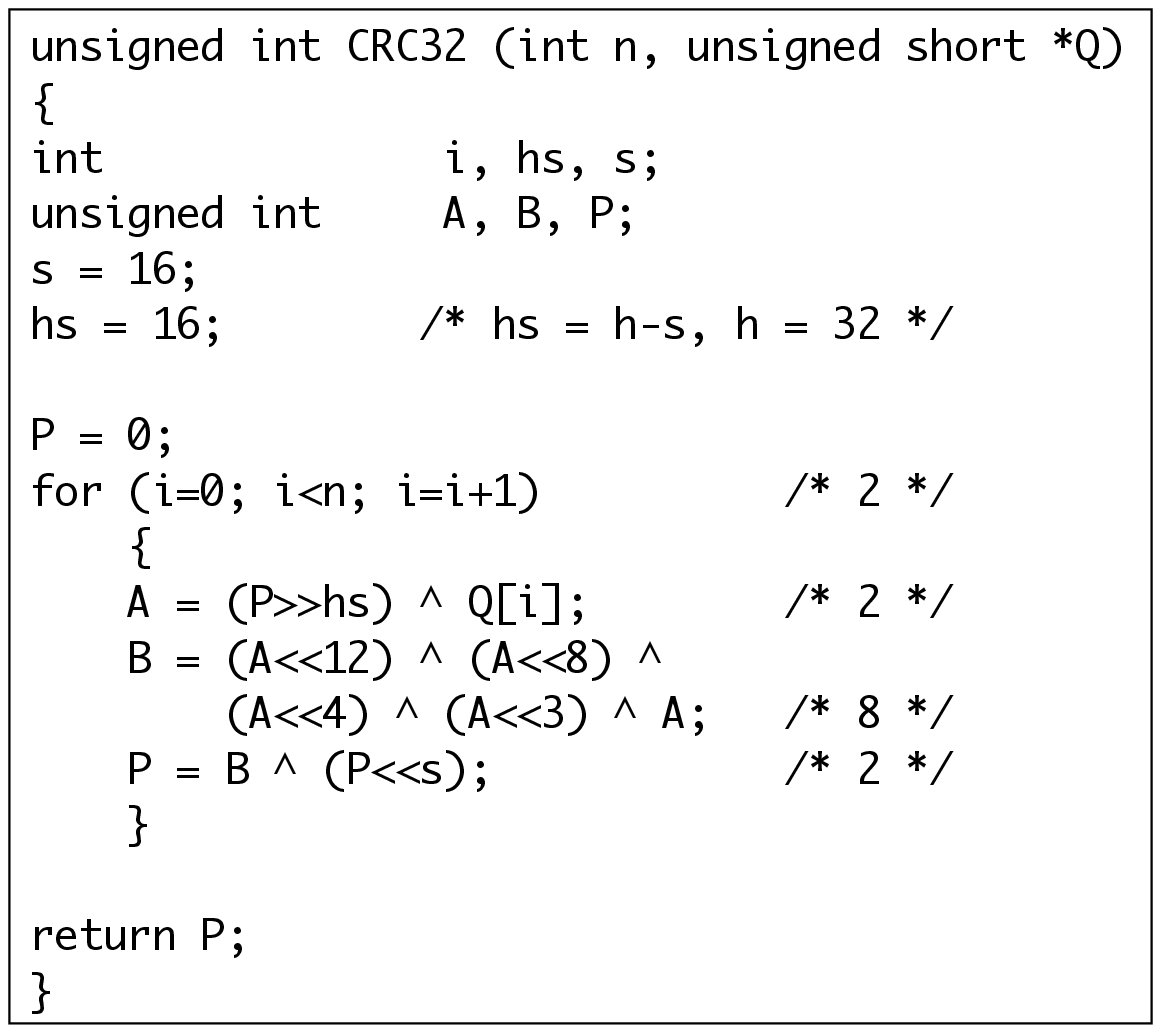} }
{\fCCodeCRCThirtytwo}{C program with $s=16$ for the 32-bit CRC generated by $ X^{32} + X^{12} + X^{8} + X^{4} + X^{3} + 1 $.}
\QED

\remark{Example 5.} Consider the 32-bit CRC generated by $ F(X) = X^{32} + X^{18} + X^{17} + X^{15} + X^{14} + 1$ (= 10006c001 in hexadecimal notation). From Fig.\fweightEqualSix, we have $l_4= 32770 $ and $l_2= 65538 $. Thus, this CRC detects (a) up to 5 errors if its total length $\le$ 32770 bits, and (b) up to 3 errors if its total length $\le$ 65538 bits. 
Here, we have $h=32$, $k=4$, $i_k = i_4 = 18$, $i_3=17$, $i_2=15$, and $i_1=14$. For implementation, we consider 3 cases: $s=8, 16$, and 32.

Case 1: $s=8 < h$. 
Then $18=i_4=i_k \le h-s=24$. Thus, the results of Section C.1.2.1 can be used in this case. From (\eDseven), we have 
$$\frac{e_b}{e} = \frac{4+5.5s}{6+2k}=\frac{4+5.5\times8}{6+2\times4} = \frac{48}{14}=3.43
$$
i.e., for bitwise implementation, the 32-bit CRC generated by $F(X)$ is 3.43 times faster than basic 32-bit CRCs. The C program for bitwise implementation and $s=8$ for this CRC is shown in Fig.\fCCodeOptimalThirtytwoBitwise.

Case 2: $s=16 < h$. 
First, consider bitwise implementation. Because $18=i_k > h-s = 16$, the results of Section C.1.2.2 can be used in this case. Further, consider Case 1 of Section C.1.2.2, which requires the existence of an $m$ such that $i_{m+1} > h-s \ge i_m$. We have $17=i_3 > h-s = 16 \ge i_2=15$. Thus, $m=2$. Using (\eDnine), we have

$$
\frac{e_b}{e} = \frac{ 4+5.5s}{6+ 5.5[i_k -(h-s)]+k+m +1} = \frac{4+5.5\times16}{6+5.5\times(18-16)+4+2+1} = \frac{92}{24} = 3.83 
$$

Using (\eFastB), we have $r=5.5[i_k-(h-s)]+k+m+1=18$, which is substituted into (\eDfiveE) to show that the operation count per input byte required for computing the CRC check tuple under the {\it bitwise} new technique is $e=12$. As shown below, by using a table of only 4 entries, $e$ can be reduced to 8.5.

We now discuss table-looup implementation for the CRC generated by $F(X) = X^{32} + X^{18} + X^{17} + X^{15} + X^{14} + 1$. We can implement the table lookup for this CRC by imitating the table-lookup implementation presented in Section A.2.2 for the fast CRCs generated by $F_h(X)=X^h+X^2+X+1$. Using the new technique (\eiiitwelve), we have 
$$
\eqalign{B(X) 
&= \MOD{A(X)(X^{18} + X^{17} + X^{15} + X^{14} + 1)}{F(X)}	\cr
&= \MOD{A(X)(X^{18} + X^{17})}{F(X)}+ A(X)(X^{15} + X^{14} + 1)
		}
$$

We now decompose $A(X)$ into 2 simpler polynomials $A_1(X)$ and $A_2(X)$:
$$A(X) = A_1(X)X^{14} + A_2(X)$$
where $\hbox{degree}(A_1(X))<2$ and  $\hbox{degree}(A_2(X))<14$. We then have

$$
\eqalign{B(X) 
&= \MOD{(A_1(X)X^{14} + A_2(X))(X^{18} + X^{17})}{F(X)}+ A(X)(X^{15} + X^{14} + 1) \cr
&= \MOD{(A_1(X)X^{14}(X^{18} + X^{17})}{F(X)}+ A_2(X)(X^{18} + X^{17})+A(X)(X^{15} + X^{14} + 1) \cr
&= \MOD{(A_1(X)X^{31}(X+1)}{F(X)}+ A_2(X)(X^{18} + X^{17})+A(X)(X^{15} + X^{14} + 1) \cr	
&= T[A_1]+ A_2(X)(X^{18} + X^{17})+A(X)(X^{15} + X^{14} + 1) \cr}
$$
where $T[\ ]$ is the table defined by
$$
T[A_1] = \MOD{(A_1(X)X^{31}(X+1)}{F(X)}
$$
where $A_1$ is a 2-tuple. Thus, table $T[\ ]$ has 4 entries, which can be shown to be: 
$$
\eqalign{
T[0] &= 0 			\cr
T[1] &= X^{31}+X^{18}+X^{17}+X^{15}+X^{14}+1 			\cr
T[2] &= X^{19}+X^{17}+X^{16}+X^{14}+X+1 			\cr
T[3] &= X^{31}+X^{19}+X^{18}+X^{16}+X^{15}+X 			\cr
}
$$
In hexadecimal notation, we have $T[1]= 8006c001$, $T[2]=b4003$, and $T[3]=800d8002$. The C program, which includes the table of 4 entries, for this CRC is shown in Fig.\fCCodeOptimalThirtytwoTable. Recall from Appendix A that~$r$ denotes the operation count required for computing $B(X)$.
From Fig.\fCCodeOptimalThirtytwoTable, we have $r=11$, which is substituted into~(\eDfiveE) to show that the operation count per input byte required for computing the CRC check tuple under the new technique with {\it table-lookup} is $e=8.5$ (as compared to $e=12$ for the case of bitwise implementation).

Case 3: $s=32 = h$. Using (\eDfiveB), we have

$$
\frac{e_b}{e} = \frac{3+5.5s}{3 + 5.5(s-h + i_k) + k} = \frac{3+5.5\times32}{3+5.5\times18+4}=\frac{179}{106} = 1.69 
$$
Thus, under the new technique, using $s=32$ is slower than using $s=8$ and 16 for the CRC generated by $ X^{32} + X^{18} + X^{17} + X^{15} + X^{14} + 1 $.
\figure{ \includegraphics[height=6cm]{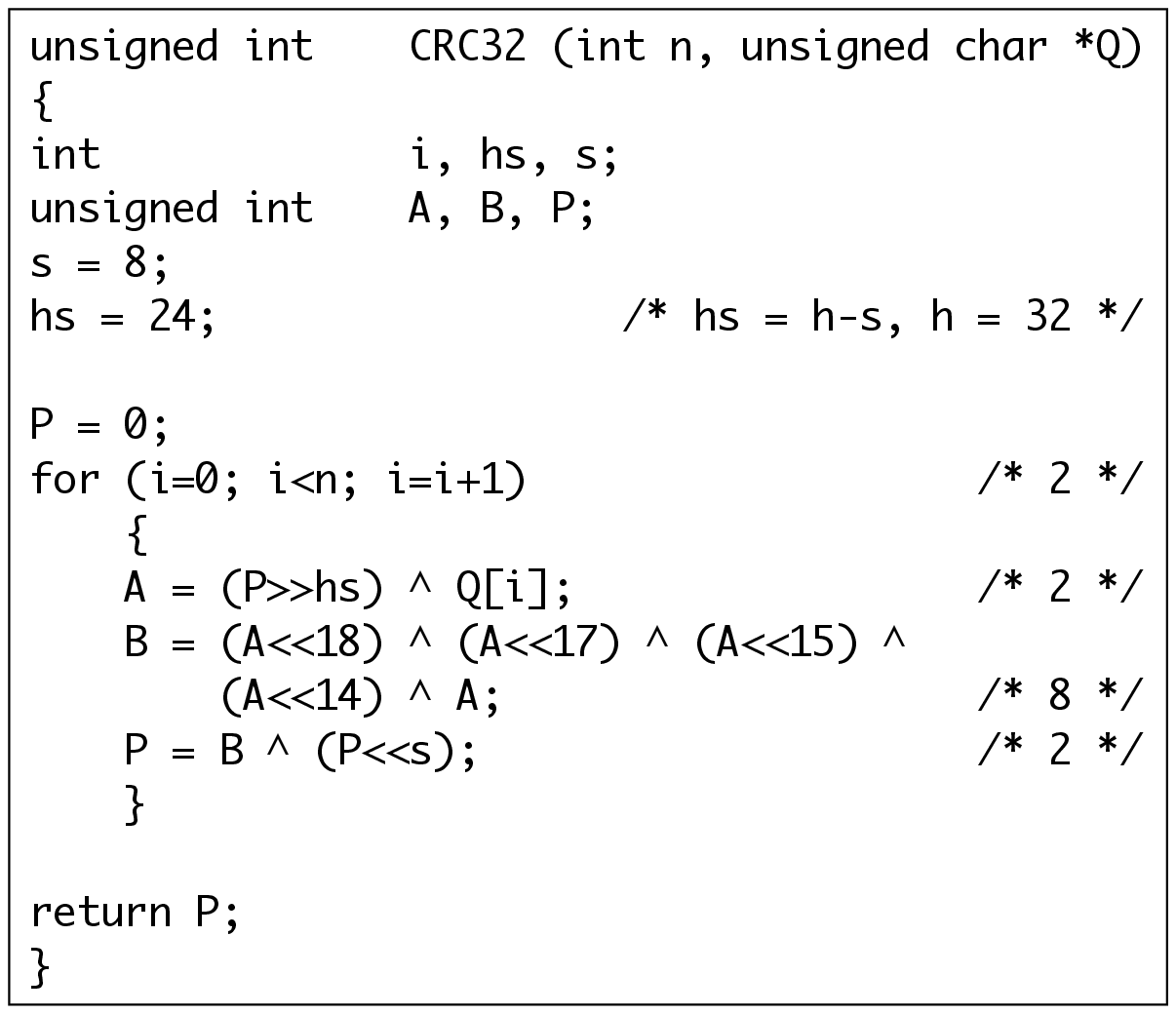} }
{\fCCodeOptimalThirtytwoBitwise}{C program ($s=8$, without table lookup) for the 32-bit CRC  generated \hbox{by $ X^{32} + X^{18} + X^{17} + X^{15} + X^{14} + 1 $} \hbox{($d_{\rm min}= 6$ if total length $\le$ 32770 bits, and $d_{\rm min}= 4$ if total length $\le$ 65538 bits).}}

\figure{ \includegraphics[height=8cm]{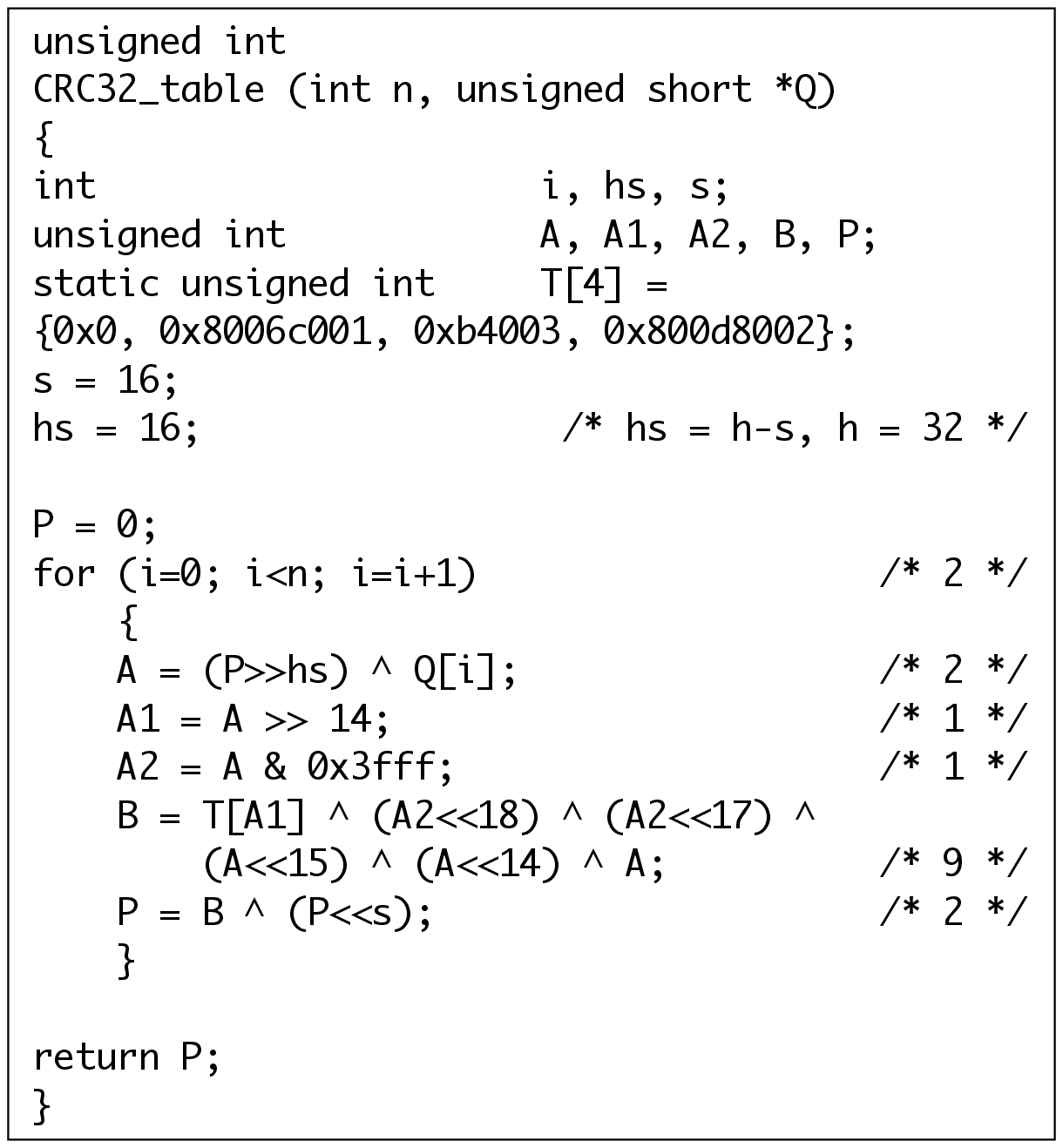} }
{\fCCodeOptimalThirtytwoTable}{C program ($s=16$, with 4-entry table lookup)  for the 32-bit CRC generated \hbox{by $ X^{32} + X^{18} + X^{17} + X^{15} + X^{14} + 1 $} \hbox{($d_{\rm min}= 6$ if total length $\le$ 32770 bits, and $d_{\rm min}\ge 4$ if total length $\le$ 65538 bits)}.}
\QED

\remark{Example 6.} Consider the 64-bit CRC generated by $ X^{64} + X^{5} + X^{3} + X^{2} + X + 1 $. From Fig.\fweightEqualSix, we have $l_4>10^5$ and $l_2=3.46\times 10^{18}$. Thus, this CRC detects (a) up to 5 errors if its total length $\le 10^5$ bits, and (b) up to 3 errors if its total length $\le 3.46\times 10^{18}$ bits.
Here, we have $h=64$, $k=4$, $i_k = i_4 = 5$. For implementation, we consider 4 cases: $s=8$, 16, 32 and 64.

Case 1: $s=8 < h$. We have $5=i_4 = i_k \le h-s=56$. Thus, the results of Section C.1.2.1 can be used in this case. From (\eDseven), we have 
$$\frac{e_b}{e} = \frac{4+5.5s}{6+2k}=\frac{4+5.5\times8}{6+2\times4} = \frac{48}{14}=3.43 
$$

Case 2: $s=16 < h$. We have $5=i_4 = i_k \le h-s=48$. Thus, the results of Section C.1.2.1 can be used in this case. From (\eDseven), we have 
$$\frac{e_b}{e} = \frac{4+5.5s}{6+2k}=\frac{4+5.5\times16}{6+2\times4} = \frac{92}{14}= 6.57 
$$

Case 3: $s=32 < h$. We have $5=i_4 = i_k \le h-s=32$. Thus, the results of Section C.1.2.1 can be used in this case. From (\eDseven), we have 
$$\frac{e_b}{e} = \frac{4+5.5s}{6+2k}=\frac{4+5.5\times32}{6+2\times4} = \frac{180}{14}= 12.86 
$$

Case 4: $s=64$. Because $s\ge h$, the results of Section C.1.1 can be used in this case. From (\eDfiveB), we have 

$$
\frac{e_b}{e} = \frac{3+5.5s}{3 + 5.5(s-h + i_k) + k} = \frac{3+5.5\times64}{3+5.5\times5+4}=\frac{355}{34.5} = 10.29
$$

Thus, under the new technique, using $s=32$ is much faster than using other values of $s$ for the CRC generated by $ X^{64} + X^{5} + X^{3} + X^{2} + X + 1 $. 
\QED

Fig.\fweightEqualSix \ shows $l_2$ and $l_4$ for CRC generator polynomials that have weight 6. Fig.\fweightGreaterSix \ shows $l_2$ and $l_4$ for CRC generator polynomials that have weights greater than 6.
Although we have $l_2>l_4$, i.e., $\min(l_2, l_4)=l_4$ for all the CRCs in Fig.\fweightEqualSix, this may not be true for all the CRCs in Fig.\fweightGreaterSix, e.g., $l_2=151$ and $l_4=152$ (i.e., $l_2<l_4$) for the CRC generated by $ X^{16} + X^{13} + X^{12} + X^{10} + X^{9} + X^{4} + X + 1 $. Note that our search also produces the ``CRC32sub8" and ``CRC32sub16" polynomials presented in [\RaK]: $ X^{32} + X^{7} + X^{6} + X^{5} + X^{2} + 1 $  (Fig.\fweightEqualSix ) and $ X^{32} + X^{13} + X^{12} + X^{10} + X^{8} + X^{6} + X^{4} + 1$ (Fig.\fweightGreaterSix). {\color{red}Without using table lookup, the CRCs generated by other CRC32sub8 and CRC32sub16 polynomials in [\RaK] can also be efficiently implemented by the fast technique (\eiiitwelve) with $s\le 24$ and $s\le 16$, respectively.}

In Fig.\fweightEqualFive, we present CRC generator polynomials of weight 5, i.e., $k=3$. These polynomials generate CRCs that have $d_{\rm min}=5$ if their total code length $\le \min\{l_2, l_3, l_4\}$ bits. Let us compare the CRCs in Fig.\fweightEqualFive \ with those in Fig.\fweightEqualSix. First, the largest values of $l_4$ for $h=16$ and 32 in Fig.\fweightEqualFive \ are almost twice of those in Fig.\fweightEqualSix. Next, while $l_m=\infty$ for the CRCs in Fig.\fweightEqualSix \ for odd  $m$, we have $l_m < \infty$ for those in Fig.\fweightEqualFive. The CRC generator polynomials of odd weights greater than  5 are given in Fig.\fweightGreaterFive. In particular, $l_4$ for $ X^{24} + X^{14} + X^{13} + X^{12} + X^{11} + X^{10} + 1 $ in Fig.\fweightGreaterFive \ is almost twice that for $ X^{24} + X^{17} + X^{12} + X^{7} + 1 $ in Fig.\fweightEqualFive.

{\color{red}%
In Fig.\fweightEqualFivePrimitive, we present fast $h$-bit CRCs that are generated by {\it primitive} polynomials of weight 5, i.e., $F(X)=X^h+X^{i_3}+X^{i_2}+X^{i_1}+1$.
We have $l_2 = 2^h-1$, because the polynomials are primitive. These CRCs have fast implementation when $s$ is chosen such that $s \le h - i_3$ or $i_3 \le h-s$ (see Section C.1.2.1).
Note that Fig.\fweightEqualFivePrimitive \ includes some polynomials in Fig.\fweightEqualFive, as well as the CRC-64-ISO polynomial $X^{64}+X^4+X^3+X+1$. 

Let us compare the polynomial $ X^{32} + X^{7} + X^{6} + X^{2} + 1 $ in Fig.\fweightEqualFivePrimitive \ with the popular CRC-32-IEEE 802.3 primitive polynomial (\eDnineA). First, both these polynomials have the same maximum period $l_2=2^{32}-1$.
Using computer search, it can be shown that the CRC-32-IEEE 802.3 polynomial has $l_4 = 3006$ and $l_3 = 91639$ [\Koo], which are smaller than $l_4=5281$ and $l_3=142741$ for $ X^{32} + X^{7} + X^{6} + X^{2} + 1 $. Thus, the 32-bit CRC generated by $ X^{32} + X^{7} + X^{6} + X^{2} + 1 $ is both faster and more effective (i.e., for patterns of 3 and 4 errors) than the CRC-32-IEEE 802.3. 
}

So far, we present polynomials that generate CRCs that have $d_{\rm min}=5$ (see Figs.\fweightEqualFive \ and \fweightGreaterFive) and $d_{\rm min}=6$ (see Figs.\fweightEqualSix \ and  \fweightGreaterSix). Generator polynomials for CRCs that have $d_{\rm min}>6$ can also be found. For example, Fig.\fweightEqualEight \ shows polynomials of weight 8, which generate CRCs that have $d_{\rm min}=8$ if their total code lengths $\le \min\{l_2, l_4, l_6\}$ bits, because $l_m=\infty$ for odd $m$. Note that these same CRCs have $d_{\rm min}\ge 4$ and $d_{\rm min}\ge 6$ if their total code lengths $\le l_2$ and $\le \min\{l_2, l_4\}$ bits, respectively. Similar to the CRCs presented in Examples~1-6, many CRCs in Fig.\fweightEqualEight \ also have fast implementation. However, they are usually not as fast as the CRCs in those examples, because they are generated by polynomials that have greater weights (see Figs.\fweightEqualSix \ and\fweightEqualEight).  

{\color{red}
There are also many other CRCs, which are not presented here, that can be efficiently implemented by the fast technique (\eiiitwelve). For example, Fig.\fweightEqualEight \ shows the 32-bit CRC generated by $ X^{32} + X^{16} + X^{15} + X^{10} + X^{6} + X^{2} + X + 1 $, which has $l_6=301$, $l_4=3298$ and $l_2=2147483644 \approx 2^{31}-1$. Not shown in Fig.\fweightEqualEight \ is an alternative 32-bit CRC generated by $ X^{32} + X^{16} + X^{15} + X^{11} + X^{6} + X^{5} + X^{2} + 1 $, which has $l_6=255$, $l_4=3509$, and $l_2=2147483647=2^{31}-1$. These 2 CRCs have almost identical $l_2$, but have different $l_6$ and $l_4$. They can also be efficiently implemented by the fast technique (\eiiitwelve) with $s=16$ or $s=8$. 
}

\remark{Remark\rIIii.}
As shown in Fig.\fweightEqualSix, the 32-bit CRC generated by $X^{32}+X^4+X^3+X^2+X+1$ has $l_4=33$, i.e., this CRC fails to detect at least one pattern of 4 errors when its total length exceeds 33 bits. Similarly, the 64-bit CRC generated by $X^{64}+X^4+X^3+X^2+X+1$ has $l_4=65$. Note that these polynomials consist of $X^h$ and {\it consecutive} powers of $X$. We now show in general that such polynomials have $l_4=h+1$.
Thus, consider the $h$-bit CRC generated by $F(X)=X^h+X^n+X^{n-1}+\cdots+X^3+X^2+X+1$, which has weight $n+2$ and consists of $X^h$ and consecutive powers of $X$, namely, $X^n, X^{n-1},\dots,X^3,X^2,X,1$. Here, we assume that $3\le n\le h-2$. We have $l_2, l_3, l_4 \ge h+1$ because $F(X)$ has weight greater than~4. The 4-error pattern $E(X)=X^{h+1}+X^h+X^{n+1}+1$ is a multiple of $F(X)$ because $F(X)(X+1) = E(X)$, which implies that $l_4 \le h+1$. Thus, $l_4 = h+1$. Note that $l_1=\infty$ because $F(X)$ is not a multiple of $X$. Thus, $\min(l_1, l_2, l_3, l_4)=l_4=h+1$.

Case 1: $n$ is even. Then $l_1=l_3=l_5=\infty$ because $F(X)$ has even weight.  The $h$-bit CRC generated by $F(X)$ then has minimum distance $d_{\rm min}=6$ when its total length = $\min(l_1, l_2, l_3, l_4, l_5)=\min(l_2, l_4)=l_4=h+1$ bits. This CRC has $d_{\rm min}\ge 4$ when its total length $\le \min(l_1, l_2, l_3)=l_2$ bits. Recall that $l_2$ is also the period of $F(X)$. 

Case 2:~$n$ is odd.~The $h$-bit CRC generated by $F(X)$ then has  $d_{\rm min}=5$ when its total length $=\min(l_1, l_2, l_3, l_4)=l_4=h+1$ bits. This CRC has  $d_{\rm min}\ge 3$ when its total length $\le \min(l_1,l_2)=l_2$~bits.~\QED

\remark{Remark\rIIiv.} Consider a CRC that is generated by $F(X)$ and has total length of $l$ bits. When a CRC codeword of length $l$ is transmitted, it is affected by errors. Let us focus on the patterns of $m$ errors. There are ${l \choose m}$ such patterns of $m$ errors. A straightforward technique to show that the CRC detects all the $m$-error patterns is to verify that it detects each of the ${l \choose m}$ $m$-error patterns.
However, we show below that, to verify that this CRC detects all the $m$-error patterns, it is sufficient to verify that it detects all the error patterns from a subset of only ${{l-1} \choose {m-1}}$ patterns of $m$ errors.
First, let $E(X)$ be an error polynomial.  We can write 
$$
E(X)=X^aE^*(X)
$$
where $a\ge 0$ and $E^*(X)$ is a polynomial whose least significant bit is 1, i.e., $E^*(0)=1$. For example, let $E(X)=X^5+X^2$. We then have $a=2$ and $E^*(X)=X^3+1$, because $E(X)=X^2(X^3+1)$. 

Next, we show that $E(X)$ is undetected iff $E^*(X)$ is undetected. Thus, suppose that the error pattern $E(X)$ is undetected, i.e., it is a codeword of the CRC generated by $F(X)$. We then have $E(X)=K(X)F(X)$ for some polynomial $K(X)$, which implies $X^aE^*(X) = K(X)F(X)$. Because we assume that $F(X)$ is not a multiple of $X$, i.e., $\hbox{gcd}(X, F(X))=1$, we must have $E^*(X) = K^*(X)F(X)$ for some polynomial $K^*(X)$. Thus, $E^*(X)$ is also a codeword, i.e., it is an undetected error pattern. Suppose now that $E^*(X)$ is undetected, i.e., $E^*(X)$ is a multiple of $F(X)$. Then $E(X)=X^aE^*(X)$ must also be undetected, because $E(X)$ is also a multiple of $F(X)$. To summarize, $E(X)$ is undetected iff $E^*(X)$ is undetected. This fact can be used to speed up the search for CRCs that can detect specified sets of error patterns.

Let ${\bf A}$ be a set of error patterns. As seen above, each $E(X) \in {\bf A}$ can be written as $E(X)=X^aE^*(X)$, for some $a \ge 0$ and $E^*(0)=1$. Let ${\bf A^*}$ be the set of all such $E^*(X)$, i.e.,
$$
{\bf A^*} = \{E^*(X) \ : \ E^*(0)=1, X^aE^*(X) \in {\bf A} \hbox{ for some } a \ge 0 \}
$$
We must have $|{\bf A^*}| \le |{\bf A}|$. However, in general, it is not necessarily that ${\bf A^*} \subseteq {\bf A}$. In particular, consider a CRC having total length of $l$ bits, and let ${\bf A}$ be the set of all patterns of $m$ errors. We then have $|{\bf A}|= {l \choose m}= \hbox{O}(l^m)$. Because ${\bf A^*}$ is the set of all patterns of $m$ errors, under the restriction that the least significant bit of each of these error patterns is 1 [i.e., $E^*(0)=1$], we must have $|{\bf A^*}|= {{l-1} \choose {m-1}}= \hbox{O}(l^{m-1})$. 

A straightforward technique to show that a CRC of length $l$ will detect all the $m$-error patterns is to verify that each $m$-error pattern in ${\bf A}$ is not a multiple of the CRC generator polynomial [\Koo]. More specifically, for each $m$-error pattern
$$
E(X) = X^{a_{m-1}} + X^{a_{m-2}} + \cdots + X^{a_1}+ X^{a_0} 
$$
in {\bf A}, we compute $\MOD{E(X)}{F(X)} = \sum_{i=0}^{m-1}\MOD{X^{a_i}}{F(X)}$. The error $E(X)$ is undetected iff $\MOD{E(X)}{F(X)}=0$. The computation can also be implemented by table lookup [\Cha]. We then have 
$$
\MOD{E(X)}{F(X)} = \sum_{i=0}^{m-1}T[a_i]
$$
where the table $T[\ ]$ is defined by $T[a]= \MOD{X^{a}}{F(X)}$. Overall, this brute-force technique has computational complexity ${l \choose m}=\hbox{O}(l^m)$. 

As seen above, $E(X) \in {\bf A}$ is undetected iff $E^*(X)\in {\bf A^*}$ is undetected. Recall that $|{\bf A}|= {l \choose m}$ and $|{\bf A^*}|= {{l-1} \choose {m-1}}$. Thus, an alternative technique to show that a CRC of length $l$ will detect all the $m$-error patterns is to verify that each $m$-error pattern in ${\bf A^*}$ is not a multiple of the CRC generator polynomial.  
This alternative technique has computational complexity ${{l-1} \choose {m-1}}= \hbox{O}(l^{m-1})$, and is faster than the brute-force technique by the factor ${l \choose m}/{{l-1} \choose {m-1}}=l/m  = \hbox{O}(l)$. \QED

\color{black}
\remark{Remark\rIIvi.} Here, we explain how the CRC generator polynomials shown in Figs.\fweightEqualSix \ and\fweightGreaterSix-\kern-1.4ex\fweightGreaterEight\ are found. We restrict our search of  $h$-bit CRC generator polynomials to a subset of polynomials ${\bf S} = \{F_0(X), F_1(X), \dots, F_{n-1}(X) \}$, for some $n$, where  each $F_j(X)$ is a polynomial of the form (\eVIIIii), i.e., it has  degree $h$ and weight $k+2$ (i.e., $F_j(X)$ has $k+2$ terms). 
For example, if we restrict ${\bf S}$ to be the set of polynomials of degree $h$ and weight 6 (i.e., $k=4$), we then have $n=|{\bf S}| = {h-1 \choose 4}$.
Note that each polynomial $F_j(X)$ can also be represented by a binary integer $F_j$ whose digits are the coefficients of $F_j(X)$, i.e., $F_j \equiv F_j(X)$. The polynomials in {\bf S} are arranged in increasing order, i.e., $F_{j_1} < F_{j_2}$ when $j_1 < j_2$.  

Consider a CRC generated by $F_j(X)$ that has the form (\eVIIIii). Because any undetected error must be a multiple of $F_j(X)$, it follows that $l_i \ge h$ for all $i\ge 1$.
In particular, $l_{k+2} = h$, because the error $E(X)=F_j(X)$, which has weight $k+2$ and length $h+1$, is undetected. We also have $l_1=\infty$. Recall that $l_2$ is also the period of the polynomial. Here, we are only interested in $l_m$ for $2 \le m \le k+1$. For example, when $k=3$, i.e., $F_j(X)$ is a polynomial of weight~5, we are only interested in $l_2, l_3$, and $l_4$ (see Fig.\fweightEqualFive). When $k$ is even, i.e., $F_j(X)$ is a polynomial of even weight, we have $l_m = \infty$ for odd $m$. In this case, we are only interested in $l_m$ for even $m$, i.e., for $m=2, 4, \dots, k$. For example, when $k=4$, i.e., $F_j(X)$ is a polynomial of weight~6, we are only interested in  $l_2$ and $l_4$ (see Fig.\fweightEqualSix).

\noindent (a) Consider Fig.\fweightEqualSix, which shows polynomials of weight~6, i.e., $k=4$ and $F_j(X)=X^h+X^{i_4}+X^{i_3}+X^{i_2}+X^{i_1}+1$.
Here, we show $l_2$ and $l_4$. Each $F_j(X)$ generates a CRC that has minimum distance $d_{\rm min} = 6$ when its total length $\le \min\{l_2, l_4 \}$ bits. The polynomials are shown in increasing values of their binary representation and increasing $l_4$, i.e., the $l_4$ of $F_{j_1}(X)$ is smaller than the $l_4$ of $F_{j_2}(X)$ for $j_1 < j_2$. Although these CRCs can be implemented by the familiar basic technique, they can be much faster implemented by the new technique~(\eiiitwelve). Using the new technique, an $h$-bit CRC with smaller $l_4$ is at least as fast as an $h$-bit CRC with larger $l_4$. Thus, as expected, there are tradeoffs between code capability and speed, i.e., CRCs with smaller $l_4$ is faster than CRCs with larger $l_4$.  

\noindent (b) As seen in Fig.\fweightEqualSix, where we impose the condition $k=4$ (i.e., the generator polynomials have weight exactly~6), the generator polynomial $ X^{16} + X^{14} + X^{11} + X^{5} + X^{2} + 1$ yields the largest $l_4= 130$ for the case of $h=16$. Can $l_4$ be improved if we allow $k>4$? Consider Fig.\fweightGreaterSix, which shows the CRC generator polynomials with even $k$ and $k>4$, i.e., each $F_j(X)$ is a polynomial of even weight greater than 6. Our purpose here is to find out if there are other CRCs that have values of $l_4$ that are larger than those in Fig.\fweightEqualSix. Such a CRC generator exists for the case $h=16$, namely, $ X^{16} + X^{13} + X^{12} + X^{10} + X^{9} + X^{4} + X + 1 $, with $l_4=152$, which is larger than the largest $l_4= 130$ in Fig.\fweightEqualSix. Note that, using the new technique~(\eiiitwelve), the CRCs in Fig.\fweightEqualSix \ are usually faster than those in Fig.\fweightGreaterSix, because they are generated by polynomials that have lower weights.

\noindent (c) Figs.\fweightEqualFive \ and\fweightGreaterFive \ show generator polynomials that have odd weights and generate CRCs that can detect 1, 2, 3, and 4 errors. We now show $l_2, l_3$ and $l_4$. In Fig.\fweightEqualFive, we require $k=3$ (i.e., the generator polynomials have weight exactly~5). In Fig.\fweightGreaterFive, we require that $k$  is odd and $k>3$ (i.e., the generator polynomials have odd weights greater than~5). 

\noindent (d) Figs.\fweightEqualEight \ and\fweightGreaterEight \ show generator polynomials that have even weights and generate CRCs that can detect 1, 2, 3, 4, 5, 6, and 7 errors. We now show $l_2, l_4$ and $l_6$. 
In Fig.\fweightEqualEight, we require $k=6$ (i.e., the generator polynomials have weight exactly~8). In Fig.\fweightGreaterEight, we require that $k$ is even and $k>6$ (i.e., the generator polynomials have even weights greater than~8).
\QED

\figure{
{\eightpoint
\vbox
{
\offinterlineskip 
\halign
{
\strut
\vrule \quad # \hss \vrule    &\quad # \vrule &\quad # \vrule &  \quad# &  \vrule  #  \cr
\noalign{\hrule} 
CRC generator polynomial & 	$l_4$ & $l_2=\hbox{period}$	& $2^{h-1}-1\over{\rm period}$ & \cr
\noalign{\hrule}
$ X^{16} + X^{6} + X^{5} + X^{4} + X^{3} + X^{2} + X + 1 $ &   17 &    30705 &         1.06716 &        \cr
 $ X^{16} + X^{7} + X^{5} + X^{4} + X^{3} + X^{2} + X + 1 $ &   104 &   3066 &  10.6872 &        \cr
 $ X^{16} + X^{9} + X^{7} + X^{6} + X^{5} + X^{4} + X^{3} + 1 $ &       128 &   254 &   129.004 &        \cr
 $ X^{16} + X^{12} + X^{11} + X^{10} + X^{6} + X^{5} + X^{4} + 1 $ &    130 &   258 &   127.004 &        \cr
 $ X^{16} + X^{13} + X^{12} + X^{10} + X^{9} + X^{4} + X + 1 $ &        152 &   151 &   217 &    \cr
\noalign{\hrule}
$ X^{24} + X^{6} + X^{5} + X^{4} + X^{3} + X^{2} + X + 1 $ &   25 &    3145722 &       2.66667 &        \cr
 $ X^{24} + X^{7} + X^{5} + X^{4} + X^{3} + X^{2} + X + 1 $ &   231 &   2796202 &       3 &      \cr
 $ X^{24} + X^{7} + X^{6} + X^{4} + X^{3} + X^{2} + X + 1 $ &   243 &   32385 &         259.028 &        \cr
 $ X^{24} + X^{7} + X^{6} + X^{5} + X^{3} + X^{2} + X + 1 $ &   388 &   3276 &  2560.62 &        \cr
 $ X^{24} + X^{7} + X^{6} + X^{5} + X^{4} + X^{3} + X + 1 $ &   453 &   1040130 &       8.06496 &        \cr
 $ X^{24} + X^{8} + X^{7} + X^{5} + X^{4} + X^{2} + X + 1 $ &   499 &   8388607 &       1 &      \cr
 $ X^{24} + X^{9} + X^{8} + X^{5} + X^{4} + X^{3} + X + 1 $ &   558 &   2046 &  4100 &   \cr
 $ X^{24} + X^{10} + X^{8} + X^{6} + X^{5} + X^{4} + X^{2} + 1 $ &      615 &   8126433 &       1.03226 &        \cr
 $ X^{24} + X^{10} + X^{9} + X^{7} + X^{5} + X^{4} + X + 1 $ &  673 &   8388604 &       1 &      \cr
 $ X^{24} + X^{11} + X^{9} + X^{8} + X^{6} + X^{5} + X^{4} + X^{3} + X + 1 $ &  831 &   32767 &         256.008 &        \cr
 $ X^{24} + X^{12} + X^{11} + X^{9} + X^{7} + X^{5} + X^{2} + 1 $ &     2048 &  4094 &  2049 &   \cr
 $ X^{24} + X^{17} + X^{16} + X^{14} + X^{10} + X^{8} + X^{7} + 1 $ &   2050 &  4098 &  2047 &   \cr
\noalign{\hrule}
$ X^{32} + X^{6} + X^{5} + X^{4} + X^{3} + X^{2} + X + 1 $ &   33 &    2139094785 &    1.00392 &        \cr
 $ X^{32} + X^{7} + X^{5} + X^{4} + X^{3} + X^{2} + X + 1 $ &   1251 &  38337390 &      56.0154 &        \cr
 $ X^{32} + X^{7} + X^{6} + X^{4} + X^{3} + X^{2} + X + 1 $ &   1442 &  66060162 &      32.508 &         \cr
 $ X^{32} + X^{7} + X^{6} + X^{5} + X^{4} + X^{2} + X + 1 $ &   4017 &  2147483647 &    1 &      \cr
 $ X^{32} + X^{11} + X^{9} + X^{5} + X^{3} + X^{2} + X + 1 $ &  4063 &  2130706305 &    1.00787 &        \cr
 $ X^{32} + X^{11} + X^{10} + X^{9} + X^{7} + X^{6} + X^{3} + X^{2} + X + 1 $ & 4085 &  2147483647 &    1 &      \cr
 $ X^{32} + X^{12} + X^{8} + X^{6} + X^{4} + X^{3} + X + 1 $ &  4241 &  28703892 &      74.8151 &        \cr
 $ X^{32} + X^{12} + X^{10} + X^{9} + X^{7} + X^{6} + X^{3} + 1 $ &     4400 &  1879048185 &    1.14286 &        \cr
 $ X^{32} + X^{12} + X^{11} + X^{9} + X^{8} + X^{7} + X^{6} + X^{5} + X^{3} + 1 $ &     5012 &  114681 &        18725.7 &        \cr
 $ X^{32} + X^{13} + X^{12} + X^{8} + X^{6} + X^{4} + X + 1 $ & 5240 &  102261126 &     21 &     \cr
$ X^{32} + X^{13} + X^{12} + X^{10} + X^{8} + X^{6} + X^{4} + 1 $ &    8222 &  253921 &        8457.29 &        \cr
$ X^{32} + X^{17} + X^{15} + X^{13} + X^{10} + X^{9} + X^{4} + X^{2} + X + 1 $ &       8224 &  253983 &        8455.23 &        \cr
 $ X^{32} + X^{18} + X^{14} + X^{13} + X^{12} + X^{11} + X^{9} + X^{5} + X^{4} + 1 $ &  16384 &         32766 &         65540 &          \cr
 $ X^{32} + X^{18} + X^{16} + X^{12} + X^{11} + X^{10} + X^{8} + X^{7} + X^{5} + X^{2} + X + 1 $ &      32768 &         65534 &         32769 &          \cr
\noalign{\hrule}
 $ X^{64} + X^{7} + X^{6} + X^{5} + X^{3} + X^{2} + X + 1 $ &   $>10^5$\hss &         $1.92\times 10^{12}$ \hss &        $4.80\times 10^{6}$ \hss &    \cr
 $ X^{64} + X^{9} + X^{8} + X^{7} + X^{6} + X^{5} + X^{3} + X^{2} + X + 1 $ &   $>10^5$\hss &         $7.20\times 10^{16}$ \hss&        128.03126 &    \cr
\noalign{\hrule}	
}
}
}
}{\fweightGreaterSix}{CRC generator polynomials of even weights greater than 6.}

\figure{
{\eightpoint
$$
\vbox
{
\offinterlineskip 
\halign
{
\strut
\vrule \quad # \hss \vrule    &\quad # \hss \vrule    &\quad # \vrule &\quad # \vrule &  \quad# &  \vrule  #  \cr
\noalign{\hrule} 
CRC generator polynomial & 	$l_4$ & $l_3$ &$l_2=\hbox{period}$	& $2^h-1\over{\rm period}$ & \cr
\noalign{\hrule}
$ X^{16} + X^{3} + X^{2} + X + 1 $ &   17 &    351 &   57337 &         1.14298 &        \cr
 $ X^{16} + X^{4} + X^{2} + X + 1 $ &   31 &    121 &   16383 &         4.00018 &        \cr
 $ X^{16} + X^{4} + X^{3} + X + 1 $ &   63 &    235 &   59055 &         1.10973 &        \cr
 $ X^{16} + X^{5} + X^{2} + X + 1 $ &   68 &    230 &   57337 &         1.14298 &        \cr
 $ X^{16} + X^{5} + X^{3} + X + 1 $ &   104 &   683 &   21845 &         3 &      \cr
 $ X^{16} + X^{5} + X^{4} + X^{2} + 1 $ &       116 &   121 &   57337 &         1.14298 &        \cr
 $ X^{16} + X^{10} + X^{5} + X^{3} + 1 $ &      126 &   317 &   65535 &         1 &      \cr
 $ X^{16} + X^{11} + X^{9} + X^{3} + 1 $ &      130 &   $\infty$ &      381 &   172.008 &        \cr
 $ X^{16} + X^{13} + X^{8} + X^{3} + 1 $ &      258 &   $\infty$ &      257 &   255 &    \cr
  \noalign{\hrule}
  $ X^{24} + X^{3} + X^{2} + X + 1 $ &   25 &    $\infty$ &      4095 &  4097 &   \cr
 $ X^{24} + X^{4} + X^{2} + X + 1 $ &   47 &    7399 &  5586603 &       3.00312 &        \cr
 $ X^{24} + X^{4} + X^{3} + X + 1 $ &   533 &   5839 &  16777215 &      1 &      \cr
 $ X^{24} + X^{5} + X^{2} + X + 1 $ &   725 &   1778 &  5586603 &       3.00312 &        \cr
 $ X^{24} + X^{6} + X^{2} + X + 1 $ &   841 &   5531 &  4194303 &       4 &      \cr
 $ X^{24} + X^{17} + X^{12} + X^{7} + 1 $ &     2050 &  $\infty$ &      6141 &  2732 &   \cr
  \noalign{\hrule}
  $ X^{32} + X^{3} + X^{2} + X + 1 $ &   33 &    351 &   469762041 &     9.14286 &        \cr
 $ X^{32} + X^{4} + X^{2} + X + 1 $ &   63 &    15873 &         268435455 &     16 &     \cr
 $ X^{32} + X^{4} + X^{3} + X + 1 $ &   2250 &  $>10^6$ &        77302995 &      55.5602 &        \cr
 $ X^{32} + X^{5} + X^{2} + X + 1 $ &   4345 &  45868 &         147436713 &     29.1309 &        \cr
 $ X^{32} + X^{7} + X^{6} + X^{2} + 1 $ &       5281 &  142741 &     4294967295 &    1 &      \cr
 $ X^{32} + X^{21} + X^{16} + X^{11} + 1 $ &    65538 & $\infty$ &         65537 &         65535 &          \cr
 \noalign{\hrule}
$ X^{64} + X^{3} + X^{2} + X + 1 $ &   65 & $> 10^6$ &         $1.01\times10^{18}$ &        18.2879 &        \cr
$ X^{64} + X^{4} + X^{3} + X + 1 $ &   $>  10^5$ & $> 10^6$ &        $2^{64}-1$ &   1 &    \cr
\noalign{\hrule}	
}
}
$$
}
}{\fweightEqualFive}{CRC generator polynomials of weight 5.}

\figure{
{\eightpoint
$$
\vbox
{
\offinterlineskip 
\halign
{
\strut
\vrule \quad # \hss \vrule    &\quad # \hss \vrule    &\quad # \vrule &\quad # \vrule &  \quad# &  \vrule  #  \cr
\noalign{\hrule} 
CRC generator polynomial & 	$l_4$ & $l_3$ &$l_2=\hbox{period}$	& $2^h-1\over{\rm period}$ & \cr
\noalign{\hrule}
$ X^{16} + X^{5} + X^{4} + X^{3} + X^{2} + X + 1 $ &   17 &    360 &   65535 &         1 &      \cr
 $ X^{16} + X^{6} + X^{4} + X^{3} + X^{2} + X + 1 $ &   95 & $\infty$ &         5115 &  12.8123 &        \cr
 $ X^{16} + X^{6} + X^{5} + X^{4} + X^{2} + X + 1 $ &   96 &    182 &   63457 &         1.03275 &        \cr
 $ X^{16} + X^{7} + X^{5} + X^{4} + X^{3} + X^{2} + 1 $ &       97 &    353 &   65535 &         1 &      \cr
 $ X^{16} + X^{8} + X^{7} + X^{3} + X^{2} + X + 1 $ &   103 &   243 &   21845 &         3 &      \cr
 $ X^{16} + X^{8} + X^{7} + X^{6} + X^{4} + X^{2} + 1 $ &       120 &   336 &   4369 &  15 &     \cr
 $ X^{16} + X^{9} + X^{7} + X^{4} + X^{3} + X^{2} + 1 $ &       130 & $\infty$ &        381 &   172.008 &        \cr
 $ X^{16} + X^{11} + X^{10} + X^{9} + X^{2} + X + 1 $ & 256 & $\infty$ &        255 &   257 &    \cr
 $ X^{16} + X^{12} + X^{11} + X^{8} + X^{5} + X^{4} + 1 $ &     258 & $\infty$ &        257 &   255 &    \cr
\noalign{\hrule}
$ X^{24} + X^{5} + X^{4} + X^{3} + X^{2} + X + 1 $ &   25 &    3275 &  25575 &         656.001 &        \cr
 $ X^{24} + X^{6} + X^{4} + X^{3} + X^{2} + X + 1 $ &   604 &   4317 &  5332341 &       3.14631 &        \cr
 $ X^{24} + X^{6} + X^{5} + X^{4} + X^{3} + X^{2} + 1 $ &       746 &   2254 &  29127 &         576.002 &        \cr
 $ X^{24} + X^{9} + X^{8} + X^{6} + X^{5} + X + 1 $ &   788 &   8703 &  16777215 &      1 &      \cr
 $ X^{24} + X^{10} + X^{9} + X^{6} + X^{5} + X^{3} + X^{2} + X + 1 $ &  790 &   5687 &  5586603 &       3.00312 &        \cr
 $ X^{24} + X^{11} + X^{9} + X^{7} + X^{6} + X^{5} + X^{4} + X^{3} + 1 $ &      901 &   10751 &         5592405 &       3 &      \cr
 $ X^{24} + X^{12} + X^{8} + X^{7} + X^{5} + X^{4} + X^{3} + X^{2} + 1 $ &      919 &   6297 &  13762455 &      1.21906 &        \cr
 $ X^{24} + X^{13} + X^{11} + X^{10} + X^{9} + X^{7} + X^{6} + X^{4} + X^{3} + X + 1 $ &        2050 & $\infty$ &       6141 &  2732 &   \cr
 $ X^{24} + X^{14} + X^{11} + X^{4} + X^{2} + X + 1 $ & 4096 & $\infty$ &       4095 &  4097 &   \cr
 $ X^{24} + X^{14} + X^{13} + X^{12} + X^{11} + X^{10} + 1 $ &  4098 & $\infty$ &       4097 &  4095 &   \cr
 \noalign{\hrule}
$ X^{32} + X^{5} + X^{4} + X^{3} + X^{2} + X + 1 $ &   33 &    $> 10^6$ &       44695211 &      96.0946 &        \cr
 $ X^{32} + X^{6} + X^{4} + X^{3} + X^{2} + X + 1 $ &   2295 &  202045 &        94972251 &      45.2234 &        \cr
 $ X^{32} + X^{6} + X^{5} + X^{4} + X^{2} + X + 1 $ &   3103 &  96097 &         4286578177 &    1.00196 &        \cr
 $ X^{32} + X^{7} + X^{5} + X^{4} + X^{2} + X + 1 $ &   3831 &  220463 &        1073741823 &    4 &      \cr
 $ X^{32} + X^{7} + X^{6} + X^{4} + X^{3} + X + 1 $ &   3960 &  92515 &         1073741823 &    4 &      \cr
 $ X^{32} + X^{8} + X^{6} + X^{4} + X^{2} + X + 1 $ &   3972 &  38335 &         3758096377 &    1.14286 &        \cr
 $ X^{32} + X^{8} + X^{6} + X^{5} + X^{4} + X^{3} + 1 $ &       4380 &  32768 &         153391689 &     28 &     \cr
 $ X^{32} + X^{9} + X^{7} + X^{5} + X^{4} + X^{2} + 1 $ &       5345 &  115188 &        4292868097 &    1.00049 &        \cr
 $ X^{32} + X^{11} + X^{7} + X^{5} + X^{4} + X + 1 $ &  5617 &  141304 &        107374182 &     40 &     \cr
 $ X^{32} + X^{12} + X^{8} + X^{5} + X^{4} + X^{3} + 1 $ &      5820 &  27707 &         402653181 &     10.6667 &        \cr
  $ X^{32} + X^{12} + X^{11} + X^{9} + X^{7} + X^{3} + X^{2} + X + 1 $ & 65536 & $\infty$ &        65535 &         65537 &          \cr
\noalign{\hrule}
$ X^{64} + X^{5} + X^{4} + X^{3} + X^{2} + X + 1 $ &   65 & $> 10^6$ &    $1.79\times10^{19}$ &        1.0323 &        \cr
$ X^{64} + X^{6} + X^{4} + X^{3} + X^{2} + X + 1 $ &   $>10^5$ & $> 10^6$ &    $2^{40}-1$ &        $1.68\times10^{7}$ &        \cr
\noalign{\hrule}	
}
}
$$
}
}{\fweightGreaterFive}{CRC generator polynomials of odd weights greater than 5.}

\figure{
{\eightpoint
$$
\vbox
{
\offinterlineskip 
\halign
{
\strut
\vrule \quad # \hss \vrule    &\quad # \hss \vrule    &\quad  # &    \vrule  #  \cr
\noalign{\hrule} 
CRC generator polynomial	&$l_4$	& $l_3$ & \cr
\noalign{\hrule}
$ X^{8} + X^{4} + X^{3} + X^{2} + 1 $ &        14 &    21 &     \cr
 $ X^{9} + X^{4} + X^{3} + X + 1 $ &    15 &    29 &     \cr
 $ X^{10} + X^{4} + X^{3} + X + 1 $ &   23 &    39 &     \cr
 $ X^{11} + X^{4} + X^{2} + X + 1 $ &   21 &    100 &    \cr
 $ X^{12} + X^{6} + X^{4} + X + 1 $ &   28 &    107 &    \cr
 $ X^{13} + X^{4} + X^{3} + X + 1 $ &   39 &    94 &     \cr
 $ X^{14} + X^{5} + X^{3} + X + 1 $ &   47 &    224 &    \cr
 $ X^{15} + X^{4} + X^{2} + X + 1 $ &   29 &    262 &    \cr
 $ X^{16} + X^{5} + X^{3} + X^{2} + 1 $ &       56 &    567 &    \cr
 $ X^{17} + X^{3} + X^{2} + X + 1 $ &   18 &    473 &    \cr
 $ X^{18} + X^{5} + X^{2} + X + 1 $ &   100 &   1347 &   \cr
 $ X^{19} + X^{5} + X^{2} + X + 1 $ &   70 &    420 &    \cr
 $ X^{20} + X^{6} + X^{4} + X + 1 $ &   175 &   3429 &   \cr
 $ X^{21} + X^{5} + X^{2} + X + 1 $ &   85 &    647 &    \cr
 $ X^{22} + X^{5} + X^{4} + X^{3} + 1 $ &       513 &   2939 &   \cr
 $ X^{23} + X^{5} + X^{3} + X + 1 $ &   191 &   5512 &   \cr
 $ X^{24} + X^{4} + X^{3} + X + 1 $ &   533 &   5839 &   \cr
 $ X^{25} + X^{3} + X^{2} + X + 1 $ &   26 &    3590 &   \cr
 $ X^{26} + X^{6} + X^{2} + X + 1 $ &   106 &   2603 &   \cr
 $ X^{27} + X^{5} + X^{2} + X + 1 $ &   689 &   19538 &          \cr
 $ X^{28} + X^{6} + X^{4} + X + 1 $ &   512 &   34033 &          \cr
 $ X^{29} + X^{4} + X^{2} + X + 1 $ &   57 &    13056 &          \cr
 $ X^{30} + X^{6} + X^{4} + X + 1 $ &   1929 &  12033 &          \cr
 $ X^{31} + X^{3} + X^{2} + X + 1 $ &   32 &    341 &    \cr
 $ X^{32} + X^{7} + X^{6} + X^{2} + 1 $ &       5281 &  142741 &         \cr
 $ X^{40} + X^5 + X^4 + X^3 + 1 $ &       24049 &   $>\!\!\!\!10^6$&         \cr
 $ X^{48} + X^8 + X^5 + X^2 + 1 $ &       96704 & $>10^6$  &         \cr
 $ X^{56} + X^7 + X^4 + X^2 + 1 $ &       $>10^5$ & $>10^6$  &         \cr
 $ X^{64} + X^4 + X^3 + X + 1 $ &       $>10^5$ & $>10^6$  &         \cr
\noalign{\hrule}	
}
}
$$
}
}{\fweightEqualFivePrimitive}{Primitive CRC generator polynomials of weight 5.}

\figure{
{\eightpoint
$$
\vbox
{
\offinterlineskip 
\halign
{
\strut
\vrule \quad # \hss \vrule    &\quad # \vrule & \quad # \vrule &\quad # \vrule &  \quad# &  \vrule  #  \cr
\noalign{\hrule} 
CRC generator polynomial & 	$l_6$ & 	$l_4$ & $l_2=\hbox{period}$	& $2^{h-1}-1\over{\rm period}$ & \cr
\noalign{\hrule}
$ X^{16} + X^{6} + X^{5} + X^{4} + X^{3} + X^{2} + X + 1 $ &   18 &    17 &    30705 &         1.06716 &        \cr
 $ X^{16} + X^{8} + X^{6} + X^{4} + X^{3} + X^{2} + X + 1 $ &   18 &    31 &    10922 &         3.00009 &        \cr
 $ X^{16} + X^{8} + X^{6} + X^{5} + X^{3} + X^{2} + X + 1 $ &   28 &    46 &    4095 &  8.00171 &        \cr
 $ X^{16} + X^{9} + X^{6} + X^{5} + X^{4} + X^{3} + X^{2} + 1 $ &       29 &    62 &    8001 &  4.09536 &        \cr
 $ X^{16} + X^{9} + X^{7} + X^{6} + X^{3} + X^{2} + X + 1 $ &   30 &    43 &    840 &   39.0083 &        \cr
 $ X^{16} + X^{12} + X^{11} + X^{10} + X^{9} + X^{4} + X + 1 $ &        33 &    32 &    31 &    1057 &   \cr
\noalign{\hrule}
$ X^{24} + X^{6} + X^{5} + X^{4} + X^{3} + X^{2} + X + 1 $ &   26 &    25 &    3145722 &       2.66667 &        \cr
 $ X^{24} + X^{8} + X^{6} + X^{4} + X^{3} + X^{2} + X + 1 $ &   26 &    228 &   2046 &  4100 &   \cr
 $ X^{24} + X^{8} + X^{6} + X^{5} + X^{3} + X^{2} + X + 1 $ &   54 &    260 &   458745 &        18.286 &         \cr
 $ X^{24} + X^{8} + X^{7} + X^{6} + X^{4} + X^{2} + X + 1 $ &   55 &    287 &   7340018 &       1.14286 &        \cr
 $ X^{24} + X^{8} + X^{7} + X^{6} + X^{5} + X^{3} + X^{2} + 1 $ &       56 &    101 &   8388607 &       1 &      \cr
 $ X^{24} + X^{9} + X^{6} + X^{4} + X^{3} + X^{2} + X + 1 $ &   64 &    188 &   7140 &  1174.87 &        \cr
 $ X^{24} + X^{9} + X^{6} + X^{5} + X^{4} + X^{2} + X + 1 $ &   73 &    195 &   3145722 &       2.66667 &        \cr
 $ X^{24} + X^{9} + X^{7} + X^{6} + X^{3} + X^{2} + X + 1 $ &   75 &    390 &   4194296 &       2 &      \cr
 $ X^{24} + X^{9} + X^{8} + X^{5} + X^{3} + X^{2} + X + 1 $ &   82 &    241 &   8388604 &       1 &      \cr
 $ X^{24} + X^{10} + X^{6} + X^{5} + X^{4} + X^{3} + X + 1 $ &  84 &    308 &   8388607 &       1 &      \cr
 $ X^{24} + X^{10} + X^{9} + X^{6} + X^{4} + X^{2} + X + 1 $ &  91 &    278 &   7340018 &       1.14286 &        \cr
 $ X^{24} + X^{10} + X^{9} + X^{6} + X^{4} + X^{3} + X + 1 $ &  98 &    500 &   8388607 &       1 &      \cr
 $ X^{24} + X^{15} + X^{14} + X^{13} + X^{11} + X^{10} + X^{8} + 1 $ &  100 &   220 &   8388607 &       1 &      \cr
 $ X^{24} + X^{19} + X^{16} + X^{13} + X^{9} + X^{6} + X^{5} + 1 $ &    254 & 129 & 508 &   16513 &          \cr
\noalign{\hrule}
$ X^{32} + X^{6} + X^{5} + X^{4} + X^{3} + X^{2} + X + 1 $ &   34 & 33 &  2139094785 &    1.00392 &        \cr
 $ X^{32} + X^{8} + X^{6} + X^{4} + X^{3} + X^{2} + X + 1 $ &   34 & 682 &  715827882 &     3 &      \cr
 $ X^{32} + X^{8} + X^{6} + X^{5} + X^{3} + X^{2} + X + 1 $ &   84 & 1350 &  14620935 &      146.877 &        \cr
 $ X^{32} + X^{8} + X^{7} + X^{6} + X^{4} + X^{2} + X + 1 $ &   197 & 2013 &  67108862 &      32 &     \cr
 $ X^{32} + X^{9} + X^{6} + X^{5} + X^{4} + X^{2} + X + 1 $ &   199 & 1140 & 40632165 &      52.8518 &        \cr
 $ X^{32} + X^{9} + X^{6} + X^{5} + X^{4} + X^{3} + X^{2} + 1 $ &       228 & 1198 & 64853054 &      33.1131 &        \cr
 $ X^{32} + X^{9} + X^{7} + X^{5} + X^{4} + X^{2} + X + 1 $ &   254 & 2868 & 33486852 &      64.1292 &        \cr
 $ X^{32} + X^{9} + X^{8} + X^{6} + X^{4} + X^{3} + X^{2} + 1 $ &       266 & 1691 & 268435454 &     8 &      \cr
 $ X^{32} + X^{11} + X^{9} + X^{7} + X^{6} + X^{5} + X^{4} + 1 $ &      270 &1943 &  587202490 &     3.65714 &        \cr
 $ X^{32} + X^{12} + X^{11} + X^{9} + X^{8} + X^{2} + X + 1 $ & 286 &2699 &  2080374753 &    1.03226 &        \cr
 $ X^{32} + X^{14} + X^{7} + X^{5} + X^{4} + X^{2} + X + 1 $ &  297 &1497 & 626349395 &     3.42857 &        \cr
 $ X^{32} + X^{14} + X^{13} + X^{9} + X^{5} + X^{4} + X + 1 $ & 300 &2347 &  2752470 &       780.202 &        \cr
 $ X^{32} + X^{16} + X^{15} + X^{10} + X^{6} + X^{2} + X + 1 $ &        301 & 3298 & 2147483644 &    1 &      \cr
 $ X^{32} + X^{20} + X^{18} + X^{15} + X^{12} + X^{6} + X^{5} + 1 $ &   310 &   2365 &  287460210 &     7.47054 &        \cr
 $ X^{32} + X^{21} + X^{19} + X^{18} + X^{17} + X^{12} + X^{6} + 1 $ &  311 &   3232 &  2146433025 &    1.00049 &        \cr
 $ X^{32} + X^{25} + X^{19} + X^{12} + X^{10} + X^{5} + X^{3} + 1 $ &   320 &   2711 &  2130706178 &    1.00787 &        \cr
\noalign{\hrule}	
}
}
$$
}
}{\fweightEqualEight}{CRC generator polynomials of weight 8.}

\figure{
{\eightpoint
$$
\vbox
{
\offinterlineskip 
\halign
{
\strut
\vrule \quad # \hss \vrule    &\quad # \vrule & \quad # \vrule &\quad # \vrule &  \quad# &  \vrule  #  \cr
\noalign{\hrule} 
CRC generator polynomial & 	$l_6$ & 	$l_4$ & $l_2=\hbox{period}$	& $2^{h-1}-1\over{\rm period}$ & \cr
\noalign{\hrule}
$ X^{16} + X^{8} + X^{7} + X^{6} + X^{5} + X^{4} + X^{3} + X^{2} + X + 1 $ &   18 &    17 &    584 &   56.1079 &        \cr
 $ X^{16} + X^{10} + X^{8} + X^{6} + X^{5} + X^{4} + X^{3} + X^{2} + X + 1 $ &  18 &    45 &    32385 &         1.0118 &         \cr
 $ X^{16} + X^{10} + X^{8} + X^{7} + X^{5} + X^{4} + X^{3} + X^{2} + X + 1 $ &  22 &    55 &    16380 &         2.00043 &        \cr
 $ X^{16} + X^{10} + X^{8} + X^{7} + X^{6} + X^{4} + X^{3} + X^{2} + X + 1 $ &  29 &    67 &    4599 &  7.12481 &        \cr
 $ X^{16} + X^{11} + X^{10} + X^{8} + X^{7} + X^{6} + X^{5} + X^{3} + X + 1 $ & 30 &    36 &    63 &    520.111 &        \cr
 $ X^{16} + X^{12} + X^{11} + X^{10} + X^{9} + X^{8} + X^{7} + X^{5} + X^{4} + X^{2} + X + 1 $ &        31 &    67 &    32767 &         1 &      \cr
 $ X^{16} + X^{14} + X^{10} + X^{9} + X^{8} + X^{7} + X^{5} + X^{4} + X^{3} + X^{2} + X + 1 $ & 33 &    32 &    31 &    1057 &   \cr
\noalign{\hrule}
$ X^{24} + X^{8} + X^{7} + X^{6} + X^{5} + X^{4} + X^{3} + X^{2} + X + 1 $ &   26 &    25 &    95480 &         87.8572 &        \cr
 $ X^{24} + X^{10} + X^{8} + X^{6} + X^{5} + X^{4} + X^{3} + X^{2} + X + 1 $ &  26 &    305 &   8388607 &       1 &      \cr
 $ X^{24} + X^{10} + X^{8} + X^{7} + X^{5} + X^{4} + X^{3} + X^{2} + X + 1 $ &  65 &    293 &   4161028 &       2.01599 &        \cr
 $ X^{24} + X^{10} + X^{9} + X^{7} + X^{5} + X^{4} + X^{3} + X^{2} + X + 1 $ &  73 &    318 &   5355 &  1566.5 &         \cr
 $ X^{24} + X^{10} + X^{9} + X^{7} + X^{6} + X^{5} + X^{3} + X^{2} + X + 1 $ &  85 &    254 &   64897 &         129.26 &         \cr
 $ X^{24} + X^{10} + X^{9} + X^{7} + X^{6} + X^{5} + X^{4} + X^{3} + X + 1 $ &  88 &    447 &   8388607 &       1 &      \cr
 $ X^{24} + X^{11} + X^{8} + X^{7} + X^{5} + X^{4} + X^{3} + X^{2} + X + 1 $ &  90 &    356 &   5115 &  1640 &   \cr
 $ X^{24} + X^{12} + X^{10} + X^{9} + X^{7} + X^{6} + X^{5} + X^{3} + X^{2} + 1 $ &     92 &    390 &   6276102 &       1.3366 &         \cr
 $ X^{24} + X^{13} + X^{10} + X^{7} + X^{6} + X^{5} + X^{4} + X^{3} + X + 1 $ & 93 &    228 &   7281799 &       1.152 &          \cr
 $ X^{24} + X^{13} + X^{10} + X^{9} + X^{8} + X^{7} + X^{6} + X^{4} + X^{3} + X^{2} + X + 1 $ & 97 &    223 &   8388607 &       1 &      \cr
 $ X^{24} + X^{15} + X^{13} + X^{12} + X^{11} + X^{9} + X^{8} + X^{6} + X + 1 $ &       129 &   130 &   381 &   22017.3 &        \cr
\noalign{\hrule}
$ X^{32} + X^{8} + X^{7} + X^{6} + X^{5} + X^{4} + X^{3} + X^{2} + X + 1 $ &   34 &    33 &    17538696 &      122.443 &        \cr
 $ X^{32} + X^{10} + X^{8} + X^{6} + X^{5} + X^{4} + X^{3} + X^{2} + X + 1 $ &  34 &    1097 &  1310715 &       1638.41 &        \cr
 $ X^{32} + X^{10} + X^{8} + X^{7} + X^{5} + X^{4} + X^{3} + X^{2} + X + 1 $ &  131 &   2747 &  268435452 &     8 &      \cr
 $ X^{32} + X^{10} + X^{8} + X^{7} + X^{6} + X^{4} + X^{3} + X^{2} + X + 1 $ &  159 &   1503 &  47500635 &      45.2096 &        \cr
 $ X^{32} + X^{10} + X^{8} + X^{7} + X^{6} + X^{5} + X^{4} + X^{2} + X + 1 $ &  237 &   3522 &  1610612733 &    1.33333 &        \cr
 $ X^{32} + X^{11} + X^{10} + X^{7} + X^{6} + X^{5} + X^{3} + X^{2} + X + 1 $ & 245 &   2400 &  2064117919 &    1.04039 &        \cr
 $ X^{32} + X^{11} + X^{10} + X^{9} + X^{8} + X^{7} + X^{5} + X^{4} + X^{2} + 1 $ &     259 &   1974 &  44389548 &      48.3781 &        \cr
 $ X^{32} + X^{12} + X^{11} + X^{10} + X^{7} + X^{6} + X^{5} + X^{4} + X^{3} + 1 $ &    272 &   1815 &  2147205122 &    1.00013 &        \cr
 $ X^{32} + X^{13} + X^{10} + X^{9} + X^{7} + X^{6} + X^{5} + X^{2} + X + 1 $ & 273 &   2621 &  125269879 &     17.1429 &        \cr
 $ X^{32} + X^{13} + X^{12} + X^{11} + X^{10} + X^{8} + X^{5} + X^{2} + X + 1 $ &       281 &   861 &   2097150 &       1024 &   \cr
 $ X^{32} + X^{14} + X^{11} + X^{10} + X^{9} + X^{6} + X^{4} + X^{3} + X^{2} + 1 $ &    286 &   3025 &  178911915 &     12.003 &         \cr
 $ X^{32} + X^{14} + X^{12} + X^{11} + X^{10} + X^{9} + X^{6} + X^{5} + X^{3} + X^{2} + X + 1 $ &       306 &   3249 &  2113929153 &    1.01587 &        \cr
 $ X^{32} + X^{15} + X^{13} + X^{12} + X^{9} + X^{8} + X^{7} + X^{6} + X^{5} + 1 $ &    313 &   1635 &  237198535 &     9.05353 &        \cr
 $ X^{32} + X^{17} + X^{14} + X^{13} + X^{12} + X^{11} + X^{8} + X^{7} + X^{6} + X^{5} + X^{2} + 1 $ &  324 &   2314 &  2113929153 &    1.01587 &        \cr
\noalign{\hrule}	
}
}
$$
}
}{\fweightGreaterEight}{CRC generator polynomials of even weights greater than 8.}

\color{red}
\section{APPENDIX D}{CRC WEIGHT DISTRIBUTIONS} \secD = \pageno

\noindent We now briefly present the computation of the weight distributions of CRCs, which are used for computing the undetected error probability of CRCs over binary symmetric channels (BSCs). A BSC is specified by the requirement $\Pr(0|1)=\Pr(1|0)$, where $\Pr(j|i)$ is the conditional probability that bit $j$ is received when bit $i$ is transmitted. The value $p=\Pr(0|1)=\Pr(1|0)$ is called the transition probability of the BSC. 

Given a code of length $l$, the sequence $(w_0, w_1, \dots, w_l)$ is called the weight distribution of the code, where $w_m$ is the number of codewords of weight $m$.  Note that $w_m=0$ when $m<d_{\rm min}$, where $d_{\rm min}$ is the minimum distance of the code. The determination of the  weight distribution of a code in general is an NP-hard problem~[\KlK]. The undetected error probability $p_u$ of a code over a BSC with transition probability $p$ is given by [\KlK,~\LiC]
$$
p_u = \sum^l_{m=1}w_m p^m(1-p)^{l-m} = \sum^l_{m=d_{\rm min}}w_m p^m(1-p)^{l-m} \eqno(\eExactPu)
$$
In the following we present CRC weight distributions obtained by computer search. Mathematical studies of the weight distributions and the undetected error probability of codes are presented in [\KlK].


Consider a CRC that has length $l$ and is generated by a polynomial $F(X)$, which is not a multiple of $X$. A polynomial $E(X)$ is a codeword of this CRC if $E(X)$ is a multiple of $F(X)$, i.e., $\MOD{E(X)}{F(X)} = 0$. This fact can be used to compute CRC weight distributions. Note that  $w_0=1$. If $F(X)$ has {\it even} weight, then a polynomial of odd weight can not be a codeword, i.e., $w_m = 0$ for odd~$m$.

For $0 < m\le l$, let ${\bf A}_m$ be the set of polynomials of degrees $<l$ and weight $m$, i.e., 

$$
{\bf A}_m = \{X^{a_{m-1}} + X^{a_{m-2}} + \cdots + X^{a_1}+ X^{a_0} \ : \ l>a_{m-1}>a_{m-2}> \cdots > a_1 > a_0 \ge 0\} \eqno(\eAm)
$$
We have  $|{\bf A}_m| = {l \choose m}$. 
Let ${\bf B}_m$ be the set of CRC codewords in ${\bf A}_m$, i.e.,
$$
{\bf B}_m = \{E(X) \in {\bf A}_m \ : \ \MOD{E(X)}{F(X)} = 0 \}	\eqno(\eBm)
$$
We have $w_m = |{\bf B}_m|$. Thus, a direct technique for computing $w_m$ is to count the number of polynomials in ${\bf A}_m$ that are multiples of $F(X)$ (cf. [\Cha]). Because $|{\bf A}_m| = {l \choose m}=\hbox{O}(l^m)$, this direct technique has computational complexity $\hbox{O}\left(|{\bf A}_m|\right)=\hbox{O}(l^m)$ (cf. [\Cha]). A faster technique for computing $w_m$, which has computational complexity $\hbox{O}(l^{m-1})$, is presented in Remark\rIIviii. Using this faster technique, we obtain the values of $w_m$ for $h$-bit CRCs as shown in Figs.\fweightDistCRCeight -\kern -0.77ex\fweightDistCRCsixtyfour \ for $m=4, 6, 8$, and $h=8, 16, 24, 32, 64$. Note that $w_m=0$ for odd $m$ and $w_2=0$, because the generator polynomials in these figures have weight 4 (i.e., even weight) and minimum distance $d=4$ at the indicated code lengths.

Fig.\fweightDistCRCeight \ shows $w_4$, $w_6$, and $w_8$ for the fast 8-bit CRC generated by $X^8+X^2+X+1$ (which is also the ATM CRC-8) and for the 8-bit CRC generated by $X^8+X^5+X^4+1$ (used in 1-Wire bus). The results show that these 2 CRCs have similar $w_4$, $w_6$, and $w_8$.

Fig.\fweightDistCRCsixteen \ shows $w_4$ for the fast 16-bit CRC generated by $X^{16}+X^2+X+1$, the CRC-CCITT generated by $X^{16}+X^{12}+X^5+1$, and the CRC-16 generated by $X^{16}+X^{15}+X^4+1$. The results show that (a) for $l \le 1000$, $w_4$ is smallest for the CRC-CCITT, largest for CRC-16, and in-between for the fast 16-bit CRC, and (b) for $l \ge 2000$, all 3 CRCs have similar $w_4$. 

Fig.\fweightDistCRCthirtytwo \ shows that $w_4$ for the fast 32-bit CRC generated by $X^{32}+X^2+X+1$ is larger than $w_4$ for the 32-bit CRC generated by $X^{32}+X^{31}+X^8+1$ (which is proposed in [\Fel]).

Fig.\fweightDistCRCsixtyfour \ shows that, for $l \ge 200$, $w_4$ for the fast 64-bit CRC generated by $X^{64}+X^2+X+1$ is smaller than $w_4$ for the 64-bit CRC generated by $X^{64}+X^{63}+X^2+1$ (which is proposed in [\Fel]).


We now consider a first-order estimate for the undetected error probability by assuming that the first term of (\eExactPu), $w_{d_{\rm min}} p^{d_{\rm min}}(1-p)^{l-d_{\rm min}}$, is much larger than all the other terms.
A simple estimate, which is reasonable when $lp << 1$, for the undetected error probability for a BSC is then 

$$
p_u \approx w_{d_{\rm min}} p^{d_{\rm min}} \eqno(\eEstimatePu)
$$

For example, suppose that the fast 32-bit CRC generated by $X^{32} + X^2 + X + 1$ is used to protect a 3000-bit codeword over a BSC with transition probability  $p$. Because $d_{\rm min}=4$ and $l=3000$, Fig.\fweightDistCRCthirtytwo \ yields $w_{d_{\rm min}}=w_4=1.855 \times 10^{5}$. Using (\eEstimatePu), we then have $p_u \approx 1.855 \times 10^{5}p^4$. In particular, $p_u \approx 1.855 \times10^{-19}$ when $p=10^{-6}$, and  $p_u \approx 1.855 \times10^{-15}$ when $p=10^{-5}$. The undetected error probability $p_u$ for the BSC can be greatly further reduced by using CRCs with $d_{\rm min}>4$. These CRCs are presented in Section~C.4, e.g., Fig.\fweightEqualSix \ shows many generator polynomials for 32-bit CRCs that have $d_{\rm min}=6$ when $l \le l_4$. Although these CRCs can be efficiently implemented by the fast technique (\eiiitwelve), they are not as fast as the fast CRC generated by $X^{32} + X^2 + X + 1$. Note that the undetected error probability $p_u$ given in (\eExactPu) and (\eEstimatePu) are for BSCs, and may not be valid for other types of channels.

\figure{
{\eightpoint
\vbox
{
\offinterlineskip 
\halign
{
\strut
\vrule \quad#  \vrule   &\quad # \vrule &  \quad# \vrule & \quad# \vrule \vrule &\quad# \vrule &\quad# \vrule &  \quad# &\vrule  #  \cr
\noalign{\hrule} 
& \multispan2 \ \ $X^8 + X^2 + X +1$ \hss  &  & \multispan2 \ \ $X^8 + X^5 + X^4 +1$ \hss & &\cr
\noalign{\hrule}
$l$ &   $w_4 $ &      $w_6 $ &            $w_8 $ &      $w_4 $ &      $w_6$ &      $w_8$ &   \cr
\noalign{\hrule} 
10 &    3.000e+00 &     0.000e+00 &     0.000e+00 &     2.000e+00 &     1.000e+00 &     0.000e+00 &      \cr
20 &    3.900e+01 &     2.870e+02 &     1.029e+03 &     4.300e+01 &     2.820e+02 &     1.011e+03 &      \cr
50 &    1.833e+03 &     1.241e+05 &     4.195e+06 &     1.813e+03 &     1.244e+05 &     4.192e+06 &      \cr
100 &   3.136e+04 &     9.304e+06 &     1.454e+09 &     3.135e+04 &     9.305e+06 &     1.454e+09 &      \cr
127 &   8.268e+04 &     4.035e+07 &     1.047e+10 &     8.268e+04 &     4.035e+07 &     1.047e+10 &      \cr
\noalign{\hrule} 	
}
}
}
}{\fweightDistCRCeight}{$w_4, w_6$, and $w_8$ for the CRCs generated by  $X^8 + X^2 + X +1$  and $X^8 + X^5 + X^4 +1$ \hbox{($w_m$ = number of codewords of weight $m$).}}

\figure{
{\eightpoint
\vbox
{
\offinterlineskip 
\halign
{
\strut
\vrule \quad #  \vrule    &\quad # \vrule &\quad # \vrule &  \quad# &  \vrule  #  \cr
\noalign{\hrule} 
  &  $X^{16}+X^2+X+1$ &      $X^{16}+X^{12}+X^5+1$ &      $X^{16}+X^{15}+X^2+1$ &      \cr
 \noalign{\hrule}
 $l$ &  $w_4$ &      $w_4 $ &      $w_4 $ &      \cr
\noalign{\hrule}
100 &   6.790e+02 &     2.870e+02 &     1.289e+03 &      \cr
200 &   3.836e+03 &     2.409e+03 &     7.523e+03 &      \cr
300 &   1.345e+04 &     9.478e+03 &     2.826e+04 &      \cr
400 &   3.839e+04 &     2.978e+04 &     6.960e+04 &      \cr
500 &   8.656e+04 &     7.587e+04 &     1.326e+05 &      \cr
600 &   1.774e+05 &     1.606e+05 &     2.568e+05 &      \cr
700 &   3.303e+05 &     3.007e+05 &     4.394e+05 &      \cr
800 &   5.630e+05 &     5.177e+05 &     6.927e+05 &      \cr
900 &   8.913e+05 &     8.344e+05 &     1.019e+06 &      \cr
1000 &  1.343e+06 &     1.276e+06 &     1.473e+06 &      \cr
2000 &  2.072e+07 &     2.050e+07 &     2.085e+07 &      \cr
3000 &  1.032e+08 &     1.030e+08 &     1.036e+08 &      \cr
4000 &  3.254e+08 &     3.252e+08 &     3.256e+08 &      \cr
5000 &  7.940e+08 &     7.938e+08 &     7.943e+08 &      \cr
6000 &  1.646e+09 &     1.646e+09 &     1.647e+09 &      \cr
7000 &  3.050e+09 &     3.050e+09 &     3.050e+09 &      \cr
8000 &  5.204e+09 &     5.204e+09 &     5.204e+09 &      \cr
9000 &  8.336e+09 &     8.337e+09 &     8.336e+09 &      \cr
10000 & 1.271e+10 &     1.271e+10 &     1.271e+10 &      \cr
11000 & 1.861e+10 &     1.861e+10 &     1.861e+10 &      \cr
12000 & 2.635e+10 &     2.635e+10 &     2.635e+10 &      \cr
13000 & 3.630e+10 &     3.630e+10 &     3.630e+10 &      \cr
14000 & 4.883e+10 &     4.883e+10 &     4.883e+10 &      \cr
15000 & 6.435e+10 &     6.435e+10 &     6.435e+10 &      \cr 
20000 & 2.034e+11 &     2.034e+11 &     2.034e+11 &      \cr
\noalign{\hrule} 	
}
}
}
}{\fweightDistCRCsixteen}{$w_4$ for the CRCs generated by  $X^{16}+X^2+X+1$, $X^{16}+X^{12}+X^5+1$, and $X^{16}+X^{15}+X^2+1$.}

\figure{
{\eightpoint
\vbox
{
\offinterlineskip 
\halign
{
\strut
\vrule \quad #  \vrule    &\quad # \vrule  &  \quad# &  \vrule#  \cr
\noalign{\hrule} 
 &  $X^{32} + X^2 + X + 1$ &      $X^{32} + X^{31} + X^8 + 1$ &      \cr
\noalign{\hrule}
 $l$ &  $w_4$ &      $w_4 $ &      \cr
\noalign{\hrule}
100 &   2.820e+02 &     1.040e+02 &      \cr
200 &   1.276e+03 &     4.560e+02 &      \cr
300 &   2.648e+03 &     1.016e+03 &      \cr
400 &   4.264e+03 &     1.816e+03 &      \cr
500 &   5.964e+03 &     2.636e+03 &      \cr
600 &   8.245e+03 &     3.736e+03 &      \cr
700 &   1.064e+04 &     4.936e+03 &      \cr
800 &   1.308e+04 &     6.136e+03 &      \cr
900 &   1.558e+04 &     7.336e+03 &      \cr
1000 &  1.809e+04 &     8.642e+03 &      \cr
2000 &  7.634e+04 &     2.946e+04 &      \cr
3000 &  1.885e+05 &     6.079e+04 &      \cr
4000 &  3.136e+05 &     9.760e+04 &      \cr
5000 &  5.024e+05 &     1.458e+05 &      \cr
6000 &  7.117e+05 &     2.007e+05 &      \cr
7000 &  9.396e+05 &     2.632e+05 &      \cr
8000 &  1.187e+06 &     3.359e+05 &      \cr
9000 &  1.567e+06 &     4.327e+05 &      \cr
10000 & 2.035e+06 &     5.555e+05 &      \cr
11000 & 2.544e+06 &     7.022e+05 &      \cr
12000 & 3.086e+06 &     8.781e+05 &      \cr
13000 & 3.668e+06 &     1.084e+06 &      \cr
14000 & 4.291e+06 &     1.332e+06 &      \cr
15000 & 4.976e+06 &     1.625e+06 &      \cr
20000 & 1.017e+07 &     4.026e+06 &      \cr
\noalign{\hrule} 	
}
}
}
}{\fweightDistCRCthirtytwo}{$w_4$ for the CRCs generated by $X^{32} + X^2 + X + 1$ and $X^{32} + X^{31} + X^8 + 1$.} %

\figure{
{\eightpoint
\vbox
{
\offinterlineskip 
\halign
{
\strut
\vrule \quad #  \vrule    &\quad # \vrule  &  \quad# &  \vrule#  \cr
\noalign{\hrule} 
 &  $X^{64} + X^2 + X + 1$ &      $X^{64} + X^{63} + X^2 + 1$ &      \cr
\noalign{\hrule}
 $l$ &  $w_4$ &      $w_4 $ &      \cr
\noalign{\hrule}
100 &   7.100e+01 &     3.600e+01 &      \cr
200 &   5.660e+02 &     5.720e+02 &      \cr
300 &   1.440e+03 &     2.200e+03 &      \cr
400 &   2.556e+03 &     4.781e+03 &      \cr
500 &   3.756e+03 &     7.939e+03 &      \cr
600 &   5.304e+03 &     1.252e+04 &      \cr
700 &   6.904e+03 &     1.771e+04 &      \cr
800 &   8.536e+03 &     2.328e+04 &      \cr
900 &   1.024e+04 &     2.920e+04 &      \cr
1000 &  1.194e+04 &     3.525e+04 &      \cr
2000 &  3.330e+04 &     1.266e+05 &      \cr
3000 &  6.188e+04 &     2.659e+05 &      \cr
4000 &  9.181e+04 &     4.253e+05 &      \cr
5000 &  1.509e+05 &     8.531e+05 &      \cr
6000 &  2.139e+05 &     1.346e+06 &      \cr
7000 &  2.786e+05 &     1.870e+06 &      \cr
8000 &  3.436e+05 &     2.408e+06 &      \cr
9000 &  4.449e+05 &     3.385e+06 &      \cr
10000 & 5.569e+05 &     4.468e+06 &      \cr
11000 & 6.689e+05 &     5.573e+06 &      \cr
12000 & 7.809e+05 &     6.685e+06 &      \cr
13000 & 8.973e+05 &     8.019e+06 &      \cr
14000 & 1.015e+06 &     9.381e+06 &      \cr
15000 & 1.133e+06 &     1.075e+07 &      \cr
20000 & 1.901e+06 &     2.034e+07 &      \cr
\noalign{\hrule} 	
}
}
}
}{\fweightDistCRCsixtyfour}{$w_4$ for the CRCs generated by $X^{64} + X^2 + X + 1$ and $X^{64} + X^{63} + X^2 + 1$.}

\remark{Remark\rIIviii.} Recall from (\eAm) that ${\bf A}_m$ is the set of polynomials of degrees $<l$ and weight $m$. Let ${\bf A}^*_m$ be the subset of polynomials in ${\bf A}_m$ whose lowest-order terms are 1, i.e., $X^{a_0}= 1$ or $a_0=0$. Thus, each member of ${\bf A}^*_m$ has the form
$$
E^*(X) = X^{a_{m-1}} + X^{a_{m-2}} + \cdots + X^{a_1}+ 1 
$$
where $l>a_{m-1}>a_{m-2}> \cdots > a_1 \ge 1$. We have $E^*(0)=1$ because $X^{a_0}= 1$. Thus,
$$
{\bf A}^*_m = \{E^*(X) \in {\bf A}_m \ : \ E^*(0)=1 \}
$$ 
We have $|{\bf A}^*_m| = {l-1 \choose m-1}$.
Let ${\bf B}^*_m$ be the set of CRC codewords in ${\bf A}^*_m$, i.e.,
$$
{\bf B}^*_m = \{E^*(X) \in {\bf A}^*_m \ : \ \MOD{E^*(X)}{F(X)} = 0 \}
$$
It can be shown that
${\bf B}^*_m = \{E^*(X) \in {\bf B}_m \ : \ E^*(0)=1 \}$, where ${\bf B}_m$ as defined in (\eBm) is the set of CRC codewords in ${\bf A}_m$.
We then have ${\bf B}^*_m \subset {\bf A}^*_m \subset {\bf A}_m$ and ${\bf B}^*_m \subset {\bf B}_m \subset {\bf A}_m$.

For each $E^*(X) \in {\bf B}^*_m$, define
$$
{\bf C}_{E^*(X)} = \{X^aE^*(X) \ : \ a=0,1,\dots, l-\hbox{degree}(E^*(X))-1  \}
$$
We have $|{\bf C}_{E^*(X)}| = l-\hbox{degree}(E^*(X))$.

Let $E(X) \in {\bf B}_m$.  We then have $E(X)=X^aE^*(X)$, for some $a \ge 0$ and some $E^*(X) \in {\bf B}^*_m$, because the generator polynomial $F(X)$ is not a multiple of $X$, i.e., $F(0)=1$. We have $0 \le a \le l-\hbox{degree}(E^*(X)) - 1$, because $l-1 \ge \hbox{degree}(E(X)) = a+\hbox{degree}(E^*(X))$. We then have
$$
{\bf B}_m \subseteq \bigcup_{E^*(X) \in {\bf B}^*_m} C_{E^*(X)}
$$
Because $C_{E^*(X)} \subseteq {\bf B}_m$ for each $E^*(X) \in {\bf B}^*_m$, we have 
$$
\bigcup_{E^*(X) \in {\bf B}^*_m} C_{E^*(X)} \subseteq {\bf B}_m
$$
Thus,
$$
{\bf B}_m = \bigcup_{E^*(X) \in {\bf B}^*_m} C_{E^*(X)}
$$
Because $E^*(X)$ is not a multiple of $X$, it can be shown that $C_{E^*(X)} \cap C_{E^*(X)'} = \emptyset$ when $E^*(X) \not= E^*(X)'$. That is, $\{C_{E^*(X)}\ : \ E^*(X) \in {\bf B}^*_m \}$ is a partition of ${\bf B}_m$. Thus, 
$$
|{\bf B}_m| = \sum_{E^*(X) \in {\bf B}^*_m} |C_{E^*(X)}| 
$$ 
Because $w_m = |{\bf B}_m|$ and $|C_{E^*(X)}| = l-\hbox{degree}(E^*(X))$, we have 
$$
w_m =  \sum_{E^*(X) \in {\bf B}^*_m}\left[ l-\hbox{degree}(E^*(X))\right]
$$ 

Thus, $w_m$ is computed by adding the numbers $\left[l-\hbox{degree}(E^*(X))\right]$ for all polynomials $E^*(X) \in {\bf B}^*_m$. Because the polynomials $E^*(X)$ are those of ${\bf A}^*_m$ that are multiples of $F(X)$, they can be found in O$(|{\bf A}^*_m|)$ steps.
Finally, $w_m$ can be computed in O$(l^{m-1})$ steps, because $|{\bf A}^*_m|={l-1 \choose m-1}=\hbox{O}(l^{m-1})$.
\QED

\color{black}
\section{APPENDIX E}{CRC PARALLEL IMPLEMENTATION} \secE = \pageno
Given a CRC, which is generated by a polynomial $M(X)$ of degree $h$, our goal is to compute the check $h$-tuple $P(X)$ to protect an input message $U(X) = (Q_0(X), \dots, Q_{n-1}(X))$, where $Q_i(X)$ is an $s$-tuple.

So far, it is implicitly assumed that the CRC algorithms are for sequential implementation. That is, the entire input message $U(X)$ is supplied to a single processor of a computer, and the output $P(X)$ is then computed by this same processor. 
Following the technique in [\JiK], we can modify these CRC algorithms for parallel implementation on~$k$ different processors of a computer, $k>1$, as follows. 

First, the input message $U(X)$ is divided into $k$ sub-messages $E_{0}(X), \dots, E_{k-1}(X)$, i.e.,
$$
U(X) = (E_{0}(X), \dots, E_{k-1}(X))
$$ 
where $E_i(X)$ consists of $n_i$ $s$-tuples. Thus,  $n=n_{0}+\cdots+n_{k-1}$. Define
$$
W_i(X) = \MOD{X^{(n_{i+1}+\cdots+n_{k-1})s}}{M(X)} \eqno(\evithirtyone)
$$
for $0 \le i \le k-2$, and $W_{k-1}(X)=1$. Note that $W_i(X)$ is computed from $X^{(n_{i+1}+\cdots+n_{k-1})s}$, which is used to determine the relative position of sub-message $E_i(X)$ in $U(X)$ (see Remark\rsevenA).

Next, for each $i=0,1,\dots, k-1$,  input sub-message $E_i(X)$ is supplied to processor $i$, which is used to compute the following $h$-tuples:
$$
P_i(X) = \MOD{E_i(X)X^h}{M(X)} \eqno(\evithirtytwo)
$$

$$
Z_i(X) = \MOD{P_i(X)W_i(X)}{M(X)} \eqno(\evithirtysix)
$$
where $W_i(X)$ is defined by (\evithirtyone). Note that $P_i(X)$ is the CRC check tuple computed by processor $i$ for sub-message $E_i(X)$. For each $i=0,1,\dots, k-1$, we assume that processor $i$ computes $P_i(X)$ and $Z_i(X)$  in~(\evithirtytwo) and (\evithirtysix), independent of other processors, i.e., the computation is done in parallel by the $k$ processors.

\remark{Theorem 5.} The tuples $Z_i(X), 0 \le i < k$, which are computed in parallel by the $k$ processors, are combined to yield the final CRC check $h$-tuple $P(X)$ for the entire input message $U(X)$, i.e., 
$$
P(X) = \sum_{i=0}^{k-1}Z_i(X) \eqno(\evithirtyeight) 
$$

\remark{Proof.} In polynomial notation, we have
$$
U(X) = \sum_{i=0}^{k-2}{E_i(X) X^{(n_{i+1}+\cdots+n_{k-1})s}} + E_{k-1}(X)
$$ 
The CRC check tuple $P(X)$ for $U(X)$ then becomes
$$
\eqalign{P(X) &= \MOD{U(X)X^h}{M(X)}	\cr
	&= \sum_{i=0}^{k-2}\MOD{E_i(X)X^h X^{(n_{i+1}+\cdots+n_{k-1})s}}{M(X)} + \MOD{E_{k-1}(X)X^h}{M(X)} 	\cr
	&= \sum_{i=0}^{k-2}\MOD{P_i(X)W_i(X)}{M(X)} + P_{k-1}(X)  \cr
	&= \sum_{i=0}^{k-1}\MOD{P_i(X)W_i(X)}{M(X)} \cr
	&= \sum_{i=0}^{k-1}Z_i(X)}
$$\QED  \goodbreak

We now determine the total CRC computation time, denoted by $t_{\rm total}$, for the parallel technique. First, let $t_{W_i}$, $t_{P_i}$, and $t_{Z_i}$ be the times for processor $i$ to compute $W_i(X)$, $P_i(X)$, and $Z_i(X)$, respectively. Let $t_P$ be the time for the computer to compute the summation (\evithirtyeight). We can consider $t_{W_i}$, $t_{Z_i}$, and $t_P$ as the overhead costs for the CRC parallel implementation. Because the $k$ processors  compute (\evithirtytwo) and (\evithirtysix) in parallel, the total time for the computer to compute the final CRC check tuple $P(X)$ is
$$
t_{\rm total} = t_{W_i} + \max \{t_{P_i}  + t_{Z_i}, 0 \le i < k \} + t_P \eqno(\evifourty)
$$ 

We now determine the speedup factor for the parallel technique under the following ideal conditions: (a) the $k$ processors have identical computational capability, (b) the sub-messages $E_i(X)$ have the same length, i.e., $n_i = n/k$, and (c) the overhead costs $t_{W_i}$, $t_{Z_i}$, and $t_P$ are negligible compared to $t_{P_i}$, i.e., $t_{W_i} + t_{P_i}  + t_{Z_i} + t_{P} \approx t_{P_i}$ (see Remark\rsevenA). From (\evifourty), we then have $t_{\rm total} \approx t_{P_i} \approx t_U/k$, where $t_U$ denotes the time for a single processor to compute the CRC check tuple $P(X)$ for the entire message $U(X)$, i.e., $t_U$ is the CRC computational time for sequential implementation. Thus, under the ideal conditions, the speedup factor is approximately $k$ for parallel implementation.

\remark{Remark\rsevenA.} Under the CRC parallel implementation, processor $i$ computes $W_i(X)$, $P_i(X)$, and $Z_i(X)$ as given in (\evithirtyone)--(\evithirtysix), $i=1, \dots, k-1$. These tuples can be computed as follows.
First, it can be shown from~(\evithirtyone) that 
$$
W_i(X) = \MOD{X^{n_{i+1}s} W_{i+1}(X)}{M(X)}
$$
with $W_{k-1}=1$. Thus, once $W_{i+1}(X)$ is known, $W_{i}(X)$ can be computed in O$(n_{i+1}s)$ steps (by Remark\rtwo). We can also write $W_i(X) = \MOD{X^{n_{i+1}s-h} W_{i+1}(X)X^h}{M(X)}$, i.e., we can view $W_i(X)$ as the output check tuple of the CRC generated by $M(X)$ when $X^{n_{i+1}s-h} W_{i+1}(X)$ is the input tuple. Thus, $W_i(X)$ can be computed by either the CRC basic technique or the CRC new technique. Suppose now that $n_0,\dots, n_{k-1}$ are known and fixed. The tuples $W_0(X), W_1(X), \dots, W_{k-1}$ can then be stored in a table defined by $T[i]=W_i(X)$, $i=0, 1, \dots, k-1$ (cf.~[\JiK]). Next, processor~$i$ can use either the basic technique or the new technique to compute the (partial) CRC check tuple $P_i(X)$. 
Further, using the technique ``Mimic long multiplication as done by hand" in [\MaS, p.~90], it can be shown that the tuple $Z_i(X)=\MOD{P_i(X)W_i(X)}{M(X)}$ can be computed in O($h$) steps. 
Finally, once $Z_0(X), \dots, Z_{k-1}(X)$ are computed by the $k$ processors, their summation in (\evithirtyeight) can be quickly computed.
Thus, for a sufficiently long sub-message $E_i(X)$ along with the use of table lookup for determining $W_i(X)$, the computational complexity of $P_i(X)$ is much greater than that of $W_i(X)$, $Z_i(X)$, and the summation (\evithirtyeight), i.e., $t_{P_i} >> t_{W_i}, t_{Z_i}, t_P$.\QED

\section{ACKNOWLEDGMENT}
This work was supported in part by the US Office of Naval Research.

\section {REFERENCES}
 
\item{[\Ber]} E.R.~Berlekamp, {\it Algebraic Coding Theory}, McGraw-Hill, 1968.

\item{[\BiR]} A.~Binstock and J.~Rex, {\it Practical Algorithms for Programmers}, Addison-Wesley, 1995.

\item{[\Cha]} T.~Chakravarty, ``Performance of Cyclic Redundancy Codes for Embedded Networks," MS Thesis, Department  of  Electrical  and  Computer  Engineering, Carnegie  Mellon  University, Dec.~2001.  

\item{[\Far]} P.~Farkas, ``Comments on `Weighted Sum Codes for Error Detection and Their Comparison with Existing Codes'," {\it IEEE/ACM Trans.\ Networking}, vol.~3, no.~2, pp.~222-223, April 1995. 

\item{[\Fel]} D.C.~Feldmeier, ``Fast Software Implementation of Error Detection Codes," {\it IEEE/ACM Trans.\ Networking}, vol.~3, no.~6, pp.~ 640-651, Dec.\ 1995. 

\item{[\Fle]} J.G.~Fletcher, ``An Arithmetic Checksum for Serial Transmissions," {\it IEEE Transactions on Communications}, vol.~30, no.~1, pp.~247-252, Jan.~1982.

\item{[\JiK]} H.M.~Ji and E.~Killian, ``Fast Parallel CRC Algorithm and Implementation on a Configurable Processor," {\it Proc.~IEEE International Conference on Communications (ICC'02)},  pp.\ 1813-1817, April/May 2002.

\item{[\KlK]} T.~Klove and V.~Korzhik, {\it Error Detecting Codes: General Theory and their Application in Feedback Communication Systems}, Kluwer Academic, 1995. 

\item{[\Knu]} D.E.~Knuth, {\it The Art of Computer Programming}, vol.~2, 3rd Ed., Addison-Wesley, 1998.

\item{[\Koo]} P.~Koopman, ``32-Bit Cyclic Redundancy Codes for Internet Applications," {\it Proc.~International Conference on Dependable Systems and Networks (DSN'02)}, pp.~459-472, June 2002.

\item{[\KoB]} M.E.\ Kounavis and F.L.~Berry, ``A Systematic Approach to Building High Performance Software-Based CRC Generators," {\it Proc.~10th IEEE Symposium on Computers and Communications (ISCC'05)}, pp.~855-862, June 2005.

\item{[\LiC]} S.~Lin and D.J.~Costello, Jr., {\it Error Control Coding: Fundamentals and Applications}, Prentice Hall, 1983.

\item{[\MaS]} F.J.\ MacWilliams and N.J.A.~Sloan, {\it The Theory of Error-Correcting Codes}, North-Holland, 1977.

\item{[\McA]} A.J.~McAuley, ``Weighted Sum Codes for Error Detection and Their Comparison with Existing Codes," {\it IEEE/ACM Trans.~Networking}, vol.~2, no.~1, pp.~16-22, Feb.~1994. 

\item {[\Ngu]} G.D.~Nguyen, ``Error-Detection Codes: Algorithms and Fast Implementation," {\it IEEE Trans.\ Computers}, vol.~54,  no.~1,  pp.~1-11,  Jan.~2005.

\item {[\NguA]} G.D.~Nguyen, ``Fast CRCs," {\it IEEE Transactions on Computers}, vol.~58, no.~10, pp.~1321-1331, Oct.~2009.

\item{[\Per]} A.\ Perez, ``Byte-wise CRC Calculations,"{\it IEEE Micro}, pp.~40-50, June 1983.

\item{[\RaG]} T.V.~Ramabadran and S.S.~Gaitonde, ``A Tutorial on CRC Computations," {\it IEEE Micro}, vol.~8, pp.~62-75, Aug.~1988.

\item{[\RaK]}J.~Ray and P.~Koopman, ``Efficient High Hamming Distance CRCs for Embedded Applications," {\it Proc.~International Conference on Dependable Systems and Networks (DSN'06)}, pp.~3-12, June 2006.

\item{[\Sar]} D.V.~Sarwate, ``Computation of Cyclic Redundancy Checks via Table-Lookup," {\it Comm.\ ACM}, vol.~31, no.~8, pp.~1008-1013, Aug.~1988.

\item{[\SGP]} J.~Stone, M.~Greenwald, C.~ Partridge, and J.~Hughes, ``Performance of Checksums and CRC's over Real Data," {\it IEEE/ACM Trans.~Networking}, vol.~6, no.~5, pp.~529-543, Oct.~1998.

\item{[\Zie]} N.~Zierler, ``On $x^n + x + 1$ over GF(2)," {\it Information and Control}, vol.~16, no.~5, pp.~502-505, 1970.

\ref = \pageno %

\vfill \eject
{\titlefont \noindent Contents}\nobreak
\bigskip \bigskip

\vbox
{\ninepoint
\offinterlineskip 
\halign
{
\strut
# \hss &\quad # \hss  \cr
{\bf 1} 	& {\bf Introduction \dotfill \the\secI} \cr
1.1 & Notation and Convention \cr
&		\cr
{\bf 2} 	& {\bf CRC Algorithms\dotfill\the\secII} \cr
2.1 & General CRC Theory \cr
2.2 & Two CRC Algorithms \cr
2.3 & Two Alternative CRC Algorithms \cr
2.4 & Basic CRC Algorithms \cr
&		\cr
{\bf 3} 	& {\bf Fast CRCs \dotfill \the\secIII} \cr
3.1 & Fast $h$-Bit CRCs \cr
3.2 & A Fast 16-Bit CRC \cr
3.3 & Error-Detection Capability of Fast CRCs \cr
&		\cr
{\bf 4} 	& {\bf CRC Software Complexity \dotfill \the\secIV} \cr
4.1 & General Complexity Analysis \cr
4.2 & CRC Complexity Under C Implementation \cr
4.3 & Other Techniques for Error-Detection Codes \cr 
&		\cr
{\bf 5} 	& {\bf Summary and Extension \dotfill \the\secV} \cr
&		\cr
{\bf A} 	& {\bf CRC Software Implementation and } \cr
& {\bf Complexity Evaluation \dotfill \the\secA} \cr
A.1 & CRC Software Implementation: Bitwise Technique (Without Table Lookup) \cr
A.1.1 & Basic CRCs  \cr 
A.1.2 & Fast CRCs \cr
A.2 & CRC Software Implementation: Table-Lookup Technique \cr
A.2.1 & Basic CRCs \cr
A.2.2 & Fast CRCs \cr
&		\cr
{\bf B} 	& {\bf Other Fast Error-Detection Codes \dotfill\the\secB} \cr
B.1 & Fast CRCs Generated by Binomials \cr
B.2 & Fast CRCs Generated by Trimomials \cr
B.3 & Fast and Optimal Error-Detection Codes \cr
&		\cr
{\bf C} 	& {\bf Application of the New Technique to} \cr
&  {\bf General CRC Generator Polynomials \dotfill\the\secC} \cr
C.1 & General CRC Generator Polynomials \cr
C.1.1 & Case: $s \ge  h$ \cr
C.1.2 & Case: $s < h$ \cr
C.1.2.1 & Case: $s < h$ and $i_k \le  h - s$ \cr
C.1.2.2 & Case: $s < h$ and $i_k > h - s$ \cr
C.2 & CRC Generator Polynomials of Weight 3 \cr
C.3 & CRC Generator Polynomials of Weight 4 \cr
C.4 & CRC Generator Polynomials of Weights Greater Than 4 \cr
&		\cr
{\bf D} 	& {\bf CRC Weight Distributions \dotfill\the\secD} \cr
&		\cr
{\bf E} 	& {\bf CRC Parallel Implementation \dotfill\the\secE} \cr
&		\cr
& {\bf References \dotfill\the\ref}  \cr
}
}

\bye